\documentclass[traditabstract]{aa}
\pdfoutput=1
\usepackage{verbatim} 
\usepackage{amsmath}
\usepackage{amssymb}
\usepackage{graphicx}
\usepackage{booktabs}
\usepackage[authoryear]{natbib}
\usepackage[colorlinks=true,linkcolor=blue,urlcolor=blue,citecolor=blue]{hyperref}
\usepackage{xspace}
\usepackage{siunitx}
\usepackage{pdflscape}
\usepackage{longtable}
\usepackage{threeparttablex}

\makeatletter

\usepackage{txfonts}\usepackage{twoopt}
\bibpunct{(}{)}{;}{a}{}{,}
\usepackage{enumitem}

\graphicspath{{./}, {figures/}}

\newcommand{\msun}{{\rm M}_{\odot}}
\newcommand{\lsun}{{\rm L}_{\odot}}
\newcommand{\rsun}{{\rm R}_{\odot}}

\newcommand{\kms}{{\rm km\,s^{-1}}}
\newcommand{\myr}{{\rm Myr}}
\newcommand{\bethe}{{\rm B}}
\newcommand{\mesa}{\mbox{\textsc{Mesa}}\xspace}
\newcommand{\eg}{e.g.\@\xspace}
\newcommand{\cf}{c.f.\@\xspace}
\newcommand{\ie}{i.e.\@\xspace}
\newcommand{\snesa}{SN~1987A\xspace}
\newcommand{\casea}{Case~A\@\xspace}
\newcommand{\caseae}{Case~A-e\@\xspace}
\newcommand{\caseal}{Case~A-l\@\xspace}
\newcommand{\caseb}{Case~B\@\xspace}
\newcommand{\casebe}{Case~B-e\@\xspace}
\newcommand{\casebl}{Case~B-l\@\xspace}
\newcommand{\casec}{Case~C\@\xspace}
\newcommand{\caseab}{Case~AB\@\xspace}

\titlerunning{Pre-SN evolution and final fate of rapidly accreting stars}
\authorrunning{F.R.N.~Schneider et al.}
\makeatother
\begin{document}
\title{Pre-supernova evolution and final fate of stellar mergers and accretors of binary mass transfer}
\author{%
    F.R.N.~Schneider\inst{\ref{HITS},\ref{ZAH}}\thanks{fabian.schneider@h-its.org}
    \and Ph.~Podsiadlowski\inst{\ref{OXFORD},\ref{HITS}}
    \and E.~Laplace\inst{\ref{HITS}}
}
\institute{%
    Heidelberger Institut f{\"u}r Theoretische Studien, Schloss-Wolfsbrunnenweg 35, 69118 Heidelberg, Germany\label{HITS}
    \and Astronomisches Rechen-Institut, Zentrum f{\"u}r Astronomie der Universit{\"a}t Heidelberg, M{\"o}nchhofstr.\ 12-14, 69120 Heidelberg, Germany\label{ZAH}
    \and University of Oxford, St Edmund Hall, Oxford, OX1 4AR, United Kingdom\label{OXFORD}
}
\date{Received 1 September 2023 / Accepted 21 February 2024}
\abstract{The majority of massive stars are expected to exchange mass or merge with a companion during their lives. This immediately implies that most supernovae (SNe) are from such post-mass-exchange objects. Here, we explore how mass accretion and merging affect the pre-SN structures of stars and their final fates. To this end, we model these complex processes by rapid mass accretion onto stars of different evolutionary stages and follow their evolution up to iron core collapse. We use the stellar evolution code \mesa and infer the outcome of core-collapse using a neutrino-driven SN model. Our models cover initial masses from $11$ to $70\,\msun$ and the accreted mass ranges from 10--200\% of the initial mass. All models are non-rotating and for solar metallicity. The rapid-accretion model offers a systematic way to approach the landscape of mass accretion and stellar mergers. It is naturally limited in scope and serves as a clean zeroth order baseline for these processes. We find that mass accretion in particular onto post-main-sequence (post-MS) stars can lead to a long-lived blue supergiant (BSG) phase during which stars burn helium in their cores. In comparison to genuine single stars, post-MS accretors have small core-to-total mass ratios, regardless of whether they end their lives as BSGs or cool supergiants (CSGs), and they can have genuinely different pre-SN core structures. As in single and binary-stripped stars, we find black-hole (BH) formation for the same characteristic CO core masses $M_\mathrm{CO}$ of $\approx7\,\msun$ and $\gtrsim13\,\msun$. In models with the largest mass accretion, the BH-formation landscape as a function of $M_\mathrm{CO}$ is shifted by about $0.5\,\msun$ to lower masses, \ie such accretors are more difficult to explode. We find a tight relation between our neutron-star (NS) masses and the central entropy of the pre-SN models in all accretors and single stars, suggesting a universal relation that is independent of the evolutionary history of stars. Post-MS accretors explode both as BSGs and CSGs, and we show how to understand their pre-SN locations in the Hertzsprung--Russell (HR) diagram. Accretors exploding as CSGs can have much higher envelope masses than single stars. Some BSGs that avoid the luminous-blue-variable (LBV) regime in the HR diagram are predicted to collapse into BHs of up to $50\,\msun$ while others explode in supernovae and eject up to $40\,\msun$, greatly exceeding ejecta masses from single stars. Both the BH and supernova ejecta masses increase to about $80\,\msun$ in our models when allowing for multiple mergers, \eg, in initial triple-star systems, and they can be even higher at lower metallicities. Such high BH masses may fall into the pair-instability-supernova mass gap and could help explain binary BH mergers involving very massive BHs as observed in GW190521. We further find that some of the BSG models explode as LBVs, which may lead to interacting SNe and possibly even superluminous SNe.
}
\maketitle
%
%
%
%
\section{\label{sec:introduction}Introduction}

Most massive stars are born in binary, triple and other multiple-star systems
\citep[\eg][]{2007ApJ...670..747K, 2009AJ....137.3358M, 2012Sci...337..444S,
2012MNRAS.424.1925C, 2013A&A...550A.107S, 2014ApJS..213...34K,
2014ApJS..215...15S, 2017ApJS..230...15M, 2023ASPC..534..275O}. For example,
almost all O-type stars are found in binary systems and the triple and
higher-order multiple-star fraction of these stars even exceeds 50\%
\citep[\eg][]{2014ApJS..215...15S, 2017ApJS..230...15M, 2023ASPC..534..275O}. As
a consequence, the majority of massive stars will exchange mass with a companion
at some point in their evolution, \ie also most supernovae (SNe) are from stars
that evolved through past mass exchange with a companion
\citep[\eg][]{1992ApJ...391..246P, 2012Sci...337..444S, 2017ApJS..230...15M}. In
about 40--50\% of the cases, mass exchange proceeds via stable Roche-lobe
overflow (RLOF), but it can also lead to stellar mergers in about 20--30\% and
common-envelope phases in 10--20\% of cases\footnote{Some stars in multiple
systems never exchange mass with a companion and evolve as effectively single
stars.} \citep[see \eg][]{2012Sci...337..444S, 2015ApJ...805...20S,
2022MNRAS.516.1406S, 2023arXiv231112124H}.

During stable mass transfer and in common-envelope events, stars lose their
hydrogen-rich envelopes. These binary-stripped stars develop distinctly
different pre-SN core structures compared to single stars, affecting their final
fate \citep[\eg][]{1996ApJ...457..834T, 1999A&A...350..148W,
2001NewA....6..457B, 2004ApJ...612.1044P, 2019ApJ...878...49W,
2020ApJ...890...51E, 2021A&A...645A...5S, 2021A&A...656A..58L,
2023A&A...671A.134A, 2023ApJ...950L...9S}. For example, they explode in SN~Ib/c
and the inherently different pre-SN core structures result in neutron-star (NS)
formation over a much larger initial mass range than in genuine single stars;
they also explode with on average higher explosion energies, which results in
larger nickel yields and faster NS kicks \citep{2021A&A...645A...5S}; the
systematically higher nickel yields are confirmed by observations
\citep[\eg][]{2019A&A...628A...7A, 2020A&A...641A.177M}. Moreover, their chemical
yields differ from those of single stars \citep{2021ApJ...923..214F,
2023ApJ...948..111F}, and they may produce BHs of universal masses of about
$9\,\msun$ and $\gtrsim16\,\msun$ over a large range of metallicities
\citep{2023ApJ...950L...9S}. First hints for the existence of such universal BH
masses may have been found in the chirp and BH mass distributions of binary BH
mergers observed by gravitational-wave detectors
\citep[\eg][]{2021ApJ...913L..19T, 2021arXiv211103634T, 2022ApJ...928..155T,
2023ApJ...946...16E}.

Compared to the situation of binary-stripped stars, less is known about the
pre-SN core structures and the final fates of the accretors of mass transfer and
stellar mergers. It is estimated that about half of all Type II SNe are from
such accretors in binary systems, with mergers being the dominant
contribution \citep[\eg][]{2019A&A...631A...5Z}. Main-sequence (MS) accretors
likely rejuvenate and adjust their structure to the new mass such that they
continue their evolution similar to that of single stars
\citep[\eg][]{1983Ap&SS..96...37H, 1995A&A...297..483B, 2007MNRAS.376...61D,
2016MNRAS.457.2355S, 2019Natur.574..211S, 2020MNRAS.495.2796S}. In contrast,
post-MS accretors can become core-helium-burning blue supergiants (BSGs) and
explode as such stars \citep[][]{1983Ap&SS..96...37H, 1984Ap&SS.104...83H,
1989Natur.338..401P, 1990A&A...227L...9P, 2011A&A...528A.131C,
2013A&A...552A.105V, 2014ApJ...796..121J, 2017MNRAS.469.4649M, 2023arXiv231105581M}. They may,
however, also evolve and explode as cool supergiants (CSGs)\footnote{In
this paper, we mainly distinguish between blue and cool supergiants. The latter
includes red and yellow supergiants, and we define cool supergiants to have
effective temperatures of $\log T_\mathrm{eff}/\mathrm{K}<3.9$.}, but with
different core masses than what is expected from single CSGs
\citep[\eg][]{2021A&A...645A...6Z}. 

The well-studied \snesa was from a BSG that is likely the result of a merger
(see \eg reviews by \citealt{1992PASP..104..717P, 2017hsn..book..635P} and
some of the latest models for \snesa by \citealt{2017MNRAS.469.4649M},
\citealt{2018MNRAS.473L.101U} and \citealt{2019MNRAS.482..438M}). More massive
BSGs from stellar mergers have further been suggested to explode as luminous
blue variables (LBVs) in interacting and possibly superluminous SNe
\citep{2014ApJ...796..121J}, and hydrogen-rich, pair-instability supernovae
\citep[PISNe;][]{2019ApJ...876L..29V}. Furthermore, very massive merged stars
with cores not massive enough for electron-positron pair production but large
envelope masses might form BHs with masses in the pair-instability mass gap
\citep[\eg][]{2020MNRAS.497.1043D, 2020ApJ...904L..13R, 2022MNRAS.516.1072C,
2023MNRAS.519.5191B}.

In this paper, we systematically model the response of stars to rapid mass
accretion on a thermal timescale in an attempt to approximate the complex
physics occurring in accretors of stable RLOF in semi-detached binaries and
stellar mergers. We follow the evolution of these models until the onset of core
collapse to understand how such accretion events may modify the core structures
and hence affect the final fates and supernova explosions of these stars. While
our models are simplified and, \eg, do not take into account rotation and the
complex mixing in stellar mergers, they nevertheless allow us to study the
resulting structures of mass accretors in a uniform and controlled way. In light
of these approximations, we caution the reader when using these models. To
explore the outcome of core collapse, we use the pre-SN stellar structures of
our accretor models as input to the neutrino-driven SN model of
\citet{2016MNRAS.460..742M}. We describe how we model accretors in the stellar
evolution code \mesa \citep{2011ApJS..192....3P, 2013ApJS..208....4P,
2015ApJS..220...15P, 2018ApJS..234...34P, 2019ApJS..243...10P} and explain the
coupling to the SN code in Sect.~\ref{sec:methods}. We present our results on
the pre-SN evolution and core structures of the accretors at core collapse in
Sect.~\ref{sec:pre-sn-evolution-and-structure}. In
Sect.~\ref{sec:final-fate-sne}, we show our findings on the final fates of
accretors with a focus on their explodability, compact-remnant and SN-ejecta
masses, and pre-SN locations in the Hertzsprung--Russell (HR) diagram. Further
explosion properties will be presented in a forthcoming publication
\citep{Schneider+2024b}. We discuss our results concerning model simplifications
and uncertainties, and wider implications, \eg, on the maximum BH and SN-ejecta
masses from accretors at solar metallicity in Sect.~\ref{sec:discussion}. A
summary of our main findings is given in Sect.~\ref{sec:conclusions}.

%
%
\section{\label{sec:methods}Methods}

The stellar models are computed with revision 10398 of the
Modules-for-Experiments-in-Stellar-Astrophysics (\mesa) software package
\citep[][]{2011ApJS..192....3P, 2013ApJS..208....4P, 2015ApJS..220...15P,
2018ApJS..234...34P, 2019ApJS..243...10P}. We use the same basic \mesa setup as
in \citet{2021A&A...645A...5S} and briefly summarise it in
Sect.~\ref{sec:methods.mesa}. In Sect.~\ref{sec:methods.binary-models}, we
explain how we model binary-star accretion. For comparison of these accretion
models with genuine single stars, we employ the single-star models of
\citet{2021A&A...645A...5S}. In Sect.~\ref{sec:sn-code}, we briefly describe how
we couple our models to a SN code to study their final fates and possible SN
explosions. An overview of all models and key properties discussed in this work
are provided in Table~\ref{tab:models}.
\mesa inlists and other settings required to reproduce our models are
published online at \url{https://zenodo.org/doi/10.5281/zenodo.10731998}.

\subsection{\label{sec:methods.mesa}Stellar-model physics}

All models are non-rotating and have the initial chemical composition of
\citet{2009ARA&A..47..481A}, \ie an initial hydrogen mass fraction of
$X=0.7155$, helium mass fraction of $Y=0.2703$ and metallicity of $Z=0.0142$.
Convection is described by mixing-length theory \citep{1965ApJ...142..841H} and
we use a mixing-length parameter of $\alpha_\mathrm{mlt}=1.8$. The Ledoux
criterion is applied and semi-convection is treated with an efficiency factor of
$\alpha_\mathrm{sc}=0.1$. We enable \mesa's MLT++ scheme that boosts convective
energy transport in some stellar envelopes. Convective-boundary mixing is only
applied to convective cores during hydrogen and helium burning, and is included
via so-called ``step overshooting'' of $0.2$ pressure-scale height. We use the
stellar-wind prescription of \citet{2021A&A...645A...5S} that closely follows
\mesa's ``Dutch'' scheme. Cool-star winds follow \citet{1990A&A...231..134N},
while hot-star winds apply the \citet{2000A&A...362..295V, 2001A&A...369..574V}
rates. For Wolf--Rayet (WR) stars, we follow \citet{2000A&A...360..227N}. The
metallicity scaling of the cool-star and WR winds are as in
\citet{2011A&A...526A.156M} and \citet{2005A&A...442..587V}, respectively. No
additional wind-mass loss is applied to luminous blue variables (LBVs). However,
we mark models that spend more than $10^{4}\,\mathrm{yr}$ on the cool side
of the diagonal S~Doradus instability strip in the Hertzsprung--Russell (HR)
diagram \citep{2004ApJ...615..475S} and/or at effective temperatures
$T_\mathrm{eff}<12500\,\mathrm{K}$ while exceeding a luminosity of
$\log\,L/\lsun = 5.5$ \citep[\ie connecting the hot-part of S~Doradus
instability strip with the absence of red supergiants (RSGs) of $\log\,L/\lsun \gtrsim
5.5$ at $T_\mathrm{eff}=12500\,\mathrm{K}$,][]{2018MNRAS.478.3138D}; such stars
likely experience significant mass loss possibly leading to the removal of the
entire hydrogen-rich envelope. Nuclear burning is implemented via the
\texttt{approx21\_cr60\_plus\_co56.net} reaction network. While this network
properly reflects energy production by nuclear burning, it cannot follow
accurately the electron fraction or proton-to-neutron ratio because of, \eg,
missing weak reactions. Reaction rates themselves are from the JINA REACLIB
database V2.2 \citep{2010ApJS..189..240C} and, for example, the important
${}^{12}\mathrm{C}(\alpha,\gamma){}^{16}\mathrm{O}$ reaction rate is taken from
\citet{2013NuPhA.918...61X}.

\subsection{\label{sec:methods.binary-models}Binary-star accretion models}

\subsubsection{\label{sec:methods.accretion}Model assumptions}

Mass transfer by stable RLOF in semi-detached binaries and stellar mergers
are events during which the masses of stars increase rapidly. Hence, to zeroth
order, these processes may be viewed as accretion events where one star, called
accretor from hereon, effectively gains mass. The accretion timescales
during these events cover the entire spectrum from dynamic-timescale
accretion in stellar mergers to thermal- and nuclear-timescale accretion in
binary mass transfer. In stable \casea RLOF in semi-detached binaries,
most mass is often accreted on a thermal timescale
\citep[\eg][]{2001A&A...369..939W}, but this can vary from case to case. In our
accretion models, we accrete a certain fraction of the initial mass of a star,
$f_\mathrm{acc}\equiv \Delta M_\mathrm{acc}/M_\mathrm{ini}$, on the
momentary thermal timescale of the accretor, \ie at a rate of
\begin{align}
    \dot{M} = \frac{M}{\tau_\mathrm{KH}} = \frac{2 R L}{G M},
    \label{eq:accretion-rate}
\end{align}
where $L$, $R$ and $M$ are the momentary luminosity, radius and mass of the
accretor, respectively, $\tau_\mathrm{KH}$ is the Kelvin-Helmholtz timescale and
$G$ is the gravitational constant. The chemical composition of the accreted mass
is always the same as that on the surface of the accretor.

We compute models for accretion fractions $f_\mathrm{acc}$ of 10\%, 25\%, 50\%,
75\%, 100\%, 150\% and 200\%. The latter two fractions are not achievable by
stellar mergers in isolated binary evolution but can be achieved by stable
RLOF\footnote{The accretors in a semi-detached binary are initially less
massive than the donor and can possibly accrete more than their initial
mass, \eg in an initially $10+3\,\msun$ binary system.}. In stellar mergers,
$f_\mathrm{acc}>1$ may be possible in triple systems and dense stellar clusters
(see Sect.~\ref{sec:formation-massive-mergers} for further discussion). To
better distinguish these cases visually, we show the $f_\mathrm{acc}=150\%$ and
200\% models in blue colour in the rest of the paper. All other accretion
fractions are possible in both stable binary mass transfer and stellar
mergers.

As with envelope stripping by RLOF \citep[\eg][]{2021A&A...645A...5S}, the
future evolution of stars depends critically on the exact evolutionary moment
when the matter is accreted. In analogy to envelope stripping, we use the terms
\casea,~B and~C, but now applied to the evolutionary stage of the
\emph{accretor}. Hence, \casea, \caseb and \casec refer to stars that
accrete mass when they burn hydrogen in their cores, have finished core hydrogen
burning but have not started core helium burning yet, and have finished core
helium burning, respectively. We have defined the end of core helium burning as
that point in evolution when the central hydrogen mass fraction drops below
$10^{-6}$. We also distinguish between ``early'' and ``late'' \casea and~B. In
our case, early and late \casea accretion refer to when the central
hydrogen mass fraction of the accretor drops below 0.35 and 0.05, respectively.
In \caseb accretors, early means that the accretor has just left the main
sequence while late is when the accretor has developed a convective envelope and
is shortly before core helium ignition.

\subsubsection{\label{sec:methods.limitations}Model limitations}

The simple accretion models cannot reproduce all the important physics of
binary mass transfer and stellar mergers, and only account for one such phase
while several mass-accretion and mass-loss episodes are possible in general
\citep[see also][]{2021A&A...645A...5S}. Nevertheless, applying such models
allows us to understand and study the resulting stellar structures in a uniform
and controlled way, and our models offer a zeroth-order approximation to these
complex physical processes. We mention some limitations of our approach here and
further discuss them in Sect.~\ref{sec:discussion}.

Our models are non-rotating while in particular accretors of binary mass
transfer are likely spun up to near critical surface rotation \citep[see
\eg][]{2012ARA&A..50..107L}. Such rapid rotation can induce mixing of chemical
elements and angular momentum that we do not take into account. The mixing can
enlarge the cores of MS accretors and thereby affect the core structure at the
pre-SN stage. Moreover, rotation changes the mass-loss history of stars.

Multi-dimensional hydrodynamic simulations show that mixing also occurs in
stellar mergers \citep[\eg][]{2002ApJ...568..939L, 2002MNRAS.334..819I,
2013MNRAS.434.3497G, 2019Natur.574..211S}. It can enlarge the effective core
sizes in mergers of two MS stars increasing the effect of rejuvenation but can
also lead to smaller cores in mergers with post-MS stars. Moreover,
nuclear-processed core material of the two stars may be mixed in the outer
layers of the merger remnant, which we do not take into account.
Consequently, our models cannot reproduce the chemical surface enrichment
observed in such simulations \citep[\cf][]{2002ApJ...568..939L, 2002PhDT........25I,
2013MNRAS.434.3497G}. Our accretion models also neglect strong magnetic fields
that may be generated in the merging process and that can affect the evolution
of merged stars, \eg, via internal angular-momentum transport and magnetic
braking \citep{2019Natur.574..211S, 2020MNRAS.495.2796S}.

Lastly, our models cannot reproduce the surface chemical abundances expected
for products of binary mass transfer and stellar mergers. For example, they do
not show the nitrogen and helium enrichment from accreting nuclear-processed
material during the later accretion phases in stable binary mass transfer
\citep[see \eg review by][]{2012ARA&A..50..107L}. In some of our models, stellar
winds have removed parts of the stellar envelope such that helium-rich layers
are visible at the surface. In these cases, we then accrete helium-enriched
material.

\subsection{\label{sec:sn-code}Semi-analytic supernova model}

Pre-SN stellar structures are analysed with the semi-analytic, parametric-SN
code of \citet{2016MNRAS.460..742M} once the \mesa models reach an iron-core
infall velocity of $950\,\kms$. The code explores the explodability of pre-SN
stellar models within the framework of neutrino-driven SNe and takes into
account the entire interior structure of our models at core collapse. It
predicts SN properties such as explosion energy, synthesized nickel masses,
compact remnant masses and NS kick velocities. We apply the same calibrations of
this neutrino-driven-explosion model as in \citet{2021A&A...645A...5S}, \ie an
average explosion energy of Type IIP SNe from single-star models of
$0.69\pm0.17\,\bethe$ and a NS kick-velocity distribution that is in agreement
with the Maxwellian with $\sigma=265\,\kms$ of \citet{2005MNRAS.360..974H}. The
maximum, gravitational NS mass is chosen to be $2\,\msun$. 

We classify SN explosions as SN\,IIb if the mass of the hydrogen-rich envelope
is smaller than $1\,\msun$ following \citet{2019ApJ...885..130S}. The mass of
the hydrogen-rich envelope is defined as the layer with a hydrogen mass fraction
$X_\mathrm{H}>0.01$ and a helium mass fraction $X_\mathrm{He}>0.1$. If the
hydrogen mass in the SN ejecta is smaller than $0.1\,\msun$
\citep{2012MNRAS.422...70H}, we say that the SN is of Type Ib/c. We do not
distinguish between SN\,Ib and SN\,Ic in this work. In all other cases with more
hydrogen and larger hydrogen-rich envelopes, we define the SN to be of Type II.

%
%
\section{\label{sec:pre-sn-evolution-and-structure}Pre-supernova evolution and structure}

Rapid accretion of mass can change the stellar structure in such a way that the
accretors (merged stars) spend most of their core-helium-burning lifetime and
immediate pre-SN evolution as BSGs \citep[\eg][]{1983Ap&SS..96...37H,
1989Natur.338..401P, 1990A&A...227L...9P, 1995A&A...297..483B,
2011A&A...528A.131C, 2013A&A...552A.105V, 2014ApJ...796..121J,
2017MNRAS.469.4649M, 2018MNRAS.473L.101U}. We describe the occurrence and structure of such BSGs in
Sect.~\ref{sec:bsg-structures}. A star's final fate is mostly determined already
by the end of core helium burning and depends on the then CO core mass
$M_\mathrm{CO}'$ and central carbon mass fraction $X_\mathrm{C}$
\citep{2020ApJ...890...43C, 2020MNRAS.499.2803P, 2021A&A...645A...5S}. In our
models, the CO core masses at the end of core helium burning are almost
identical to those at core collapse, $M_\mathrm{CO}$. We show how core masses
such as $M_\mathrm{CO}$ and the important central carbon mass fraction
$X_\mathrm{C}$ are affected by mass accretion in
Sect.~\ref{sec:core-final-masses} and Sect.~\ref{sec:central-carbon-abundance},
respectively, before presenting the structures of our models at core collapse in
Sect.~\ref{sec:pre-sn-structure}.

\subsection{\label{sec:bsg-structures}Blue supergiants}

\begin{figure}
    \includegraphics{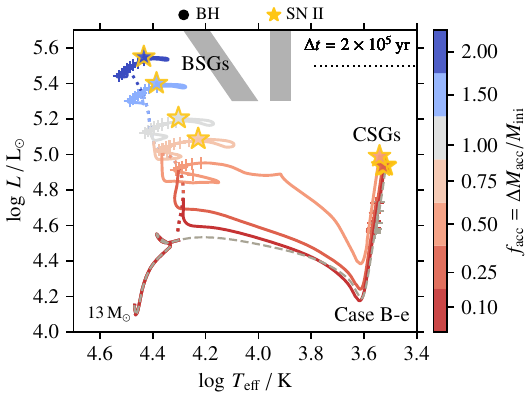}
    \caption{Hertzspung--Russell diagram of initially $13\,\msun$ stars accreting a fraction $f_\mathrm{acc}$ of their initial mass (color coded) shortly after they finished core hydrogen burning (early \caseb). The dotted parts of the tracks show the accretion phase and the grey dashed evolutionary track is for a single star. Plus symbols are added every $2\times10^5\,\mathrm{yr}$ after finishing core hydrogen burning to indicate in which region of the HR diagram stars burn helium in the core. The observationally-inferred maximum luminosity of RSGs of $\log L/\lsun=5.5$ \citep{2018MNRAS.478.3138D} is shown by a dotted black line. Explosion sites are indicated as well as the likely supernova type. The grey bands show the hot and cool part of the S~Doradus instability strip \citep{2004ApJ...615..475S}. For $f_\mathrm{acc}\gtrsim0.50$, the models burn helium as a BSG and for $f_\mathrm{acc}\gtrsim0.75$ they also explode as a BSG.}
    \label{fig:hrd-case-be-m13}
\end{figure}

In Fig.~\ref{fig:hrd-case-be-m13}, we show an exemplary Hertzsprung--Russell
(HR) diagram of initially $13\,\msun$ stars accreting mass shortly after
finishing core hydrogen burning (early \caseb\footnote{HR diagrams for \casea,
late \caseb and \casec accretors can be found in
Figs.~\ref{fig:hrd-case-ae-al-m13} and~\ref{fig:hrd-case-bl-c-m13},
respectively.}). We find three qualitatively different outcomes of mass
accretion: the accretor burns helium in its core (i) as a CSG
($f_\mathrm{acc}=0.10$, $0.25$), (ii) as a BSG but explodes as a CSG
($f_\mathrm{acc}=0.50$) and (iii) as a BSG and explodes as a BSG
($f_\mathrm{acc}\geq0.75$). Here, we define BSGs as supergiants with
effective temperatures of $\log\,T_\mathrm{eff}/\mathrm{K}>3.9$ while CSGs are
cooler than that. When burning helium in the core as a BSG, stars do so for a
nuclear timescale, \ie for several $10^5\text{--}10^6\,\mathrm{yr}$ (see
Figs.~\ref{fig:mini-facc-overview} and~\ref{fig:mini-facc-bsg-lifetime},
and Table~\ref{tab:models}), and we refer to them as long-lived
BSGs\footnote{Our single star models only spend a thermal timescale as
BSGs, which is 1--2 orders of magnitude shorter than these long-lived BSGs from
the accretor models.}. The required amount of accreted mass to turn stars into
BSGs depends on the mass and evolutionary stage of the accretors. For example,
none of our early \casea accretors turn into BSGs and only one model does so in
late \casea accretion ($f_\mathrm{acc}=0.25$;
Fig.~\ref{fig:hrd-case-ae-al-m13}). Most of the \casea models fully rejuvenate
and the accretors continue their evolution similarly to single stars with a
higher mass (see below for more details on the rejuvenation). In late \caseb
accretion, the models with $f_\mathrm{acc}=1.0$, $1.5$ and $2.0$ are BSGs during
core helium burning and also explode as BSGs, while this only occurs for
$f_\mathrm{acc}=1.5$ and $2.0$ in the \casec accretors
(Fig.~\ref{fig:hrd-case-bl-c-m13}). In general, we find that the more evolved
the stars are, the more mass needs to be accreted to turn them into BSGs. 

We have not considered the mixing of helium-rich material out of the core into
the envelope. It is well known that the dredge-up of helium favours the
occurrence of BSG structures \citep{1988ApJ...332..247B, 1988ApJ...331..388S}.
Therefore, in the context of a stellar merger, the combination of the addition
of matter to the envelope and the dredge-up of part of the helium-rich core,
which increases the helium abundance in the evelope, but also further decreases
the core to total mass ratio, is particularly favourable for producing BSGs
\citep[\eg][]{1989A&A...219L...3H, 1990A&A...227L...9P, 1992ApJ...391..246P,
2002Ap&SS.281..191I, 2017MNRAS.469.4649M, 2017arXiv170203973P}. The more recent,
systematic study of such mergers by \citet{2017MNRAS.469.4649M} also shows that
the ultimate supernova explosions resulting from such mergers can reproduce the
properties of peculiar supernovae, such as \snesa, much better than single BSG
progenitors \citep{2019MNRAS.482..438M}. Similarly, merger products also
better reproduce the observed population of BSGs \citep[][]{2023A&A...677A..50B,
2023arXiv231105581M}. Since we do not include these dredge-up effects, this
means that the accretion thresholds for BSG formation mentioned above are upper
limits when compared to actual stellar mergers. Interior mixing from stellar
mergers also affects \casea mergers, which is not accounted for in our models
either but, \eg, can further enhance complex morphologies in star clusters above
and around the MS turn-off region \citep[see \eg][]{2020ApJ...888L..12W}. In
\caseb and~C accretors of binary mass transfer, such mixing is not expected to
occur, and the extra mixing and growth of the convective core in \casea
accretors is covered by our models. We show an overview of our accretor model
grid in Fig.~\ref{fig:mini-facc-overview} that illustrates and summarises which
of our models evolve and end their lives as CSGs and BSGs. For further,
general considerations of stellar structures that may spend time as a BSG, see
\eg \citet{2019A&A...625A.132S}, \citet{2022MNRAS.512.4116F} and
\citet{2023A&A...680A.101S}.

\begin{figure*}
    \centering
    \includegraphics[width=0.9\textwidth]{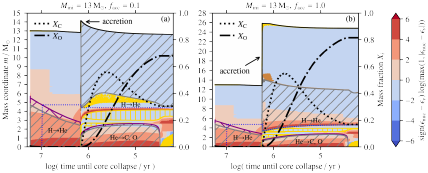}
    \caption{Kippenhahn diagrams of initially $13\,\msun$ early \caseb systems accreting (a) 10\% and (b) 100\% of their initial mass. The moment of mass accretion is indicated by an arrow. Energy generation by nuclear burning, $\epsilon_\mathrm{nuc}$, and loss by neutrinos, $\epsilon_\nu$, are color-coded. The right $y$-axes show the mass fractions of carbon, $X_\mathrm{C}$, and oxygen, $X_\mathrm{O}$, in the center of the stars. Carbon starts being produced when core helium burning begins. The grey hatched areas show convective regions, purple hatchings are for convective overshooting, yellow stands for thermohaline mixing and brown is for semi-convection. The blue dotted lines show where the hydrogen mass fraction drops below 0.5, \ie they trace the helium core mass. Hydrogen shell burning in the $f_\mathrm{acc}=0.1$ model is radiative and the star evolves into a CSG. In the $f_\mathrm{acc}=1.0$ model, hydrogen shell burning proceeds convectively and the star evolves as a BSG (\cf Fig.~\ref{fig:hrd-case-be-m13}).}
    \label{fig:kipp-comparison-m13}
\end{figure*}

The interior structures of the long-lived BSGs are characterised by hydrogen
shell burning driving a large convective region
(Fig.~\ref{fig:kipp-comparison-m13}). This occurs if the accreted mass is so
large that the temperature at the base of the hydrogen-rich envelope becomes
high enough to drive convection by large nuclear energy generation from hydrogen
shell burning (\cf Sect.~\ref{sec:hrd-explosion-sites}). For
$f_\mathrm{acc}=0.1$, the accretion modifies the temperature stratification at
the base of the hydrogen-rich envelope such that hydrogen shell burning is
pushed to lower mass coordinates but does not turn convective
(Fig.~\ref{fig:kipp-comparison-m13}a). Contrarily, for $f_\mathrm{acc}=1.0$,
shell burning becomes convective (Fig.~\ref{fig:kipp-comparison-m13}b). The
structure of such BSGs may be viewed as that of a (relatively) unevolved MS star
on top of a helium star. In this picture, the convective hydrogen-burning shell
corresponds to the core of the MS star. This analogous picture also explains
intuitively why the envelope of such BSGs remains radiative and why the extent
of the convective hydrogen-burning shell decreases during core helium
burning\footnote{Because of a continuously decreasing electron-scattering
opacity from converting hydrogen into helium, \ie for the same reason why the
convective cores of MS stars recede during hydrogen burning.}
(Fig.~\ref{fig:kipp-comparison-m13}b).

\begin{figure}
    \centering
    \includegraphics[width=1.0\linewidth]{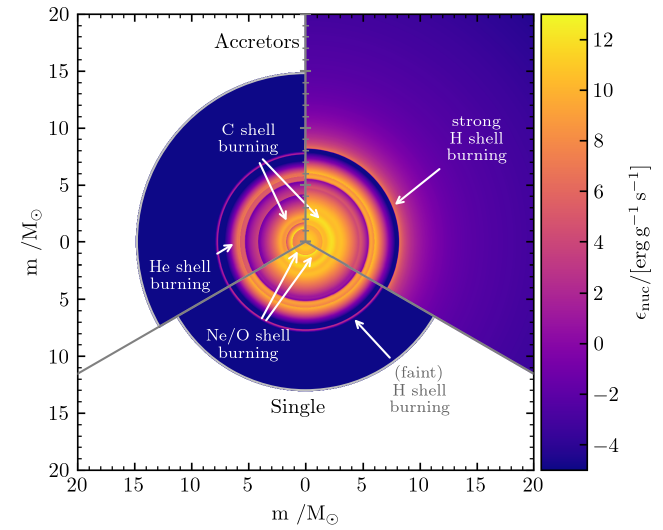}
    \caption{TULIPS diagram \citep{2022A&C....3800516L} of the nuclear energy generation rate $\epsilon_\mathrm{nuc}$ at the end of core oxygen burning (central oxygen mass fraction $<10^{-4}$). Shown are the interiors of a single star of $M_\mathrm{ini}=20\,\msun$ (bottom), an early \caseb accretor model of $M_\mathrm{ini}=20\,\msun$ and $f_\mathrm{acc}=0.1$ (top left), and an early \caseb accretor model of $M_\mathrm{ini}=23\,\msun$ and $f_\mathrm{acc}=1.0$ (top right). The three models have a similar CO core masses of $M_\mathrm{CO}=5.62\,\msun$, $5.64\,\msun$ and $5.89\,\msun$, respectively. The $f_\mathrm{acc}=0.1$ accretor model is a CSG while the $f_\mathrm{acc}=1.0$ model is a BSG, which can be recognized by the strong hydrogen-shell burning that drives a thick convection zone (not indicated here for clarity, but \cf Figs.~\ref{fig:kipp-comparison-m13} and~\ref{fig:kipp-comparison-m30}).}
    \label{fig:tulips-comparison-same-MCO}
\end{figure}

To further illustrate the structural differences between stars evolving
through a long-lived BSG phase and those becoming CSGs, we compare in
Fig.~\ref{fig:tulips-comparison-same-MCO} the interior nuclear burning at the
end of core oxygen burning of an initially $20\,\msun$ single star and two early
\caseb accretors with initial masses of $20\,\msun$ and $23\,\msun$, and
accretion fractions of $f_\mathrm{acc}=0.1$ and $f_\mathrm{acc}=1.0$,
respectively. All models have similar CO core masses of
$M_\mathrm{CO}=5.62\,\msun$, $5.64\,\msun$ and $5.89\,\msun$, respectively, but
evolve differently. In particular, the accretor model with $f_\mathrm{acc}=1.0$
evolves as a long-lived BSG while the other two models become CSGs. The
largely different hydrogen-shell burning in the BSG $f_\mathrm{acc}=1.0$
accretor is visible (\cf Fig.~\ref{fig:kipp-comparison-m13}) as well as markedly
different strengths, mass coordinates and extensions in mass of the nuclear
reaction rates due to carbon and neon burning (helium burning is also
different). Carbon and neon burning are particularly important for the final
core structures (see also Sect.~\ref{sec:central-carbon-abundance}). As is shown
later, such differences can distinguish between successful SN explosions or
direct collapse into a BH.

\begin{figure}
    \centering
    \includegraphics{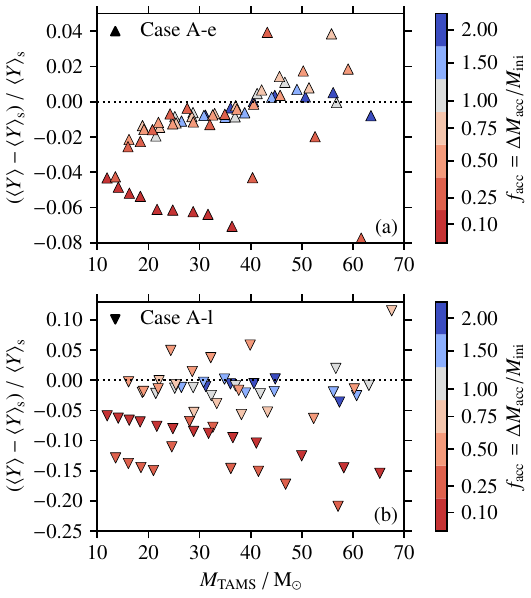}
    \caption{Rejuvenation of early (panel a) and late (panel b) \casea accretors. The rejuvenation is quantified as the difference of the average helium mass fraction $\left<Y\right>$ of the accretors at the terminal-age main sequence (TAMS) relative to that of single stars. The relative differences are plotted as a function of mass at the TAMS, $M_\mathrm{TAMS}$. Negative differences mean that the accretors have burnt less hydrogen into helium compared to single stars, \ie that they have less massive convective cores. This is a sign that accretors did not (fully) adjust their structure to the new total mass after accretion, \ie that they did not (fully) rejuvenate.}
    \label{fig:rejuvenation-case-a}
\end{figure}

Whether \casea accretors can become long-lived BSGs depends on whether they
adjust their interior structure to the new mass, \ie whether they rejuvenate. If
an accreting MS star fully rejuvenates, the convective core will have the same
size as that of a single star with the same mass. This means that also the
helium core masses at the terminal-age main sequence (TAMS) will be the same and
thus also the average helium mass fraction $\left<Y\right>$. The relative
deviation of $\left<Y\right>$ at the TAMS between our \casea accretors and
genuine single stars shows that early \casea accretors (almost) fully rejuvenate
for accretion fractions larger than ${\approx}\,10\%$
(Fig.~\ref{fig:rejuvenation-case-a}a). The same is true for late \casea
accretors for $f_\mathrm{acc}\gtrsim25\%$ (Fig.~\ref{fig:rejuvenation-case-a}b).
These findings and in particular that late \casea accretors need to accrete more
mass than early \casea accretors to fully rejuvenate agree qualitatively with
the results of \citet{1995A&A...297..483B}. As explained by
\citeauthor{1995A&A...297..483B}, semi-convection regulates whether a convective
core can grow and mix through chemical gradients upon mass accretion. The
chemical gradients are steeper at a later MS phase such that more accretion is
needed to allow semi-convection to mix through the gradient and the convective
core to grow to a value found in genuine single stars of the same mass. Compared
to the models of \citet{1995A&A...297..483B}, semi-convection with
$\alpha_\mathrm{sc}=0.1$ is much more efficient in our models such that most of
them tend to fully rejuvenate. According to \citet{2019A&A...625A.132S},
semi-convection may even be more efficient than assumed here such that all
\casea accretors would fully rejuvenate.

In our models, not all \casea accretors fully rejuvenate
(Fig.~\ref{fig:rejuvenation-case-a}). However, this does not automatically mean
that they can also burn helium in their cores as BSGs. To evolve into long-lived
BSGs, they must also reach conditions to drive convective hydrogen shell
burning. The initially $13\,\msun$ late \casea accretor with
$f_\mathrm{acc}=0.25$ does not fully rejuvenate
(Fig.~\ref{fig:rejuvenation-case-a}b) and has accreted sufficient mass such that
it becomes a BSG during core helium burning
(Fig.~\ref{fig:hrd-case-ae-al-m13}b). Contrarily, the $f_\mathrm{acc}=0.10$
model also does not fully rejuvenate but accretes too little mass to develop
convective hydrogen shell burning (Fig.~\ref{fig:hrd-case-ae-al-m13}b).

\subsection{\label{sec:core-final-masses}Helium core, carbon-oxygen core and final stellar masses}

In the evolution of stars, the core mass plays an important role. Here, we
concentrate on the helium and carbon-oxygen (CO) core masses $M_\mathrm{He}$ and
$M_\mathrm{CO}$, respectively, that are mostly set by the extent of convection
during core hydrogen and helium burning. Both core masses can vary during the
evolution (\eg due to shell burning or dredge-ups) and we measure them at the
onset of core collapse in this work. We define $M_\mathrm{He}$ as the mass coordinate
where the hydrogen mass fraction $X_\mathrm{H}$ drops below $<10^{-5}$ while
$M_\mathrm{CO}$ is defined by the helium mass fraction $X_\mathrm{He}$ becoming
smaller than $0.5$.

\begin{figure*}
    \centering
    \includegraphics{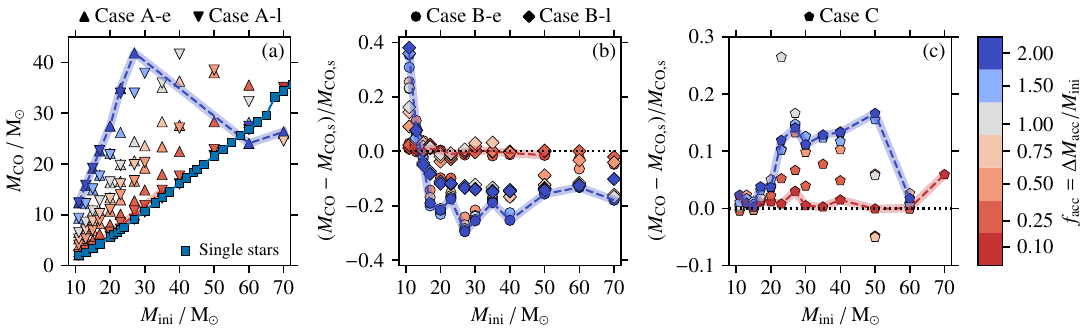}
    \caption{CO core masses $M_\mathrm{CO}$ at core collapse as a function of initial mass $M_\mathrm{ini}$ for \casea (panel a), \caseb (panel b) and \casec (panel c) accretors. Colours show the accretion fraction $f_\mathrm{acc}$ of the models, and single stars are shown in blue squares for comparison in panel (a). Panels (b) and (c) show the difference in $M_\mathrm{CO}$ relative to the corresponding single-star core masses. To guide the eye, data points for $f_\mathrm{acc}=0.1$ (mostly CSGs) and $2.0$ (BSGs) are connected by dashed lines and highlighted.}
    \label{fig:MCO-vs-Mini}
\end{figure*}

\begin{figure*}
    \centering
    \includegraphics{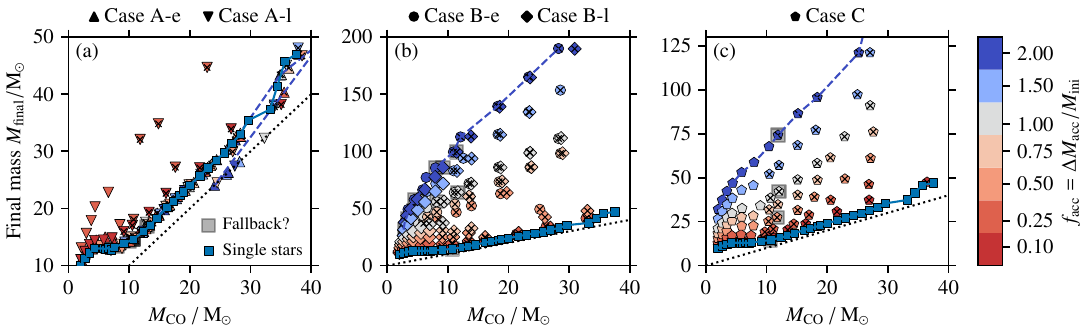}
    \caption{Similar to Fig.~\ref{fig:MCO-vs-Mini} but showing the final masses of stars as a function of the CO core mass $M_\mathrm{CO}$ at core collapse. In all three panels, single-star models are plotted as blue squares. We further indicate models that experience SN fallback by grey squares while crosses mark those models that should have experienced enhanced LBV-like mass loss, which is not accounted for in our models (see Sect.~\ref{sec:methods.mesa}). The black dotted line is a one-to-one relation.}
    \label{fig:MCO-vs-Mf}
\end{figure*}

In Fig.~\ref{fig:MCO-vs-Mini}, we show $M_\mathrm{CO}$ as a function of initial
mass $M_\mathrm{ini}$ for single stars and \casea, B and~C accretors. The
trends for the helium core masses are qualitatively similar to those of the CO
core masses and are therefore not shown in the main text but only in
Fig.~\ref{fig:MHe-vs-Mini}. Most of our \casea accretors fully adjust their
interior structures to the new mass and then continue the evolution as if they
were born with that high mass, \ie most of them fully rejuvenate (see
Fig.~\ref{fig:rejuvenation-case-a} and Sect.~\ref{sec:bsg-structures}). This is
exemplified by the increasing helium and CO core masses at constant initial mass
and increasing accretion fractions. The rejuvenation can even be seen more
directly in Fig.~\ref{fig:MCO-vs-Mf}, where we plot the masses of stars at core
collapse, $M_\mathrm{final}$, as a function of $M_\mathrm{CO}$. \casea accretors
mostly follow the trend of single stars, \ie they produce the same core mass for
a given total final mass, except for the models that do not fully rejuvenate and
have higher final masses (\cf Fig.~\ref{fig:rejuvenation-case-a}). This
deviation from the single-star final masses is larger in the late than the early
\casea models. In particular, some non-rejuvenating, late \casea accretors
(partly) burn helium in their cores as BSGs and not CSGs
\citep[][]{1995A&A...297..483B}. In turn, they avoid the stronger CSG wind
mass loss and retain higher final masses at core collapse.

Some \casea accretor models with $f_\mathrm{acc}\gtrsim0.75$ at
$M_\mathrm{CO}>20\,\msun$ appear to be offset to higher $M_\mathrm{CO}$ relative
to the single stars (Figs.~\ref{fig:MCO-vs-Mini}a and
~\ref{fig:MCO-vs-Mf}a). These models are so massive after accretion that their
strong winds erode their envelopes and the final mass is the same as the CO core
mass. At $M_\mathrm{CO}>10\,\msun$, winds have removed most of the hydrogen-rich
envelope, explaining the almost linear relation of single stars and \casea
accretors in Fig.~\ref{fig:MCO-vs-Mf}a.

\caseb and~C accretors cannot rejuvenate as is visible by the relative
difference of their helium and CO core masses in comparison to single stars
(Figs.~\ref{fig:MCO-vs-Mini}b and~\ref{fig:MHe-vs-Mini}b). We find that they
differ at most by an order of $\mathcal{O}(10\%)$. This immediately implies
that -- when comparing to single stars -- such accretors roughly keep their core
masses constant while mass is mostly added to their envelopes. One may thus
expect that, to zeroth order, the cores of such accretors evolve similarly to
core collapse as those of the corresponding single stars with the same initial
mass. This expectation is met in accretors of binary mass transfer but not
in fully realistic stellar mergers. In the latter, parts of the core are mixed into the
envelope of the merged star, which reduces the core mass. This effect is not
covered by our models (see Sect.~\ref{sec:methods.limitations}), \ie the core
masses of our \caseb and~C accretors are upper limits in comparison to what is
expected from fully realistic stellar mergers.

In \caseb accretors, the CSG and BSG sequences can be distinguished in
Fig.~\ref{fig:MCO-vs-Mini}b and~\ref{fig:MHe-vs-Mini}b: models with
$f_\mathrm{acc}=0.1$ are generally CSGs while those with
$f_\mathrm{acc}=2.0$ always become BSGs (see
Fig.~\ref{fig:mini-facc-overview} for a detailed overview of which minimum
accretion fractions are required in our models to turn accretors into BSGs).
Post-accretion CSGs hardly deviate from the core masses of the single stars
while the BSGs deviate by a few tens of percent in $M_\mathrm{He}$ and
$M_\mathrm{CO}$; they follow the indicated trend for $f_\mathrm{acc}=2.0$. For
$M_\mathrm{ini}\lesssim15\,\msun$, the \caseb accretors have helium and CO core
masses at core collapse that are larger than those of single stars while they
are smaller for $M_\mathrm{ini}\gtrsim15\,\msun$\footnote{Because of
overshooting, the models studied here show no drastic second dredge-up. Thus,
the larger $M_\mathrm{He}$ and $M_\mathrm{CO}$ of the accretors compared to
single stars are not related to a second dredge-up in the single stars which is
naturally absent in BSG accretors without a deep convective envelope
\citep[\cf][]{2004ApJ...612.1044P}.}. For all initial masses, the convective
cores of accretors \emph{at the beginning of core helium burning} are less
massive than those of single stars and the location of the hydrogen-burning
shell is at lower mass coordinates (\ie smaller $M_\mathrm{He}$ at this time in
evolution as can be seen in Fig.~\ref{fig:kipp-comparison-m13}). Because of the
hydrogen shell burning, the helium core and with it the convective core grow in
mass. In \caseb accretors, the convective core grows relatively more in less
massive stars. For example, for $M_\mathrm{ini}\gtrsim30\,\msun$, the convective
core masses remain almost constant during core helium burning
(Fig.~\ref{fig:kipp-comparison-m30}) while they grow significantly for
$M_\mathrm{ini}\lesssim15\,\msun$ (Fig.~\ref{fig:kipp-comparison-m13}).

In \casec accretors, there is not enough remaining time before core collapse for
hydrogen and helium shell burning to add significantly to $M_\mathrm{He}$ and
$M_\mathrm{CO}$. This is indeed borne out by the helium core masses being almost
unchanged (they change by an order of $\mathcal{O}(1\%)$) regardless of
whether stars evolve as CSGs or BSGs (Fig.~\ref{fig:MHe-vs-Mini}c). The
picture is seemingly different for the CO core masses at
$M_\mathrm{ini}\gtrsim20\,\msun$ that can exceed the single-star $M_\mathrm{CO}$
by an order of $\mathcal{O}(10\%)$. This increase is not caused by helium
shell burning but by a brief revival of core helium burning\footnote{Note that
$X_\mathrm{He}=10^{-6}$ at the point of mass accretion and core helium burning
has almost ceased but is still ongoing.} because of the mass accretion (see also
Sec.~\ref{sec:central-carbon-abundance}). The nuclear burning drives a transient
convective core that mixes in fresh helium fuel, thereby leading to more massive
$M_\mathrm{CO}$. For $M_\mathrm{ini}\lesssim20\,\msun$, we do not find such a
strong revival of core helium burning and $M_\mathrm{CO}$ does not grow as much.

The final masses of rejuvenated \casea accretors closely follow those of single
stars of the same CO core mass, which is not the case in non-rejuvenating
\casea and most \caseb accretors (Fig.~\ref{fig:MCO-vs-Mf}). In the latter
cases, stars retain more mass and we find final stellar masses of up to
$100\text{--}200\,\msun$ in our models. So while $M_\mathrm{He}$ and
$M_\mathrm{CO}$ change at most by some tens of percent in comparison to single
stars, the final core-to-stellar mass ratios can be reduced by up to an order of
magnitude: \caseab accretors have comparably small cores for their total stellar
mass.

As mentioned in Sect.~\ref{sec:methods}, our models do not account for LBV-like
mass loss. Even disregarding all models that likely experience significant LBV
mass loss (marked by crosses in Fig.~\ref{fig:MCO-vs-Mf}), \caseb accretors have
the largest final stellar masses, followed by \casec and \casea accretors (and
single stars). \casec accretors generally have smaller final masses than \caseb
accretors for the following reasons. Firstly, when accretion starts, \casec
models have been subject to strong wind mass loss on the RGB branch while \caseb
accretors can avoid this entirely as BSGs and thus reach larger final masses.
Secondly, even if \caseb accretors do not evolve through a long-lived BSG phase
but also become CSGs, they have smaller helium and CO core masses than \casec
accretors and single stars. As we will show below, the luminosity of CSGs scales
with their core mass such that \casec accretors and single stars have stronger
wind mass loss and smaller final masses because of their higher luminosities. We
will further discuss the final stellar masses in relation to BH and SN-ejecta
masses in Sect.~\ref{sec:compact-remnants-ejecta-masses}.

\subsection{\label{sec:central-carbon-abundance}Central carbon abundance after core helium burning}

As explained above, the central carbon mass fraction $X_\mathrm{C}$ after core
helium burning is an important parameter that sets the initial conditions for
the later nuclear-burning stages and thereby the pre-SN core structure. It is
determined by helium burning that proceeds in two steps. First, the
triple-$\alpha$ reaction produces $^{12}\mathrm{C}$. Second, once
$\alpha$-particles become rarer and $^{12}\mathrm{C}$ nuclei have been formed,
the $^{12}\mathrm{C}(\alpha,\gamma)^{16}\mathrm{O}$ nuclear reaction eventually
becomes dominant and consumes $^{12}\mathrm{C}$ to its final abundance
$X_\mathrm{C}$ \citep[see \eg][and Figs.~\ref{fig:kipp-comparison-m13}
and~\ref{fig:kipp-comparison-m30}]{1989A&A...210...93L, 1993ApJ...411..823W,
2001NewA....6..457B, 2020ApJ...890...43C, 2020MNRAS.492.2578S,
2021A&A...656A..58L}. How much carbon is depleted depends on the central
density\footnote{The triple-$\alpha$ reaction is proportional to the third power
of the $\alpha$-particle number density while the
$^{12}\mathrm{C}(\alpha,\gamma)^{16}\mathrm{O}$ reaction scales linearly with
this density. More massive stars have denser cores such that the latter reaction
becomes dominant earlier during helium burning than in lower-mass stars, and
more carbon is depleted and converted into oxygen.}, \ie the (core) mass of the
star, and how much the convective core grows during helium burning, which can
mix in fresh $\alpha$-particles that are then captured onto $^{12}\mathrm{C}$ to
produce $^{16}\mathrm{O}$.

\begin{figure*}
    \centering
    \includegraphics{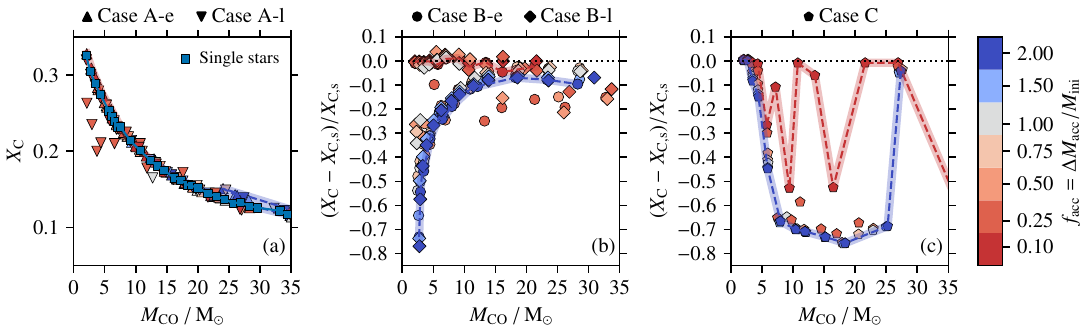}
    \caption{Similar to Fig.~\ref{fig:MCO-vs-Mini} but showing the central carbon mass fraction at the end of core helium burning $X_\mathrm{C}$ as a function of the CO core mass $M_\mathrm{CO}$.}
    \label{fig:XC-vs-MCO}
\end{figure*}

We show the carbon mass fraction $X_\mathrm{C}$ at the end of core helium
burning of our models as a function of the CO core mass in
Fig.~\ref{fig:XC-vs-MCO} (a similar diagram as a function of the initial
mass can be found in Fig.~\ref{fig:XC-vs-Mini}). As discussed in
Sect.~\ref{sec:core-final-masses}, most \casea accretors fully rejuvenate and
adjust their interior structure to the new total mass such that they have the
same $X_\mathrm{C}$ as in single stars of the same $M_\mathrm{CO}$
(Fig.~\ref{fig:XC-vs-MCO}a). We also find again that those \casea accretors,
which spend most of their time burning helium in their cores as BSGs, deviate
from the single-star trend and deplete relatively more $^{12}\mathrm{C}$. The
reason is the same as in \caseb accretors discussed next.

In \caseb accretors, we find that the CSG and BSG sequences differ considerably
in $X_\mathrm{C}$ (Fig.~\ref{fig:XC-vs-MCO}b). The post-accretion BSG
models roughly follow the $f_\mathrm{acc}=2.0$ models and end core helium
burning with less carbon than single stars while the CSG models follow the
single-star trends. We find that \caseb accretors with
$M_\mathrm{CO}\lesssim15\,\msun$ ($M_\mathrm{ini}\lesssim40\,\msun$) can have up
to 80\% less carbon in their cores than single stars while this is ${<}\,10\%$
for $M_\mathrm{CO}\gtrsim15\,\msun$. 

The differences are due to different growths of the convective cores during
helium burning. At $M_\mathrm{ini}=13\,\msun$
(Fig.~\ref{fig:kipp-comparison-m13}), the convective core driven by helium
burning still grows considerably more in mass in the $f_\mathrm{acc}=1.0$ model
after reaching a maximum carbon mass fraction of ${\approx}\, 0.55$ than in the
$f_\mathrm{acc}=0.1$ model. The larger core growth during this phase mixes more
$\alpha$-particles into the convective core and the
$^{12}\mathrm{C}(\alpha,\gamma)^{16}\mathrm{O}$ nuclear reaction thus burns more
$^{12}\mathrm{C}$ into $^{16}\mathrm{O}$. The $f_\mathrm{acc}=1.0$ model then
ends core helium burning with $X_\mathrm{C}\approx0.13$ while the
$f_\mathrm{acc}=0.1$ model has a larger abundance of $X_\mathrm{C}\approx0.30$
(Fig.~\ref{fig:kipp-comparison-m13}). For a higher initial mass of $30\,\msun$,
the convective cores in the $f_\mathrm{acc}=0.1$ and $f_\mathrm{acc}=1.0$ models
grow similarly after reaching the maximum core carbon mass fraction of
${\approx}\, 0.55$ (Fig.~\ref{fig:kipp-comparison-m30}). Consequently, similar
amounts of $\alpha$-particles are mixed into the convectively helium-burning
cores such that both models reach similar final core carbon mass fractions of
about $0.2$.

The carbon mass fractions $X_\mathrm{C}$ of the $f_\mathrm{acc}=2.0$ early
\caseb accretors never reach the same values in single stars as a function
of the CO core mass (Fig.~\ref{fig:XC-vs-MCO}b) while they are the same as in
single stars as a function of initial mass for
$M_\mathrm{ini}\,{\gtrsim}\,35\,\msun$ (Fig.~\ref{fig:XC-vs-Mini}b). At first
sight, this may sound contradictory but it is not. The accretor models follow a
different $M_\mathrm{CO}$--$M_\mathrm{ini}$ relation than single stars. For
$M_\mathrm{ini}>15\,\msun$, they have ${\approx}\,20\%$ smaller CO cores than
single stars of the same initial mass (Fig.~\ref{fig:MCO-vs-Mini}b). Generally,
$X_\mathrm{C}$ drops with increasing $M_\mathrm{CO}$ and $M_\mathrm{ini}$
(Figs.~\ref{fig:XC-vs-MCO} and~\ref{fig:XC-vs-Mini}). At the same initial
mass, both single stars and \caseb accretors have the same $X_\mathrm{C}$.
Translating into a $X_\mathrm{C}$--$M_\mathrm{CO}$ relation, the $X_\mathrm{C}$
of the accretor models are then compared to those of single stars of
${\approx}\,20\%$ smaller $M_\mathrm{CO}$, \ie higher $X_\mathrm{C}$.

In \casec accretors, we find $X_\mathrm{C}$ values reduced by ${\approx}\,70\%$
for $7\lesssim M_\mathrm{CO}/\msun \lesssim 25$ and $f_\mathrm{acc}\gtrsim0.1$
(Fig.~\ref{fig:XC-vs-MCO}c). In these models, accretion increases the
star's central temperature. Helium burning is revived and drives a transient
convective core that mixes in fresh helium fuel. As
$^{12}\mathrm{C}(\alpha,\gamma)^{16}\mathrm{O}$ is the dominant reaction in
these models at this moment of evolution, significantly more carbon is converted
into oxygen compared to the equivalent single-star models. Once core helium
burning ceases entirely, $X_\mathrm{C}$ is reduced to 4--6\%. In the other
models with different $M_\mathrm{CO}$, this mechanism does not operate and
$X_\mathrm{C}$ remains close to that found in the corresponding single-star
models.

As explained above, fully realistic \caseb and~C stellar mergers have smaller cores
than in our accretion models. The same mechanism of mixing fresh helium
into the helium-burning core will also operate in such stars. Hence, we expect
that also such stars show different $X_\mathrm{C}$ branches depending on whether
the merged star becomes a long-lived BSG or evolves into a CSG.

\subsection{\label{sec:pre-sn-structure}Pre-SN stellar structure}

The different starting conditions for the later nuclear-burning stages, \ie
$M_\mathrm{CO}$ and $X_\mathrm{C}$ (see Sects.~\ref{sec:core-final-masses}
and~\ref{sec:central-carbon-abundance}), can lead to inherently different core
structures during the advanced burning stages and at core collapse. We summarise the pre-SN stellar structures of our models by their iron core
mass, $M_\mathrm{Fe}$, the central specific entropy, $s_\mathrm{c}$, and the
so-called compactness parameter,
\begin{align}
    \xi_M = \frac{M/\msun}{R(M)/1000\,\mathrm{km}},
    \label{eq:compactness}
\end{align}
which is the ratio of a mass coordinate $M$ and the radius at this location
$R(M)$ \citep[][]{2011ApJ...730...70O}. Usually, $M=2.5\,\msun$, and
$\xi_{2.5}{\gtrsim}\,0.45$ indicates that stars are unable to explode in SNe and
form BHs \citep[\eg][]{2011ApJ...730...70O, 2012ApJ...757...69U,
2014ApJ...783...10S, 2016ApJ...818..124E, 2016ApJ...821...38S,
2016MNRAS.460..742M, 2020ApJ...890...43C, 2020MNRAS.499.2803P,
2021A&A...645A...5S, 2023ApJ...950L...9S}. $M_\mathrm{Fe}$, $s_\mathrm{c}$ and
$\xi_\mathrm{2.5}$ are closely related and they provide similar insights into
the pre-SN structure \citep[see also][]{2014frap.confE...4F, 2021A&A...645A...5S,
2023ApJ...945...19T, 2024A&A...682A.123T}. For example, the effective Chandrasekhar mass
that an iron core approaches when it starts to collapse is a function of the
electron fraction $Y_\mathrm{e}$ and electronic entropy $s_\mathrm{e,c}$
\citep[the average electronic entropy is roughly
$\left<s_\mathrm{e,c}\right>\approx \left<s_\mathrm{c}\right>/3$ but depends on
each case; see \eg][]{1990ApJ...353..597B, 1996ApJ...457..834T},
\begin{align}
    M_\mathrm{Ch} \approx 1.457\,\msun\,\left(\frac{Y_\mathrm{e}}{0.5}\right)^2 \left[ 1 + \left(\frac{s_\mathrm{e,c}}{\pi Y_\mathrm{e} k_\mathrm{B} N_\mathrm{A}}\right)^2 \right].
    \label{eq:eff-Mch}
\end{align}
This relation directly links $s_\mathrm{c}$ to $M_\mathrm{Fe}$ at core collapse
and $s_\mathrm{c}$ is further a measure of the mass-radius relation of the iron
core, \ie it relates to $\xi_\mathrm{2.5}$ \citep{2021A&A...645A...5S}. Also,
the gravitational binding energy of the material exterior of the iron core could
have been used as a summary proxy of the structure of stars at core collapse
\citep[\eg][]{2014ApJ...783...10S, 2024A&A...682A.123T}. All of these quantities follow
similar trends such as peaks and troughs at the same $M_\mathrm{ini}$ and
$M_\mathrm{CO}$.

\begin{figure*}
    \centering
    \includegraphics{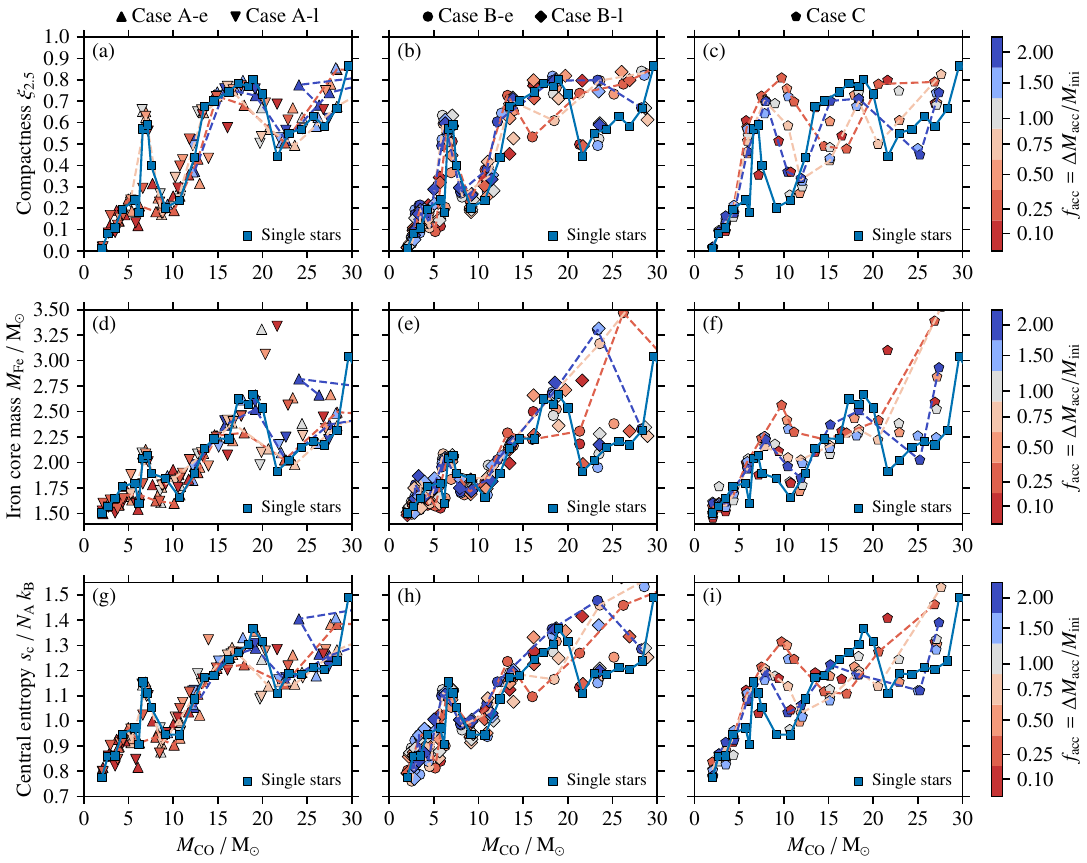}
    \caption{Core compactness $\xi_{2.5}$ (panels a--c), iron core mass $M_\mathrm{Fe}$ (panels d--f) and central specific entropy $s_\mathrm{c}$ (panels g--i) as a function of CO core mass $M_\mathrm{CO}$ at core collapse. As in other figures, we distinguish between \casea, B and~C accretors and color-code the mass accretion fraction $f_\mathrm{acc}$. Accretion fractions of 25\%, 75\% and 200\% are connected by dashed lines to better visualize trends and systematic shifts.}
    \label{fig:MCO-xi-MFe}
\end{figure*}

The evolution of massive stars beyond core helium burning is driven by thermal
neutrino losses and the delicate balance between these losses, energy gain from
nuclear burning, and energy release and consumption from changes in the
gravitational potential of stars. For example, the central carbon abundance
$X_\mathrm{C}$ at the end of core helium burning generally decreases with
increasing $M_\mathrm{CO}$ (Fig.~\ref{fig:XC-vs-MCO}). With fewer carbon
atoms available for carbon burning, there is a critical $M_\mathrm{CO}$ beyond
which energy loss from neutrinos overcomes energy generation from core carbon
burning. Consequently, the carbon burning proceeds no longer under convective
but radiative conditions, the evolution speeds up and the stellar cores at
core-collapse become more compact, reach higher $M_\mathrm{Fe}$ and have larger
entropy. With further increasing $M_\mathrm{CO}$, the next nuclear burning
stages ignite more quickly after the beginning of core carbon burning, become
more vigorous and reduce $\xi_{2.5}$, $M_\mathrm{Fe}$ and $s_\mathrm{c}$ at core
collapse. For yet higher $M_\mathrm{CO}$, also neon burning becomes neutrino
dominated, turns radiative and again leads to increased $\xi_{2.5}$,
$M_\mathrm{Fe}$ and $s_\mathrm{c}$ at core collapse
\citep[\eg][]{2020MNRAS.499.2803P, 2020MNRAS.492.2578S, 2021A&A...645A...5S,
2023ApJ...950L...9S, Laplace+2023}.

This sequence of changes in the nuclear burning conditions in the deep interior
of stars leads to peaks and troughs in $\xi_{2.5}$, $M_\mathrm{Fe}$ and
$s_\mathrm{c}$ at core collapse with $M_\mathrm{CO}$
(Fig.~\ref{fig:MCO-xi-MFe}). In our single-star models, a peak in $\xi_{2.5}$,
$M_\mathrm{Fe}$ and $s_\mathrm{c}$ is found at $M_\mathrm{CO}\approx7\,\msun$
and they increase and reach generally high values for
$M_\mathrm{CO}\gtrsim13\,\msun$. We find an indistinguishable $\xi_{2.5}$,
$M_\mathrm{Fe}$ and $s_\mathrm{c}$ landscape with $M_\mathrm{CO}$ compared to
single stars in our \casea accretors (Fig.~\ref{fig:MCO-xi-MFe}a, d, g); even in
the non-rejuvenated \casea systems, there is no obviously deviating trend. We
attribute the latter to the fact that also the non-rejuvenated \casea accretors
show the same relation between $X_\mathrm{C}$ and $M_\mathrm{CO}$ as the single
stars (Fig.~\ref{fig:XC-vs-MCO}a).

In the \caseb accretors, the $\xi_{2.5}$, $M_\mathrm{Fe}$ and $s_\mathrm{c}$
landscapes are qualitatively similar to those of single stars but we find a
systematic shift of the peaks and troughs to lower $M_\mathrm{CO}$
(Fig.~\ref{fig:MCO-xi-MFe}b, e, h). The magnitude of the shift depends on the
amount of accreted mass. For $f_\mathrm{acc}=2.0$, the compactness peak is
shifted by about $0.5\,\msun$ to a lower $M_\mathrm{CO}$ of
${\approx}\,6.5\,\msun$ and the second major increase of compactness is shifted
by about $1\,\msun$ to $M_\mathrm{CO}{\approx}\,12\,\msun$. In these models, the
carbon abundance at the end of core helium burning is reduced
(Fig.~\ref{fig:XC-vs-MCO}) such that core carbon and core neon burning
become neutrino dominated at slightly lower $M_\mathrm{CO}$. This finding is
analogous but in the opposite direction to binary-stripped stars where the
removal of the hydrogen-rich envelope results in a larger $X_\mathrm{C}$ at the
end of core helium burning and thus in a systematic shift of the compactness
landscape to higher $M_\mathrm{CO}$ \citep[][]{2019ApJ...878...49W,
2021A&A...645A...5S, 2023ApJ...950L...9S, 2021A&A...656A..58L}. In the example
in Fig.~\ref{fig:tulips-comparison-same-MCO}, the accretor model with
$f_\mathrm{acc}=1.0$ falls into the shifted compactness peak for such models
($\xi=0.63$) while the single star and $f_\mathrm{acc}=0.1$ accretor have rather
low compactness parameters of $0.24$ and $0.12$, respectively. This illustrates
further that accretor models can achieve inherently different core structures
compared to single stars despite similar CO core masses.

The relation of $\xi_{2.5}$, $M_\mathrm{Fe}$ and $s_\mathrm{c}$ as a function of
$M_\mathrm{CO}$ in \casec accretors is qualitatively different from those of the
single stars and \caseab accretors at $M_\mathrm{CO}\gtrsim5\,\msun$
(Fig.~\ref{fig:MCO-xi-MFe}c, f, i). The much reduced $X_\mathrm{C}$ in these models
(Fig.~\ref{fig:XC-vs-MCO}c) results in a shortened core-carbon-burning
phase and generally larger $\xi_{2.5}$, $M_\mathrm{Fe}$ and $s_\mathrm{c}$ at
$M_\mathrm{CO}\gtrsim5\,\msun$ compared to the single-star models. The first
compactness peak occurs at lower $M_\mathrm{CO}$ and is broadened with respect
to that of the single stars and we find a dependence on the amount of accreted
mass. The compactness peak at $M_\mathrm{CO}\approx7\,\msun$ is broadened the
most for $f_\mathrm{acc}=0.25$ while it is broadened to a lesser extent for
higher accretion fractions. Because of this, the second major increase in
compactness is pushed to $M_\mathrm{CO}\approx18\,\msun$ in the
$f_\mathrm{acc}=0.25$ models and to $M_\mathrm{CO}\approx14\,\msun$ for
$f_\mathrm{acc}=2.0$. The $f_\mathrm{acc}=0.1$ models, where we found no clear
trend in the reduction of $X_\mathrm{C}$ (Fig.~\ref{fig:XC-vs-MCO}c),
also seem to show no clear trend in $\xi_{2.5}$, $M_\mathrm{Fe}$ and
$s_\mathrm{c}$.

The mixing of core material into the envelope in fully realistic
stellar mergers may affect both the envelope and the pre-SN core structures.
However, the qualitative trends of $\xi_{2.5}$, $M_\mathrm{Fe}$ and
$s_\mathrm{c}$ with $M_\mathrm{CO}$ are expected to be similar. We expect fully realistic
mergers that end their lives as CSGs (BSGs) to follow similar relations of
$\xi_{2.5}$, $M_\mathrm{Fe}$ and $s_\mathrm{c}$ with $M_\mathrm{CO}$ as is found
in pre-SN CSGs (BSGs) from our accretors. However, the mapping of
$M_\mathrm{CO}$ to $M_\mathrm{ini}$ will differ and, as explained above, which
$f_\mathrm{acc}$ thresholds result in long-lived BSGs from \caseb and~C stellar
mergers and hence which mergers correspond to the most strongly shifted
$\xi_{2.5}$, $M_\mathrm{Fe}$ and $s_\mathrm{c}$ curves.

\section{\label{sec:final-fate-sne}Final fate and SN explosions}

Applying the SN code of \citet{2016MNRAS.460..742M} to our models at
core-collapse (Sect.~\ref{sec:methods}), we find that a compactness threshold of
$\xi_{2.5}\,{<}\,0.44$ for successful SN explosions agrees with the model of
\citet{2016MNRAS.460..742M} in more than 95\% of cases. Similarly, the
two-parameter explodability criterion of \citet{2016ApJ...818..124E}, $\mu_4 <
k_1 M_4\mu_4 + k_2$ with constants $k_1=0.200$ and $k_2=0.072$
\citep[see][]{2023ApJ...950L...9S}, agrees in 89\% of cases. In the
\citeauthor{2016ApJ...818..124E} criterion, $\mu_4=\mathrm{d}m/\mathrm{d}r$ is
the radial mass derivative in units of $\msun/1000\,\mathrm{km}$ and $M_4$ the
mass in $\msun$ at a specific entropy of $s=4$. In particular, all explodability
criteria predict BH formation in the compactness peak at
$M_\mathrm{CO}\,{\approx}\,7\,\msun$ and generally also for high compactness at
$M_\mathrm{CO}\,{\gtrsim}\,13\,\msun$. The $\xi_{2.5}$-landscape found in this
work is thus a good proxy for whether stars may explode in neutrino-driven
SNe\footnote{Similarly, one can define explodability criteria based on
$M_\mathrm{Fe}$, $s_\mathrm{c}$ and the gravitational binding energy outside the
iron core.} \citep[see also the review by][]{heger2023a}. This agrees
with the expectation that stars with high compactness, large iron core mass and
high gravitational binding energy exterior to the iron core are more prone to
collapsing into BHs \citep[see \eg][]{2014ApJ...783...10S, heger2023a}.
Note, however, that there are core-collapse simulations that show shock revival
in stars with high core compactness \citep[see \eg][]{2018ApJ...855L...3O,
2020MNRAS.491.2715B, 2018ApJ...852L..19C, 2020MNRAS.495.3751C,
2018MNRAS.477L..80K, 2020ApJ...890..127C, 2020MNRAS.494.4665P}, but this does
not immediately imply a successful SN explosion \citep[][]{heger2023a}.

We first present the pre-SN locations of our models in the HR diagram in
Sect.~\ref{sec:hrd-explosion-sites} and then the resulting compact remnant and
SN ejecta masses using the SN model of \citet{2016MNRAS.460..742M} in
Sect.~\ref{sec:compact-remnants-ejecta-masses}.

\subsection{\label{sec:hrd-explosion-sites}Pre-SN locations in the HR diagram}

\begin{figure}
    \centering
    \includegraphics[width=\linewidth]{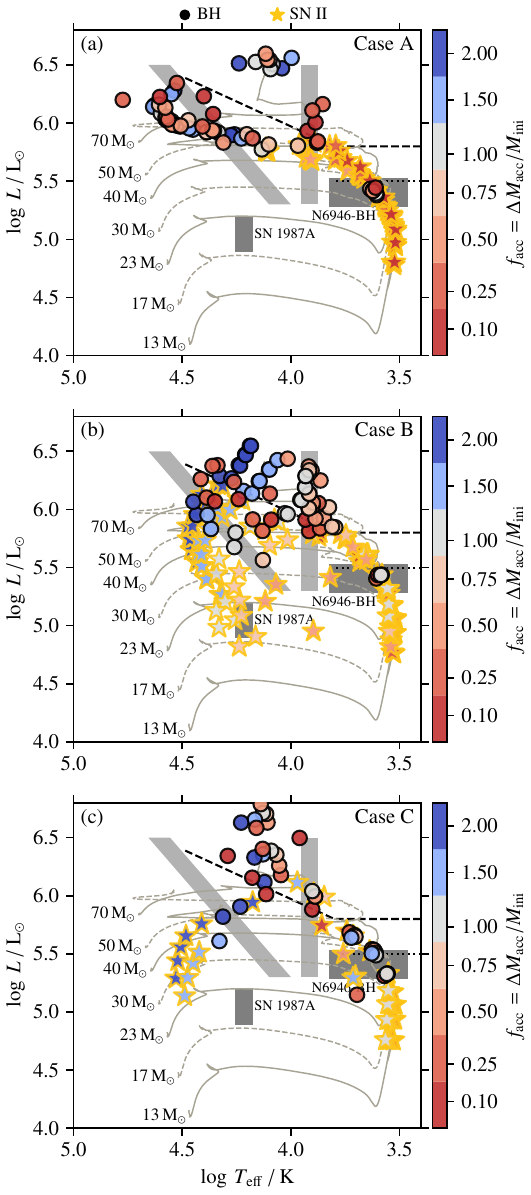}
    \caption{Pre-SN locations of the \casea, B and~C accretors in the HR diagram. The dotted line at $\log\,L/\lsun=5.5$ \citep{2018MNRAS.478.3138D} is the maximum luminosity of RSGs and the grey shaded regions indicate the hot and cool parts of the S~Doradus instability strip. The dashed line is the Humphreys--Davidson limit \citep{1979ApJ...232..409H}. The dark-grey boxes indicate the locations of the RSG N6946-BH that likely collapsed into a BH without a bright SN \citep{2015MNRAS.450.3289G, 2017MNRAS.468.4968A, 2017MNRAS.469.1445A, 2021MNRAS.508.1156B}, and the BSG progenitor of \snesa \citep{1988ApJ...330..218W}. The grey evolutionary tracks are single-star models with the masses given by the labels.}
    \label{fig:hrd-2-case-a-b-c}
\end{figure}

We show the pre-SN locations of all our models in the HR diagram in
Fig.~\ref{fig:hrd-2-case-a-b-c}. Because we do not apply LBV-like mass loss, our
models at $\log\,L/\lsun\,{\gtrsim}\,5.5$ do not lose large fractions of their
envelopes and do not become WR stars. Moreover, we do not model stars with
initial masses ${<}\,13\,\msun$, which explains the lowest pre-SN luminosity of
$\log\,L/\lsun\, {\approx}\, 4.7$.

\begin{figure*}
    \centering
    \includegraphics{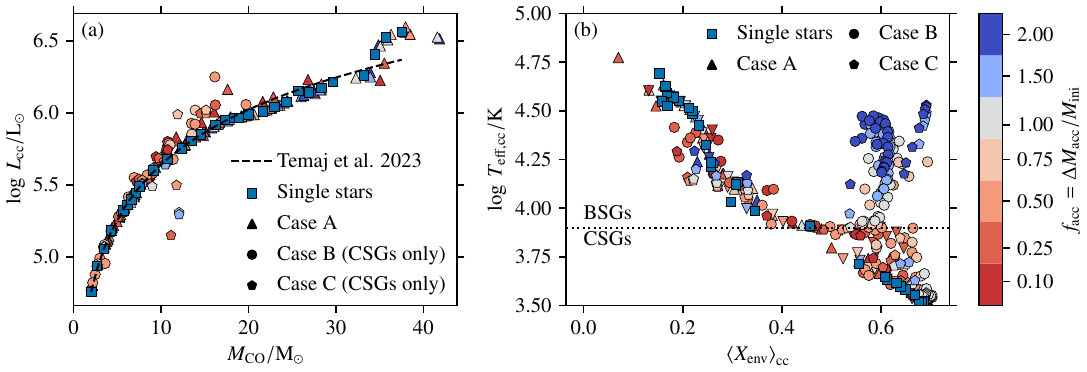}
    \caption{Core-mass--luminosity relation (panel a) and the relationship between the effective temperature and the average hydrogen mass fraction in the envelope (panel b) of single stars and \casea, B and~C accretors at core collapse. In the core-mass--luminosity relation, only CSG \caseb and~C accretors are shown as the BSG accretors with their thick hydrogen-burning shell follow a different mass-luminosity relation (see main text and Fig.~\ref{fig:hrd-bsg-explanation}). The black dashed curve in panel (a) is the fit of the core-mass--luminosity relation from \citet{2024A&A...682A.123T}.}
    \label{fig:hrd-2-case-a-explanation}
\end{figure*}

The single-star models reach core collapse at the same sites in the HR diagram
as the \casea accretors, and are not shown for clarity
(Fig.~\ref{fig:hrd-2-case-a-b-c}a). Both reach core collapse along a
well-defined branch in the HR diagram. Moreover, also those \caseb and~C
accretors that have no thick hydrogen-burning shell at core-collapse follow the
same branch. The cores of these pre-SN stars after core helium burning evolve
independently from the envelope because of a drop in pressure by many orders of
magnitude at the core--envelope boundary. The photon luminosity is produced by
nuclear burning which takes place in and near the core, \ie these stars follow a
core-mass--luminosity relation (Fig.~\ref{fig:hrd-2-case-a-explanation}a).
The long-lived BSGs with their thick hydrogen-burning shell do not follow
such a relation as shown below. For the core mass, we take $M_\mathrm{CO}$ at
core collapse but using $M_\mathrm{He}$ gives a similar result. For comparison,
we show the fit function $\log\,L/\lsun = 4.372 +
1.268\log\,M_\mathrm{CO}/\msun$ from \citet{2024A&A...682A.123T} in
Fig.~\ref{fig:hrd-2-case-a-explanation}a. This function has been obtained by
fitting the pre-SN luminosities of single stars with different convective core
overshooting parameters. It also represents well the pre-SN luminosities of
\casea, B and~C accretors that do not explode as BSGs. The luminosities
of the CSG \casec and~B accretors fit well up to
$M_\mathrm{CO}\,{\approx}\,10\,\msun$ and for larger $M_\mathrm{CO}$ the slope
of the relation steepens slightly from $1.268$ to ${\approx}\,1.350$
(Fig.~\ref{fig:hrd-2-case-a-explanation}a).

The effective temperature of CSGs (\ie the Hayashi line) is set by opacity
and also the convective mixing-length efficiency, $\alpha_\mathrm{MLT}$. With an
increasing mass of stars, more and more of the hydrogen-rich envelope is lost,
and stars become hotter and appear as BSGs until they are eventually hot helium
stars. We find a close relationship between the effective temperature of single
stars and \casea, B and~C accretors at core collapse and the average
hydrogen mass fraction in their envelope,
$\left<X_\mathrm{env}\right>_\mathrm{cc}$ at effective temperatures larger
than those of CSGs (\ie $\log\,T_\mathrm{eff,cc}/\mathrm{K}\,{\gtrsim}\,3.9$)
and for $\left<X_\mathrm{env}\right>_\mathrm{cc}\lesssim 0.5$
(Fig.~\ref{fig:hrd-2-case-a-explanation}b). As expected, stars with lower
hydrogen content in their envelopes have a hotter effective temperature.
Together with the core-mass--luminosity relation, this explains the location of
pre-SN single stars, the \casea accretors and some \caseb and~C
accretors in the HR diagram. The effective temperatures of BSGs with a
thick hydrogen-burning shell and stars on the Hayashi line do not depend on
$\left<X_\mathrm{env}\right>_\mathrm{cc}$.

Our models develop high compactness at $M_\mathrm{CO}\,{\approx}\,7\,\msun$ and
$M_\mathrm{CO}\,{\gtrsim}\,13\,\msun$ (Fig.~\ref{fig:MCO-xi-MFe}a, d) that
results in BH formation (see also
Sect.~\ref{sec:compact-remnants-ejecta-masses}). Because of the core
mass-luminosity relation, this translates into luminosities of
$\log\,L/\lsun\,{\approx}\,5.4$ and $\log\,L/\lsun\,{\gtrsim}\,5.8$ where BHs
are formed by stars that avoided the aforementioned BSG phase caused by the
thick convective hydrogen burning shell. To date, N6946-BH is the only star
known that may have collapsed into a BH without giving rise to a
SN\footnote{After submission of this work, new JWST observations revealed an
infrared source at the location of N6946-BH that, \eg, could be the dust
enshrouded missing RSG or dust debris from a failed SN
\citep{2023arXiv230916121B, 2024ApJ...962..145K}. Hence, only future
observations will finally clarify whether this RSG indeed collapsed into a BH or
not.} \citep{2015MNRAS.450.3289G, 2017MNRAS.468.4968A, 2017MNRAS.469.1445A,
2021MNRAS.508.1156B}. Interestingly, its location in the HR diagram seems to
agree with where we predict BH formation from the compactness peak
\citep[Fig.~\ref{fig:hrd-2-case-a-b-c}; see also][]{2020MNRAS.492.2578S}.
Progenitors of SN IIP are observed up to luminosities of
$\log\,L/\lsun\,{\approx}\, 5.1$ \citep{2015PASA...32...16S}, but these
luminosities remain uncertain (see \eg \citealt{2018MNRAS.478.3138D},
\citealt{2020MNRAS.493..468D}, \citealt{2024A&A...682A.123T} and also the vastly
different luminosities of $\log\,L/\lsun\,{=}\, 4.74\pm0.07$ and
$\log\,L/\lsun\,{=}\, 5.1\pm0.2$ inferred for the progenitor of SN~2023ixf by
\citealt{2023ApJ...952L..23K} and \citealt{2023ApJ...952L..30J}, respectively).

\begin{figure*}
    \centering
    \includegraphics{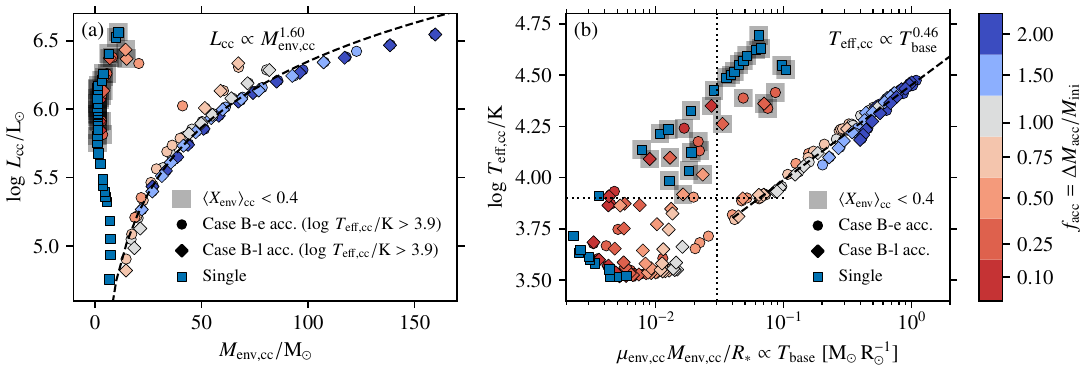}
    \caption{Envelope mass-luminosity relation (panel a) and H-envelope base temperature-effective temperature relation (panel b) for \caseb accretors. Single stars are shown for comparison and models with an average hydrogen mass fraction in their envelopes of $\left<X_\mathrm{env}\right>_\mathrm{cc}<0.4$ are marked. These stars may appear as BSGs at core collapse but their hot $T_\mathrm{eff,cc}$ is due to the low hydrogen content in their envelopes and not due to a thick convective hydrogen burning shell. Power-law fits are added and the dotted lines in panel (b) indicate $\log\,T_\mathrm{eff,cc}/\mathrm{K}=3.9$ and the transition base temperature at $\mu_\mathrm{env,cc}M_\mathrm{env,cc}/R_*\approx0.02\,\msun\,\rsun^{-1}$ where we find the BSG solutions. \casebe and~B-l refer to early and late \caseb accretors, respectively.}
    \label{fig:hrd-bsg-explanation}
\end{figure*}

The long-lived BSG structures in some of the \caseb and~C accretors
are characterised by a thick convective shell driven by hydrogen burning and we
describe it in Sect.~\ref{sec:bsg-structures} as if a main sequence star
with a convective core was added on top of a helium star. The surface photon
luminosity of these stars is primarily given by the luminosity produced by
hydrogen shell burning such that they follow an envelope mass-luminosity
relation and no longer a core-mass--luminosity relation
(Fig.~\ref{fig:hrd-bsg-explanation}a). The mass-luminosity exponent is larger at
low ($M_\mathrm{env,cc}\lesssim30\,\msun$) and smaller at high envelope masses
($M_\mathrm{env,cc}\lesssim80\,\msun$) than the value of $1.60$ found by fitting
the entire mass range.

Convection during hydrogen burning occurs if the produced energy is so large
that it cannot be transported away by photons \citep{1994sse..book.....K}. In
general, energy generation by nuclear burning has a very strong temperature
sensitivity and this tends to increase for isotopes with higher nucleon
numbers\footnote{In hydrogen burning via the CNO cycle, the energy generation
rate $\epsilon_\mathrm{CNO}$ scales with $T^{13\ldots23}$ for
$T=10\ldots50\times10^6\,\mathrm{K}$ \citep{1994sse..book.....K}.}. Convective
hydrogen shell burning is thus expected to happen if the temperature at the base
of the envelope exceeds a certain threshold. From homology relations of main
sequence stars, one finds that the central temperature scales as $T_\mathrm{c}
\propto \mu M/R$, where $\mu$ is the mean molecular weight, $M$ the total mass
of a star and $R$ the radius. Within this analogy, the temperature at the base
of the hydrogen-rich envelope then scales as $T_\mathrm{base} \propto
\mu_\mathrm{env} M_\mathrm{env}/R_\mathrm{env}$. The helium core of such stars
has a negligible radius ($\mathcal{O}(1\,\rsun)$) compared to the envelope
($\mathcal{O}(10\text{--}100\,\rsun)$) such that $R_\mathrm{env}$ can be
approximated by the radius of the entire star, $R_*$. Indeed, the pre-SN stars
with thick convective hydrogen shell burning follow a close relation in the
$T_\mathrm{eff}$--$T_\mathrm{base}$ diagram
(Fig.~\ref{fig:hrd-bsg-explanation}b). Beyond the critical value of
\begin{align}
    T_\mathrm{base} \propto \frac{\mu_\mathrm{env,cc}M_\mathrm{env,cc}}{R_*} \approx 0.02\,\msun\,\rsun^{-1},
    \label{eq:T-base-critical}
\end{align}
stars have a hot enough $T_\mathrm{base}$ such that hydrogen shell burning is
convective and the stars reach core-collapse as BSGs with effective temperatures
exceeding $\log\,T_\mathrm{eff,cc}/\mathrm{K}\,{\approx}3.9$.

We also find high compactness in \caseb and \casec accretors for
$M_\mathrm{CO}\,{\approx}\,7\,\msun$ and $M_\mathrm{CO}\,{\gtrsim}\,13\,\msun$
that leads to BH formation and no SN explosion (Figs.~\ref{fig:MCO-xi-MFe}
and~\ref{fig:MCO-Mrm}). In the HR diagram, BH formation occurs close to the hot
side of the S~Doradus instability strip because of the compactness peak at
$M_\mathrm{CO}\,{\approx}\,7\,\msun$ and at cooler effective temperatures and
larger luminosities because of high compactness at
$M_\mathrm{CO}\,{\gtrsim}\,13\,\msun$ (Fig.~\ref{fig:hrd-2-case-a-b-c}b, c).
Hence, some of our accretor models may explode in SNe while appearing as LBVs
and they may have experienced significant mass loss before their SN. Other
models in a very similar location in the HR diagram may just collapse into BHs
but also these stars could have experienced significant mass loss before their
collapse. Accretor models that are too hot to enter the LBV regime in the HR
diagram are all predicted to explode in SNe. 

While our models generally predict exploding BSGs in a region of the BSG
progenitor of \snesa, a \casec model is required to explain this particular
object \citep[see \eg][]{1992PASP..104..717P, 2017hsn..book..635P,
2017MNRAS.469.4649M, 2018MNRAS.473L.101U}. None of our current \casec accretors
reproduce its location in the HR diagram (Fig.~\ref{fig:hrd-2-case-a-b-c}c), but
these models do not dredge up core material and mix it throughout the envelope
(see Sect.~\ref{sec:methods.binary-models}). Models including such mixing can
explain the BSG location and further observed characteristics (see discussion in
Sect.~\ref{sec:discussion}). Moreover, all our long-lived BSG accretors are
likely systematically biased to too-cold effective temperatures compared to what
is expected from fully realistic mergers because of this missing mixing of core material
into the envelope.

\subsection{\label{sec:compact-remnants-ejecta-masses}Compact-remnant and SN-ejecta masses}

\begin{figure*}
    \centering
    \includegraphics{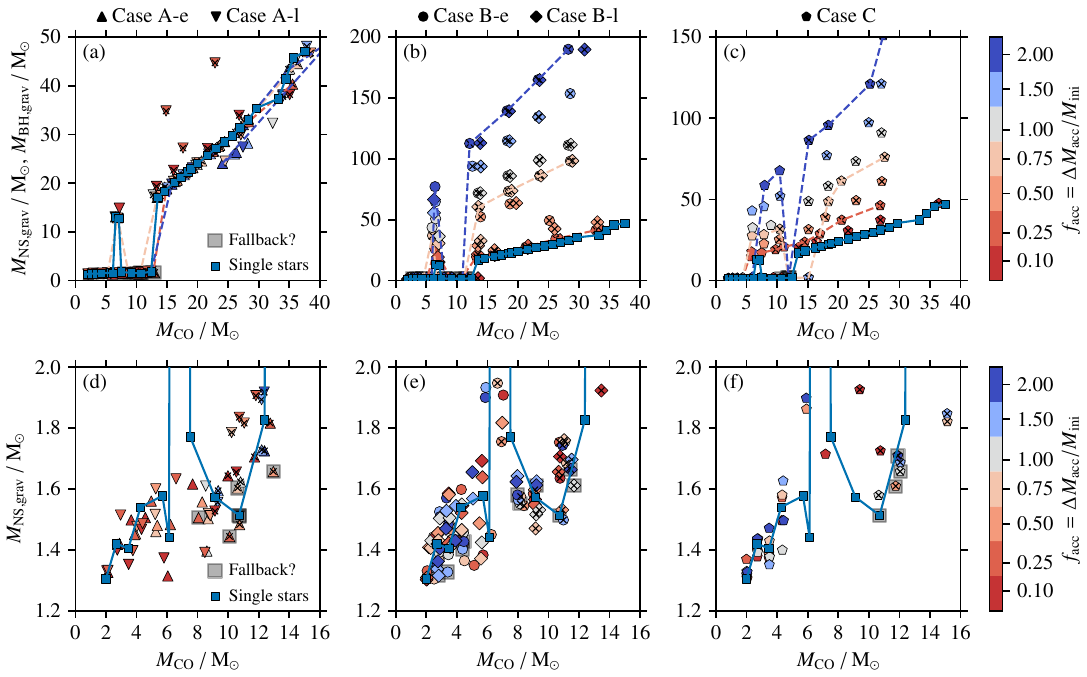}
    \caption{Black-hole and neutron-star masses of \casea (panels a, d), \caseb (panels b, e) and \casec (panels d, f) accretors as a function of the final CO core mass $M_\mathrm{CO}$. The bottom row is a zoom into the NS-mass regime. The NS and BH masses of the single-star models are shown for comparison and accretor models with $f_\mathrm{acc}\,{=}\,0.25$, $0.75$ and $2.00$ are connected by lines to guide the eye. Systems likely experiencing SN fallback are marked by grey boxes and models that should have experienced LBV-like mass loss are indicated by crosses (\cf Sect.~\ref{sec:methods.mesa}).}
    \label{fig:MCO-Mrm}
\end{figure*}

We show the gravitational NS and BH masses in Fig.~\ref{fig:MCO-Mrm}. The NS
masses appear uncorrelated with the amount of accreted mass and the
mass-accretion Cases~A, B and~C. Generally, the NS masses are on average more
massive in stars with larger CO core masses, and the largest NS masses are found
at CO core masses close to regions of the highest compactness
(Fig.~\ref{fig:MCO-Mrm}d, e, f), where the models are on the brink of not to
result in a successful neutrino-driven explosion.

\begin{figure*}
    \centering
    \includegraphics{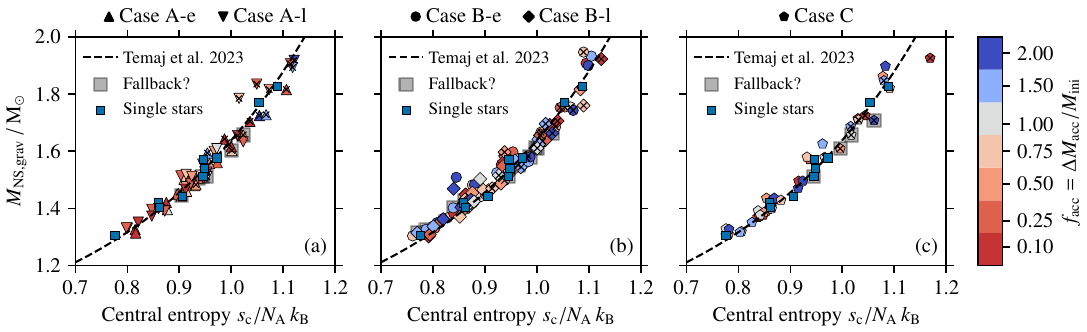}
    \caption{Gravitational neutron star mass $M_\mathrm{NS,grav}$ as a function of central specific entropy $s_\mathrm{c}$ of the progenitor stars for \casea (panel a), \caseb (panel b) and \casec (panel c) accretion. The black dashed line is the exponential fit to the neutron star masses of single stars computed for different convective boundary mixing found by \citet{2024A&A...682A.123T}.}
    \label{fig:MNS-entropy}
\end{figure*}

As shown by \citet{2024A&A...682A.123T} for single stars with different degrees of
convective boundary mixing, the NS masses follow a tight relation with the
central specific entropy $s_\mathrm{c}$ of the SN progenitors. We find a similar
relation in our models and the exponential best fit of \citeauthor{2024A&A...682A.123T},
\begin{align}
    M_\mathrm{NS,grav}/\msun=0.048\, \exp\left(2.76\,s_\mathrm{c}/N_\mathrm{A}\,k_\mathrm{B} \right) + 0.880,
    \label{eq:NS-mass-entropy} 
\end{align}
also fits our data well despite the different evolutionary histories of their
and our models (Fig.~\ref{fig:MNS-entropy}). In particular, the neutron star
mass--entropy relation is the same in our \casea, B and~C accretors. Both
studies apply the same SN code of \citet{2016MNRAS.460..742M}, where shock
reversal and hence SN explosions are triggered outside the iron core near a mass
coordinate of $M_4$, \ie within the oxygen shell \citep[\cf figure~A.1
in][]{2021A&A...645A...5S}. Hence, the resulting NS masses are directly
proportional to $M_4$ and thus $M_\mathrm{Fe}$. Because $M_\mathrm{Fe}$ and
$s_\mathrm{c}$ are closely related (see Eq.~\ref{eq:eff-Mch} and
Fig.~\ref{fig:MCO-xi-MFe}), it is not surprising that we find a strong
correlation between $M_\mathrm{NS,grav}$ and $s_\mathrm{c}$, but it could
not be foreseen that this relation is so tight and does not carry information on
the evolutionary history of the stars (Fig.~\ref{fig:MNS-entropy}). It thus
seems that $s_\mathrm{c}$ is a good indicator of that mass coordinate at which
neutrino-driven SNe within the model of \citet{2016MNRAS.460..742M} are
launched. Whether such a connection is more general and can also be made in
other neutrino-driven SN models remains unclear and warrants further
investigation.

\begin{figure*}
    \centering
    \includegraphics{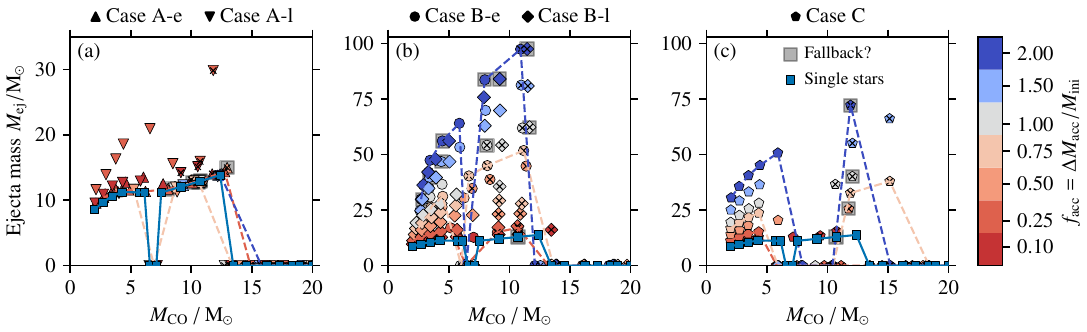}
    \caption{Same as Figs.~\ref{fig:MCO-vs-Mf} and~\ref{fig:MCO-Mrm} but showing the SN ejecta mass $M_\mathrm{ej}$. Models marked by crosses (\ie also those with the highest SN ejecta mass) are expected to have experienced LBV-like mass loss such that these ejecta masses are upper limits.}
    \label{fig:MCO-Mej}
\end{figure*}

The BH masses (Fig.~\ref{fig:MCO-Mrm}a, b, c) are closely linked to the final
stellar masses (Fig.~\ref{fig:MCO-vs-Mf}), \ie to the mass loss history of
stars. Because of rejuvenation, most \casea accretors evolve as single stars of
higher initial masses and thus collapse to BHs of similar mass for the same
$M_\mathrm{CO}$. Only the non-rejuvenating accretors have smaller cores compared
to their total mass and lose less mass, thus forming more massive BHs
(Fig.~\ref{fig:MCO-Mrm}a). In \caseb and~C accretors and regardless of whether
stars explode as red or blue supergiants, we find that all BH masses are larger
than those of single stars with the same $M_\mathrm{CO}$ because of the larger
final stellar masses (Fig.~\ref{fig:MCO-Mrm}b, c; see also
Sect.~\ref{sec:core-final-masses}). The compactness peak at
$M_\mathrm{CO}\,{\approx}\,7\,\msun$ results in BH formation and this island of
BH formation is shifted by about $0.5\,\msun$ to lower $M_\mathrm{CO}$ in \caseb
models that accreted enough mass to evolve through a long-lived BSG phase
(Fig.~\ref{fig:MCO-Mrm}b; \cf Sect.~\ref{sec:pre-sn-structure}). The generally
higher compactness at core-collapse in \casec accretors is reflected by BH
formation over a larger $M_\mathrm{CO}$ range compared to \caseb accretors (\eg
at $M_\mathrm{CO}\,{\approx}\,10\,\msun$, some \casec accretors collapse into
BHs while this is not the case in the \caseb models). Neglecting enhanced mass
loss in massive CSGs and LBVs, BH masses are in the range of
$12.5\text{--}50\,\msun$, $15\text{--}190\,\msun$ and $15\text{--}150\,\msun$ in
our \casea accretors/single stars, \caseb accretors and \casec accretors,
respectively. The minimum BH masses (${\approx}\,12.5\,\msun$) are for stars in
the compactness peak. 

The SN-ejecta masses (Fig.~\ref{fig:MCO-Mej}) are similar to those of BHs
(Fig.~\ref{fig:MCO-Mrm}): if a star does not collapse into a BH, it explodes and
ejects a mass that we assume is given by the final total stellar mass minus the
(baryonic) mass of the formed NS. Stars in our models explode and eject
significant mass for $M_\mathrm{CO}\,{\lesssim}\,13\,\msun$ and outside the
compactness peak at $M_\mathrm{CO}\,{\approx}\,7\,\msun$
(Fig.~\ref{fig:MCO-Mej}). Fully rejuvenating \casea accretors eject the same
mass as single stars of the same $M_\mathrm{CO}$ while the non-rejuvenating
models retain higher final mass and eject correspondingly more mass in their SN
explosions
(Fig.~\ref{fig:MCO-Mej}a). The ejecta masses of \caseb and~C accretors scale
directly with the amount of accreted mass and greatly exceed those of single
stars (Fig.~\ref{fig:MCO-Mej}b, c). While single stars eject
$8.6\text{--}14.0\,\msun$, the ejecta masses reach values of up to $30\,\msun$,
$100\,\msun$ and $70\,\msun$ in our (non-rejuvenated) \casea, B and~C accretors,
respectively. 

\begin{figure}
    \centering
    \includegraphics{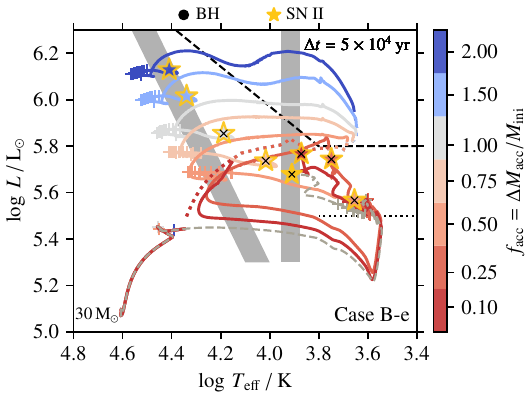}
    \caption{Same as Fig.~\ref{fig:hrd-case-be-m13} but for initially $30\,\msun$ stars. The black dashed line is the Humphreys--Davidson limit \citep{1979ApJ...232..409H}. Cross symbols indicate models that may experience enhanced, LBV-like mass loss and may thus rather explode at hotter effective temperatures as stripped-envelope SNe of Type Ib/c. Plus symbols are separated by $\Delta t\,{=}\, 5\times10^{4}\,\mathrm{yr}$.}
    \label{fig:hrd-case-be-m30}
\end{figure}

As with the final masses (Fig.~\ref{fig:MCO-vs-Mf}), some BH and SN ejecta
masses (Figs.~\ref{fig:MCO-Mrm} and~\ref{fig:MCO-Mej}, respectively) are
rather upper limits because of unaccounted for mass loss. For example, no RSG
has been observed today at $\log\,L/\lsun\,{\gtrsim}\,5.5$
\citep{2018MNRAS.478.3138D} and this might be because of enhanced mass loss in
such luminous RSGs that erode the hydrogen-rich envelope and turn stars into
hotter, \eg WR, stars. Similarly, eruptive mass loss, \eg, because of envelope
instabilities in LBVs may result in the loss of the entire hydrogen-rich
envelope. Typical averaged LBV mass-loss rates at
$\log\,L/\lsun\,{\gtrsim}\,6.0$ are $10^{-4}\,\msun\,\mathrm{yr}^{-1}$
\citep{2014ARA&A..52..487S}, suggesting that such LBVs would lose $1\,\msun$
within $10^{4}\,\mathrm{yr}$. We identify all models that would be subject to
such LBV-like mass loss for longer than this time, \ie those models that likely
lose more than $1\,\msun$ before core collapse. For example, none of our
initially $13\,\msun$ early \caseb models are expected to be subject to such
enhanced mass loss (\eg Fig.~\ref{fig:hrd-case-be-m13}) while some of the
initially $30\,\msun$ models are (\eg Fig.~\ref{fig:hrd-case-be-m30}).

The initially $30\,\msun$ early \caseb accretors burn helium in their cores as
BSGs for $f_\mathrm{acc}\,{\gtrsim}\,0.50$ and otherwise as CSGs
(Fig.~\ref{fig:hrd-case-be-m30}). Only the $f_\mathrm{acc}\,{=}\,1.0$ and $2.0$
models enter the LBV instability regime for less than $10^{4}\,\mathrm{yr}$ and
are thus probably not much affected by LBV-like mass loss. The other models may
encounter enhanced mass loss as an LBV star just before the SN explosion. Models
with $f_\mathrm{acc}\,{=}\,0.50$ and $0.75$ explode in the S~Dor instability
strip and may be LBV SN progenitors. The models evolving into CSGs (including
the initially $30\,\msun$ single star) are expected to lose significant mass
that we do not account for in our models. They may even lose their entire
hydrogen-rich envelope and then rather explode as a stripped-envelope supernova
(SN~Ib/c).

\begin{figure}
    \centering
    \includegraphics{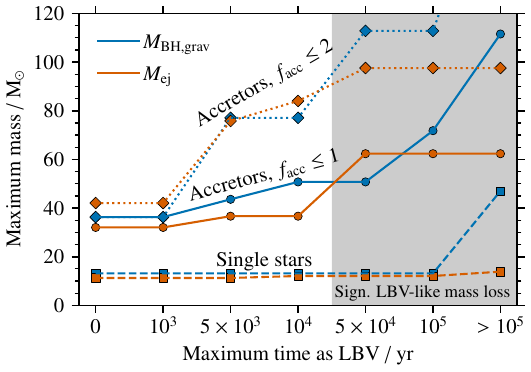}
    \caption{Maximum BH and SN ejecta mass of single stars and accretors with $f_\mathrm{acc}\,{\leq}\,1.0$ and $f_\mathrm{acc}\,{\leq}\,2.0$ as a function of the maximum time spent in the LBV-region of the HR diagram. The grey-shaded region indicates models for which we expect significant LBV-like mass loss such that the maximum BH and SN ejecta masses should be considered upper limits.}
    \label{fig:max-bh-mej}
\end{figure}

To better understand the possible maximum BH and SN ejecta masses of our models,
we show them as a function of the maximum time spent in the LBV region of the HR
diagram in Fig.~\ref{fig:max-bh-mej}. Models indicated by the grey-shaded areas
are the same as those marked by cross symbols throughout this paper, \ie for
which considerable mass loss is expected. Generally, the maximum BH masses are
larger than the SN ejecta masses and, if a single-star model evolves into the
LBV region in the HR diagram, it spends most of its core-helium burning in these
locations (\ie ${>}10^5\,\mathrm{yr}$). Considering $10^4\,\mathrm{yr}$ as the
limiting time a star can spend as an LBV in the HR diagram before it loses
considerable fractions of its mass, we find maximum BH (SN ejecta) masses of
about $13$, $50$ and $77\,\msun$ ($12$, $37$ and $84\,\msun$) for our single
stars and accretors with $f_\mathrm{acc}\,{\leq}\,1.0$ (\eg mergers in isolated
binary stars) and $f_\mathrm{acc}\,{\leq}\,2.0$ (\eg allowing also for repeated
mergers in triples), respectively. These values reduce to $13$, $36$ and
$36\,\msun$ ($11$, $32$ and $42\,\msun$) for models that avoid the LBV-region in
the HR diagram altogether -- they can thus be considered lower limits.
Neglecting LBV-like mass loss, we find maximum BH (SN ejecta) masses of $47$,
$112$ and $190\,\msun$ ($14$, $62$ and $98\,\msun$), respectively.

The exact BH and ejecta masses depend on a multitude of modelling
uncertainties and the limitations of our models to fully represent mass gainers
of binary mass transfer and stellar mergers (see discussion in
Sect.~\ref{sec:model-uncertainties}). However, regardless of the uncertainties
and limitations, accretors of binary mass transfer and stellar mergers lead to a
much larger range of BH and SN ejecta masses that single stars can only
reproduce at much lower metallicities (\cf Sect.~\ref{sec:maximum-bh-mass}).

%
%
\section{\label{sec:discussion}Discussion}

\subsection{\label{sec:model-uncertainties}Main model uncertainties and limitations}

Our effective merger and binary mass-transfer prescription of accreting certain
fractions $f_\mathrm{acc}$ of the initial mass of stars over the accretor's
thermal timescale (see Sect.~\ref{sec:methods.binary-models}) is simplified and
cannot fully catch the complex physics of these phases (see
Sect.~\ref{sec:methods.limitations}). In the following, we discuss how our
assumptions of the chemical composition of the accreted mass, the absence of
mixing of parts of the helium core in mergers and the dynamic mass loss in
mergers affect our results. We also touch upon the role of wind mass loss,
rotation of our accretors, possible magnetic fields produced in stellar
mergers, and multiple and extended mass exchange episodes in binary stars.
Moreover, we discuss to which extent the findings from our solar-metallicity
models can be generalised to accretors of other metallicities. The role of
semi-convection is discussed already in Sect.~\ref{sec:bsg-structures}. As
discussed in the following and mentioned previously, our assumptions for
the merger models are such that one should expect to find a higher fraction of
post-merger BSGs than what is predicted by our models (\ie BSG solutions occur
at lower $f_\mathrm{acc}$ than predicted by our current models).

\subsubsection{\label{sec:discussion-accretion}Chemical composition of accreted mass} 
We assume that the accreted mass has a chemical composition of that of the
surface of the accretor, \ie almost the primordial chemical composition of our
models. However, the accreted material is certainly enriched in helium in
stellar mergers and possibly also during binary mass-transfer events. In
\caseb and~C accretors, a larger mean molecular weight in the accreted
material will increase the temperature at the base of the hydrogen-rich
envelope, $T_\mathrm{base}\propto \mu_\mathrm{env} M_\mathrm{env}/R_*$; this
might also occur in \casea accretors that do not fully rejuvenate. Hence, we
expect to find BSGs for smaller $f_\mathrm{acc}$ compared to our current models
such that more models evolve as BSGs and less as CSGs (\cf
Eq.~\ref{eq:T-base-critical}). The BSGs themselves will then have a higher
luminosity and cooler effective temperature because the luminosities and
effective temperatures of the BSGs mostly depend on the hydrogen-rich envelopes
which evolve analogously to main-sequence stars (Sects.~\ref{sec:bsg-structures}
and~\ref{sec:hrd-explosion-sites}). In RSGs, the larger mean molecular weight of
accreted mass decreases the opacity and thereby increases the effective
temperature according to the new location of the Hayashi line in the HR diagram
\citep{1961PASJ...13..442H}. If enough helium-rich material is accreted, such
stars may no longer look like RSGs but more like yellow super-/hypergiants, and
show a clear surface helium and nitrogen enrichment.

\subsubsection{\label{sec:discussion-core-mixing}Core mixing in stellar mergers and \snesa}
In all stellar mergers (\ie \casea, B and~C), parts of the
helium-enriched core of the more evolved star can be dredged up and
replenished by hydrogen-rich material \citep[see \eg][]{1992PASP..104..717P,
2002MNRAS.334..819I, 2002Ap&SS.281..191I, 2007AIPC..937..125P,
2013MNRAS.434.3497G, 2017MNRAS.469.4649M, 2019Natur.574..211S}. This mechanism
may operate during the dynamic merger phase but also during the later thermal
relaxation of the merger product \citep[see also][]{2020MNRAS.495.2796S}. In
\casea mergers, the mixing can prolong the core hydrogen burning phase and
thereby extend the MS turn-off to cooler temperatures and larger luminosities.
In \caseb and~C mergers, the partial mixing of the helium core out into the
envelope has two main consequences: (i) the helium core becomes effectively less
massive and the envelope correspondingly more massive, and (ii) the dredged-up
helium results in a higher mean molecular weight of the envelope. The effect of
the larger mean molecular weight in the envelope is already discussed above, and
the smaller helium core mass/larger envelope mass effectively has similar
consequences as an increased mean molecular weight. Because of a larger
$M_\mathrm{env}$, $T_\mathrm{base}\propto \mu_\mathrm{env} M_\mathrm{env}/R_*$
is hotter and stars more easily become BSGs. Moreover, a larger $M_\mathrm{env}$
also increases the star's luminosity and effective temperature (see
Sect.~\ref{sec:hrd-explosion-sites}). This will also affect the structures
of stars at core collapse and their explodability, and can lead to vastly
different outcomes of core collapse because of the non-monotonic relation of
various interior properties (\eg compactness and central entropy) with the
stellar core mass.

In models for the progenitor of \snesa \citep[see \eg][]{1989A&A...219L...3H,
1990A&A...227L...9P, 1992PASP..104..717P, 2017MNRAS.469.4649M,
2017arXiv170203973P}, the dredge-up of helium-rich material from the core of the
more evolved star is required to explain observations for two main reasons.
First, the triple-ring nebula is enriched in helium and nitrogen, \ie
by-products of hydrogen burning \citep[\eg][]{1989ARA&A..27..629A,
1996ApJ...464..924L}. Second, the expansion timescale of the rings surrounding
\snesa is only about $20,000\,\mathrm{yr}$ \citep[\eg][]{1995ApJ...452..680B},
\ie the proposed merger must have occurred after core-helium burning as a \casec
system. The mixing of helium-rich material into the envelope of the merged star
helps the progenitor models to become a BSG (see the discussion in
Sect.~\ref{sec:bsg-structures}). Indeed, we do not find BSGs from \casec
accretors with $f_\mathrm{acc}\leq1.0$ that can explain the position of \snesa's
progenitor in the HR diagram (see \eg Fig.~\ref{fig:hrd-2-case-a-b-c}). Most
recently, this has again been demonstrated by the merger models of
\citet{2017MNRAS.469.4649M} and \citet{2018MNRAS.473L.101U} that show a good
match to the BSG progenitor of \snesa. Such models can then also reproduce the
light curve of \snesa, along with other peculiar Type II SNe
\citep{2019MNRAS.482..438M}. Moreover, \caseb mergers with similar physics
can also explain several properties of observed BSGs
\citep{2023arXiv231105581M}.

\subsubsection{Mass loss via binary mass transfer and mergers}
We do not take into account mass loss from non-conservative binary mass transfer
or dynamic stellar mergers. In this regard, our accretion fractions are
``effective'' fractions of how much mass is truly accreted by stars. Hence, mass
loss is only relevant when linking our different accretion scenarios to the
initial binary-star configurations. Binary mass transfer can be fully
conservative for \casea systems but also almost entirely non-conservative when
the donor star is more evolved \citep[\eg][]{1992ApJ...391..246P,
1994A&A...290..119P, 2001A&A...369..939W, 2007A&A...467.1181D,
2023arXiv231112124H}. In stellar mergers, \citet{2013MNRAS.434.3497G} find that
up to ten percent of the total binary mass is dynamically ejected and the
ejection fraction depends on the binary mass ratio $q=M_2/M_1$,
$f_\mathrm{ej}\,{\approx}\,0.3 q/(1+q)^2$ ($M_1>M_2$). The authors conduct
head-on collisions where the ejection fraction depends strongly on the exact
collision setup \citep[\eg collision energy and impact parameter, see
also][]{2005MNRAS.358.1133F} and similar mergers in binary systems on Keplerian
orbits find lower ejection fractions of less than one percent
\citep{2019Natur.574..211S}. In general, the mass loss fraction depends on the
evolutionary state of the merging stars, and mergers of more evolved stars lead
to higher mass loss because their binding energies are lower. While mergers of
two main-sequence stars may eject less than one percent of their total mass
\citep[\eg][]{2019Natur.574..211S}, mergers of giant stars may proceed more like
traditional common-envelope phases and could lead to full envelope ejection, \ie
ejection fractions of 50\% and higher \citep[\eg][]{2022A&A...667A..72M}.
The triple-ring nebula of \snesa is an example of left-over debris from a
stellar merger \citep[see \eg][]{2007Sci...315.1103M, 2009MNRAS.399..515M}.

\subsubsection{\label{sec:discussion.mass-loss}Mass loss via winds and LBV-like outflows}
Stellar wind mass-loss rates are still substantially uncertain \citep[easily by
a factor of 2--3 depending on the stellar type, see \eg][]{2014ARA&A..52..487S}.
In \casea accretors, wind mass loss directly influences the core evolution and
stars adjust their structure to the modified total mass. In \casec accretors,
the core and envelope are already decoupled such that mass loss mainly reduces
the envelope mass and leaves the further core evolution untouched. This is not
true for \caseb accretors where a reduced envelope mass can still affect the
further core evolution. With a lower envelope mass (\ie higher mass loss), BSG
formation becomes more difficult as discussed in
Sects.~\ref{sec:discussion-accretion} and~\ref{sec:discussion-core-mixing} and
some initial BSGs might evolve to the red giant branch when hydrogen shell
burning turns radiative.

In our hydrogen-rich accretors, WR wind mass loss is negligible, and the wind
mass loss applied to hot ($T_\mathrm{eff}\,{\geq}\, 11,000\,\mathrm{K}$) and
cool ($T_\mathrm{eff}\,{\leq}\, 10,000\,\mathrm{K}$) stars mostly affects our
results. The applied cool wind mass loss is generally higher than that applied
to our hot stars such that models that remain hot for most of their evolution
lose considerably less mass than models that evolve into CSGs. Hence,
models that end as CSGs tend to be more affected by uncertainties in the applied
wind mass-loss rates than those that end as BSGs. \citet{2014ARA&A..52..487S}
suggest that the mass-loss rates as applied in our models are an overestimate.
If true, our models would lose less mass and could result in higher SN ejecta
masses and more massive BHs.

In massive stars, winds are mostly line-driven and therefore scale with
metallicity. This is likely not true for LBV-like winds and eruptions. These
have been suggested to be connected to helium opacity in the outer envelopes of
massive stars \citep{2018Natur.561..498J, 2021A&A...647A..99G} and such a
mechanism would operate at all metallicities (albeit less severely at lower
metallicity with generally lower opacity and luminosity-to-mass ratios). This
seems to be confirmed observationally: \citet{2018MNRAS.478.3138D} find that
their re-evaluated Humphreys--Davidson limit for cool supergiants is the same in
the Milky Way, LMC and SMC, and LBVs have been found in low-metallicity galaxies
\citep[\eg SMC;][]{1994PASP..106.1025H, 2020Galax...8...20W}. So while
line-driven winds of hot and cool stars become less strong at lower metallicity,
LBV-like mass loss might be always present. This is particularly relevant for
our massive accretors that evolve through or close to the S~Dor instability
strip and that may lose a significant fraction of their hydrogen-rich envelopes.

Our models employ \mesa's MLT++ scheme. Switching this off, stars would develop
more strongly inflated envelopes and would more easily evolve into the LBV
region in the HR diagram. This could potentially lead to enhanced mass loss.

\subsubsection{\label{sec:discussion.rotation}Rotation}
Accretors of binary mass transfer can be easily spun up to critical rotation if
tides are not efficient enough to slow them down \citep[see
\eg][]{2013ApJ...764..166D}. Merger products may also show unusual rotation
velocities. Traditionally, it is assumed that such stars would rotate critically
because of the large amount of angular momentum in the pre-merger binary orbit.
However, \citet{2019Natur.574..211S, 2020MNRAS.495.2796S} have shown that merged
stars may actually spin slowly. They attribute this to the efficient coupling of
the merged star to a toroidal accretion structure right after the merger that
could transport away angular momentum and to an unusual moment of inertia of the
merged star right after the coalescence that does not allow the merged star to
gain high spin angular momentum as a solid body rotator. The merger of two stars
gives birth to a new, more massive star and this situation might be comparable
to the star-formation process. Also, most young massive stars do not seem to
rotate unusually rapidly \citep[\eg][]{2013A&A...550A.109D, 2013A&A...560A..29R,
2015A&A...580A..92R} and this may well apply to merger products. Moreover,
merged stars could lose angular momentum in a variety of ways, \eg, through
magnetic braking and mass ejection episodes \citep[see below and
also][]{2020MNRAS.495.2796S}. In lower-mass blue straggler stars originating
from collisions of stars, slow rotation seems to be favored over rapid rotation
\citep[\eg][]{1995ApJ...447L.121L, 2005MNRAS.358..716S, 2023NatCo..14.2584F},
further supporting our hypothesis that merged massive stars may actually not
rotate rapidly. 

While most of the above is true for MS stars and hence \casea accretors,
less is known for \caseb and~C accretors. The larger binary orbits of such
systems imply an overall larger orbital angular momentum which could enhance the
likelihood of obtaining more rapidly rotating \caseb and~C accretors than
those from \casea accretion. However, as shown by \citet{2018MNRAS.473L.101U},
most of the available angular momentum of their merger models must be lost by
some mechanism such that the merged star can become massive enough to explain
the BSG progenitor of \snesa. This would also speak in favour of a mechanism
that can efficiently remove angular momentum from \caseb and~C mergers.

Our models are for non-rotating stars. While this assumption may be adequate for
some of our accretors as described above, some will certainly rotate rapidly.
Rotation could lead to enhanced mass loss, internal mixing, reduced ejecta
masses and spin up of the remnants. As discussed in
Sect.~\ref{sec:discussion.mass-loss}, the enhanced mass loss makes BSG formation
more difficult. The additional mixing of chemical elements has the opposite
effect and enhances the likelihood of BSG formation because it likely mixes
helium-enriched material out into the envelope.

\subsubsection{Magnetic fields}
Observationally, strong, large-scale surface magnetic fields are observed in
about 5--10\% of massive stars \citep[\eg][]{2009ARA&A..47..333D,
2014IAUS..302..265W, 2015SSRv..191...77F, 2015A&A...582A..45F,
2017A&A...599A..66S, 2017MNRAS.465.2432G}. It has been shown that such magnetic
fields can be produced in the merger of two stars \citep{2019Natur.574..211S}.
In stars accreting mass by binary mass transfer, strong, large-scale magnetic
fields are probably not being produced \citep{2014IAUS..302....1L,
2016MNRAS.457.2355S} as exemplified by the absence of detected magnetic fields
in Be stars of which at least some are expected to be post-mass-transfer
accretors \citep[\eg][]{2014IAUS..302..265W, 2015IAUS..305...61N}. Hence,
only the accretor models representing stellar mergers may lack physics
ingredients to account for the effects of magnetic fields.

The magnetic fields in merger products can influence the further evolution in
various ways \citep[see \eg][]{2011A&A...525L..11M, 2012MNRAS.424.2358P,
2015A&A...584A..54P, 2017MNRAS.466.1052P, 2018MNRAS.477.2298Q,
2019MNRAS.485.5843K, 2020MNRAS.495.2796S}. Most importantly, magnetic fields may
effectively reduce wind mass loss by confinement, lead to enhanced angular
momentum transport in the stellar interior, inhibit convective boundary mixing,
and magnetic braking can spin stars down. The reduced mass loss and, in
particular, the inhibition of convection can make it more likely that merged
stars evolve as BSGs \citep{2015A&A...584A..54P}. Given that our accretor
models are non-rotating, they may be viewed as models with extremely efficient
magnetic braking.

\subsubsection{Multiple and extended binary mass-transfer phases}
In the evolution of binary stars, there can be multiple phases of mass
exchange with a companion before a star reaches core collapse. Each phase can
leave its characteristic fingerprint on the final stellar structure and may
influence the outcome of core collapse. For example, a star that accreted mass
in the first mass-transfer episode may be stripped off its envelope in a second
mass-transfer phase. For example, \citet{2023ApJ...942L..32R} modelled a
rejuvenated former accretor that evolves into a supergiant with a reduced
envelope binding energy because of the past mass accretion. The supergiant then
engulfs a companion and initiates a common-envelope phase whose outcome may now
be affected by the previous mass accretion. Similarly, previous mass exchange in
a binary can affect the structures of pre-merger stars \citep[see
\eg][]{2023arXiv231105581M}. Such additional mass-exchange phases are not
covered by our models. They may affect the accretor models representing mass
gainers of binary mass transfer and stellar mergers (the pre-merger stars may
have already evolved through a phase of binary mass transfer).

As detailed in Sect.~\ref{sec:methods.binary-models}, our models accrete
mass on a thermal timescale but this timescale can be much longer, \eg on a
nuclear timescale in case of accretion from an MS donor star (classical \casea
binary systems, \emph{not} to be confused with a \casea accretor). In such
systems, we expect continued rejuvenation. The evolution of such systems can
differ from our models, most notably by the chemical composition and specific
angular momentum of the accreted matter. Also in the case of stellar mergers,
there can be a preceding mass-transfer episode that modifies the structures of
the pre-merger stars and then gives rise to a different merger outcome. Such
effects are not accounted for in our models.

\subsection{\label{sec:metallicity}Metallicity}

Our models are for the solar metallicity of \citet{2009ARA&A..47..481A}, but
to which degree can our results be generalised to other metallicities $Z$? While
the quantitative results will differ for models of another metallicity, the
qualitative picture should remain the same. For example, the general landscape
of $\xi_{2.5}$, $M_\mathrm{Fe}$ and $s_\mathrm{c}$ with $M_\mathrm{CO}$ in
binary-stripped stars is robust for a large range of metallicities
\citep{2023ApJ...950L...9S} and we expect the same outcome for accretors of
binary mass transfer and stellar mergers. Similarly, long-lived BSGs may form
from merging at various metallicities. These stars will always lose the least
mass in stellar winds, hence they will produce the most massive BHs and the
largest SN ejecta masses for a given $Z$. As such they could naturally explain
some high-mass BHs in the observed population of binary BH mergers
\citep{2021arXiv211103606T, 2021arXiv211103634T}. Furthermore, the accretor
models may always have larger final masses than single stars of the same $Z$ and
larger envelope-to-core-mass ratios.

The metallicity enters stellar evolution mainly through the opacity of the
stellar matter and the strength of line-driven stellar winds. At lower
metallicities, stars are more compact and have weaker winds. This leads to the
general picture that wind mass loss is most important for the evolution of
single stars at high $Z$ while rotation and rotationally-induced mixing become
more important at low $Z$ where less angular momentum is lost through winds. As
discussed in Sect.~\ref{sec:discussion.mass-loss}, weaker winds at lower $Z$
increase the likelihood of the formation of long-lived BSGs (and vice versa at
higher $Z$ with stronger winds). Because of the increased importance of rotation
at low $Z$, our accretor models for mass gainers of binary mass transfer become
less and less applicable because they are for non-rotating stars. This may not
apply to stellar mergers as discussed in Sect.~\ref{sec:discussion.rotation}.

\subsection{\label{sec:maximum-bh-mass}Maximum BH and SN ejecta mass}

The maximum BH and SN ejecta masses are both closely connected to the maximum
final mass of our models and thus to wind mass loss
(Sects.~\ref{sec:core-final-masses}
and~\ref{sec:compact-remnants-ejecta-masses}). Wind mass-loss rates of CSGs and
WR stars are stronger than those of stars on the main sequence
\citep[\eg][]{2014ARA&A..52..487S}. Hence, stars that undergo core helium
burning as BSGs can avoid the strong CSG and WR winds, and, if they are also not
subject to LBV-like mass loss, they will retain the highest mass. A natural way
to obtain such stars is by stellar mergers as shown here\footnote{It is also
possible to tune the physics of stars to obtain core-helium-burning BSGs from
single stars as explored in detail when trying to understand the BSG progenitor
of \snesa \citep[see \eg the review in section~3 of][]{1992PASP..104..717P}.
\citet{2021MNRAS.504..146V} show that this is also possible for stars with
initial masses of $90\text{--}100\,\msun$. In many cases, however, the choice of
physics might be in tension with other observations of stars (\eg unusually
little or no convective boundary mixing).} \citep[see
also][]{2014ApJ...796..121J}.

Our models are for solar-like metallicity. Even when assuming that the entire
hydrogen-rich envelope is lost when stars enter the LBV region in the HR
diagram, we find BHs and SN ejecta masses of almost $40\,\msun$ and $35\,\msun$,
respectively, from stellar mergers in binaries (\ie
$f_\mathrm{acc}\,{\leq}\,1.0$). These numbers go up to about $50\,\msun$ and
$40\,\msun$, respectively, when assuming that LBV-like mass loss can be
neglected for stars spending less than $10^4\,\mathrm{yr}$ in the LBV region of
the HR diagram. At lower metallicities, the final and thus also the BH and SN
ejecta masses of all our models would be even larger.

Already at solar metallicity and for rather conservative assumptions on LBV-like
mass loss, we predict BH masses that may populate the so-called PISN BH mass gap
at about $45\text{--}130\,\msun$ \citep{2002RvMP...74.1015W,
2017ApJ...836..244W, 2019ApJ...887...53F, 2020ApJ...902L..36F,
2020A&A...640L..18M, 2020MNRAS.493.4333R, 2022ApJ...937..112F}. Avoiding the
pair instability in merger models that become BSGs is possible because such
stars have relatively small cores and large envelope masses. For example, an
initially $27\,\msun$ star turns into a BSG upon accreting with
$f_\mathrm{acc}\,{=}\,1.0$ as an early \caseb system and has a CO core mass of
$M_\mathrm{CO}\,{\approx}\,6.5\,\msun$ at core collapse. This $M_\mathrm{CO}$
places the star in the compactness peak and a BH of
$M_\mathrm{BH,grav}\,{\approx}\,50\,\msun$ forms in our models (it spends some
$9000\,\mathrm{yr}$ as an LBV in the HR diagram). Such a core mass is far from
the ${\gtrsim}\,35\,\msun$ required to trigger a PISN. In the event of
multiple/repeated mergers (\eg $f_\mathrm{acc}\,{=}\,2.0$, see
Sect.~\ref{sec:formation-massive-mergers}), the final BH mass even reaches
$80\,\msun$ in our models (in this case, the model evolves for
${\approx}\,4500\,\mathrm{yr}$ as an LBV in the HR diagram).

With the observation of the binary BH merger GW190521 with component masses of
${\approx}\,85\,\msun$ and ${\approx}\,66\,\msun$ \citep{2020PhRvL.125j1102A,
2020ApJ...900L..13A}, two BHs in the PISN BH mass gap have been observed. This
raises the question of how such massive BHs can form. Unless we neglect LBV-like
mass loss in our models and/or consider mergers with $f_\mathrm{acc}\,{>}\,1.0$,
we do not predict the existence of such massive BHs at solar-like metallicity. A
lower metallicity and/or mergers with $f_\mathrm{acc}\,{>}\,1.0$ are required.
This is generally consistent with other studies involving the merger of massive
stars to obtain BSGs that avoid significant mass loss and retain a large final
mass to produce BHs in the PISN mass gap \citep[\eg][]{2019MNRAS.487.2947D,
2020ApJ...904L..13R, 2022MNRAS.516.1072C, 2023MNRAS.519.5191B}. As noted by
\citet{2020ApJ...904L..13R}, however, the maximum possible BH mass depends
crucially on how strong the LBV-like wind mass loss is and whether the merged
stars can retain their large hydrogen mass. As shown here, some BH-forming,
merged BSGs may be able to avoid the LBV regime, but their total mass is limited
by the fact that these stars have to have CO core masses which fall into the
compactness peak where we anticipate BH formation. Hence, even at lower
metallicity, it seems that multiple/repeated mergers are required to obtain BHs
of ${\gtrsim}\,66\text{--}85\,\msun$ when assuming that LBV-like mass loss can
erode most of the hydrogen-rich envelope of stars.

The maximum SN ejecta mass in our models is closely related to the maximum BH
mass if we assume that models lose almost their entire envelope by LBV-like mass
loss. In this case, the most massive BH forms from a \caseb accretor with a core
mass falling into the compactness peak while the SN with the largest ejecta mass
is a model that has a core just a little bit less massive such that it does not
reach high core compactness and explodes in a SN. Consequently, we predict that
the maximum SN ejecta mass is a little bit less massive than the most massive BH
(\cf Fig.~\ref{fig:max-bh-mej}). Our accretors with high SN ejecta mass explode
as Type II SNe. Observationally, \citet{2003ApJ...582..905H} report SN ejecta
masses in the range $14\text{--}56\,\msun$ for SN~IIP, and SN ejecta masses of
$10\text{--}60\,\msun$ are reported for long-rising \snesa-like Type~II SNe
\citep{2016A&A...588A...5T, 2023MNRAS.tmp..837P}. In our models at solar
metallicity, single stars have SN ejecta masses of ${\lesssim}\,15\,\msun$ and
can thus not account for the observed large SN ejecta masses \citep[see
also][]{2021A&A...645A...6Z}. Contrarily, the accretor models can have larger
final stellar masses and thus better match these observations.

\subsection{\label{sec:formation-massive-mergers}Formation channels for $f_\mathrm{acc}\,{>}\,1$ mergers}

In the following, we denote the primary star in a binary system as the initially
more massive and thus more evolved star at the beginning of the first phase of
binary mass exchange (stable mass transfer or stellar merger). In a merger, it
is expected that its core forms the core of the merger remnant because it is
denser and less buoyant\footnote{In almost equal-mass merger, however, the core
of the secondary often forms the core of the merger product \citep[see
\eg][]{2013MNRAS.434.3497G, 2019Natur.574..211S}.}. In this picture, the less
massive secondary star accretes onto the more massive primary star and
$f_\mathrm{acc}\,{\leq}\,1$ in our models of standard stellar mergers in
isolated binary stars. 

In binary mass transfer, $f_\mathrm{acc}\,{>}\,1$ is possible, because now the
initially less massive secondary accretes mass from the initially more massive
primary. In most cases, the accretor will be a main sequence star (
accretor). \caseb and \casec accretors, \ie accreting post-main sequence
stars, are only possible in binaries with initial mass ratios close to one
because the post-main sequence lifetime is short compared to that of the main
sequence. Hence, \caseb and \casec accretors are also only possible up to
$f_\mathrm{acc}\,{\approx}\,1$ in the first mass transfer phase of binary stars. 

Triples and other multiples are common among massive stars \citep[${>}\,50\%$ of
all systems, see \eg][]{2014ApJS..215...15S, 2017ApJS..230...15M,
2023ASPC..534..275O}. In triples and higher order multiples, mergers with
$f_\mathrm{acc}\,{>}\,1$ are possible. For example, the initially most massive
star could merge with a close binary system whose individual masses do not
exceed the mass of the primary but its total mass does, possibly leading to
\casea, B and~C mergers with $f_\mathrm{acc}\,{>}\,1.0$. Similarly, an initially
hierarchical triple/multiple system can be rendered unstable if, \eg, mass
transfer in a close binary leads to orbital widening \citep[see \eg merger model
in an initial triple for the formation of $\eta$~Car,][]{2021MNRAS.503.4276H}.
Furthermore, \citet{2014ApJ...796..121J} estimate that 10\% of early \caseb
mergers might later accrete additional mass from an outer tertiary companion,
effectively leading to accretors with $f_\mathrm{acc}\,{>}\,1.0$.

In star clusters and other dense stellar systems, repeated mergers of stars are
possible on a short timescale \citep[\eg][]{2002ApJ...576..899P,
2004Natur.428..724P, 2016MNRAS.459.3432M, 2018MNRAS.478.2461G,
2019MNRAS.487.2947D, 2020MNRAS.497.1043D}. This mechanism can easily explain
effective accretion fractions of $f_\mathrm{acc}\,{>}\,1.0$ and even
$f_\mathrm{acc}\,{>}\,2.0$ that are not covered here. Such mergers have in fact
been invoked to explain how BHs may populate the PISN mass gap
\citep[\eg][]{2019MNRAS.487.2947D, 2020MNRAS.497.1043D} and they can also form
in sequential mergers in triples and higher-order multiples
\citep{2021ApJ...907L..19V}.

%
%
\section{\label{sec:conclusions}Summary and conclusions}

Envelope stripping because of binary mass exchange is known to affect the
structures of stars at core collapse and generally allows stars to explode in
supernovae over a larger initial mass range than what is expected from single
stars \citep{1996ApJ...457..834T, 1999A&A...350..148W, 2001NewA....6..457B,
2004ApJ...612.1044P, 2019ApJ...878...49W, 2020ApJ...890...51E,
2021A&A...645A...5S, 2021A&A...656A..58L, 2023ApJ...950L...9S}. In this work, we
have explored how mass \emph{accretion} as such in stable Roche-lobe
overflow or in a stellar merger affects the further evolution of stars and the
stellar structures just prior to core collapse. To this end, we assume that
stars accrete a fraction $f_\mathrm{acc}$ of their initial mass on a thermal
timescale at various points in their evolution. Our main findings can be
summarised as follows.
\begin{itemize}
    \item Beyond certain accretion thresholds and regardless of when stars
    accrete mass in their evolution, they can evolve through a long-lived BSG
    phase during which they burn helium in their cores. Some also reach core
    collapse as BSGs and thereby avoid the CSG phase altogether. The BSG phase
    is related to a thick, convective, hydrogen-burning shell in stars with a
    sufficiently large envelope mass compared to their core mass such that
    hydrogen-shell burning becomes convective.
    \item The accretion thresholds depend on when stars accrete mass in their
    evolution, the initial mass, and also the amount of mixing of core material
    into the envelope (the latter is only relevant for stellar mergers and has
    not been considered here). Most of our \casea accretors fully rejuvenate
    such that they evolve just as single stars of higher mass. At low accretion
    fractions, however, for which stars do not fully rejuvenate, we also find
    long-lived BSG phases. Generally, the more evolved the post-main-sequence
    accretors are, the more mass they need to accrete to become long-lived BSGs. 
    \item At core collapse, all (non-rejuvenated) accretors have a core size
    within 10--20\% of that of single stars of the original initial mass but a
    possibly much higher envelope mass. Because of the larger envelope mass,
    accretors end core helium burning with a lower central carbon abundance than
    genuine single stars. These fewer carbon atoms change the burning episodes
    beyond core-helium burning and generally lead to pre-SN structures that are
    more difficult to explode than those of single stars with the same final
    core mass. This is the opposite trend of what is found in binary-stripped
    stars.
    \item The final masses of non-rejuvenated \casea, and all \caseb/C accretors
    are larger than those of single stars, with those of \caseb accretors being
    the largest. In particular, accretors that evolve through a long-lived BSG
    phase can avoid the higher wind mass-loss rates as CSGs and
    thereby retain the most mass.
    \item We find that the central entropy is a particularly useful summary
    quantity of the pre-SN structures because it directly links to the effective
    Chandrasekhar mass at which core collapse occurs and contains information on
    the mass-radius relation of the inner core region. For these reasons, we
    prefer this quantity over the often-used compactness parameter.
    \item The trend of central entropy (and compactness) at core collapse with
    the CO core mass $M_\mathrm{CO}$ is qualitatively similar to that found in
    single and binary-stripped stars: there are two main characteristic
    $M_\mathrm{CO}$ of about $7\,\msun$ and $13\,\msun$ at which we find high
    central entropy/compactness. These core masses are linked to central carbon
    and neon burning becoming radiative as a consequence of a low carbon
    abundance after core-helium burning. While the overall trend is shifted to
    larger $M_\mathrm{CO}$ in binary-stripped stars, it is shifted to lower
    $M_\mathrm{CO}$ in (non-rejuvenated) accretors. Stars that evolve through a
    long-lived BSG phase show the largest shift towards lower $M_\mathrm{CO}$ of
    about $0.5\,\msun$.
    \item When analysing our pre-SN interior structures with the neutrino-driven
    SN model of \citet{2016MNRAS.460..742M}, we find that the explodability
    closely follows the trend of central entropy with core mass. We find BH
    formation for the high central entropy models with $M_\mathrm{CO}$ of
    $\approx7\,\msun$ and $\gtrsim13\,\msun$, successful SN explosions with NS
    formation for $M_\mathrm{CO}\lesssim 7\,\msun$, and partial fallback SNe with
    likely BH formation in some models with $M_\mathrm{CO}\approx
    7\text{--}13\,\msun$.
    \item The positions in the HR diagram of accretors that do not reach core
    collapse in the long-lived BSG phase, are in the same region as those of
    single stars. Because of the island of BH formation at
    $M_\mathrm{CO}\approx7\,\msun$, accretors ending their lives as CSGs
    collapse into BHs at luminosities of $\log L/\lsun\approx5.4$, the same
    luminosity as in BH-forming CSGs from genuine single stars. Accretors with
    $M_\mathrm{CO}\approx7\,\msun$ that end their lives as BSGs and collapse
    into BHs, populate a region along the hot side of the S~Dor instability
    strip.
    \item Some of the accretors evolving as BSGs explode inside the S~Dor
    instability strip, \ie our models predict that some LBVs may be SN
    progenitors. If the supernova explodes into recently ejected material from
    an LBV eruption, interacting and possibly superluminous SNe seem possible.
    \item In all of our accretor models, we find a tight relation between their
    central entropy at core collapse and the resulting NS mass. A similar
    relation has already been found in single stars with varying degrees of
    convective boundary mixing \citep{2024A&A...682A.123T} and the very same relation
    also fits our models. 
    \item Because accretors evolving and ending their lives as BSGs have high
    final stellar masses, they can also result in large BH and SN ejecta masses.
    Accounting only for models that likely do not experience strong LBV-like
    mass loss and restricting $f_\mathrm{acc}$ to values of $\leq1$ (\ie values
    achievable by stellar mergers in binary stars), we find maximum BH and SN
    ejecta masses of up to $50\,\msun$ and $40\,\msun$, respectively. The ejecta
    masses greatly exceed those from genuine single stars. Allowing for
    repeated/multiple mergers (\ie $f_\mathrm{acc}>1$), both masses increase to
    about $80\,\msun$. This could potentially result in BH masses from stars at
    solar metallicity that fall into the PISN mass gap.
\end{itemize}
We have shown here that accretor models approximating mass gainers in binary
mass transfer and products of stellar mergers reach core collapse with interior
structures that can be profoundly different from single stars. While our
models cannot fully account for the complex physics of binary mass transfer and
stellar mergers, they serve as a zeroth-order baseline. To obtain a more
complete picture of stellar mergers and specifically the mixing in such events,
further detailed multi-dimensional, hydrodynamic simulations and a systematic
exploration of the relevant parameter space are warranted. This will be explored
in upcoming work. We have focussed on the core structures at core collapse, the
explodability of models, and their final BH and SN ejecta masses. In a
forthcoming publication \citep{Schneider+2024b}, we will present further
explosion properties and implications for the zoo of supernova transients.

\begin{acknowledgements}
We thank the reviewer for their careful reading of our manuscript and helpful and constructive comments.
The authors acknowledge support by the Klaus Tschira Foundation.
This work has received funding from the European Research Council (ERC) under the European Union’s Horizon 2020 research and innovation programme (Grant agreement No.\ 945806) and is supported by the Deutsche Forschungsgemeinschaft (DFG, German Research Foundation) under Germany’s Excellence Strategy EXC 2181/1-390900948 (the Heidelberg STRUCTURES Excellence Cluster).
This research made use of NumPy \citep{oliphant2006numpy}, SciPy
\citep{2020NatMe..17..261V}, Matplotlib \citep{hunter2007matplotlib} and Jupyter
Notebooks \citep{kluyver2016jupyternotebook}.
\end{acknowledgements}

\bibliographystyle{aa}

\begin{thebibliography}{194}
\expandafter\ifx\csname natexlab\endcsname\relax\def\natexlab#1{#1}\fi

\bibitem[{{Abbott} {et~al.}(2020{\natexlab{a}}){Abbott}, {Abbott}, {Abraham},
  {Acernese}, {Ackley}, {Adams}, {Adhikari}, {Adya}, {Affeldt}, {Agathos},
  {Agatsuma}, {Aggarwal}, {Aguiar}, {Aich}, {Aiello}, {Ain}, {Ajith}, {Akcay},
  {Allen}, {Allocca}, {Altin}, {Amato}, {Anand}, {Ananyeva}, {Anderson},
  {Anderson}, {Angelova}, {Ansoldi}, {Antier}, {Appert}, {Arai}, {Araya},
  {Areeda}, {Ar{\`e}ne}, {Arnaud}, {Aronson}, {Arun}, {Asali}, {Ascenzi},
  {Ashton}, {Aston}, {Astone}, {Aubin}, {Aufmuth}, {AultONeal}, {Austin},
  {Avendano}, {Babak}, {Bacon}, {Badaracco}, {Bader}, {Bae}, {Baer}, {Baird},
  {Baldaccini}, {Ballardin}, {Ballmer}, {Bals}, {Balsamo}, {Baltus},
  {Banagiri}, {Bankar}, {Bankar}, {Barayoga}, {Barbieri}, {Barish}, {Barker},
  {Barkett}, {Barneo}, {Barone}, {Barr}, {Barsotti}, {Barsuglia}, {Barta},
  {Bartlett}, {Bartos}, {Bassiri}, {Basti}, {Bawaj}, {Bayley}, {Bazzan},
  {B{\'e}csy}, {Bejger}, {Belahcene}, {Bell}, {Beniwal}, {Benjamin}, {Bentley},
  {Bergamin}, {Berger}, {Bergmann}, {Bernuzzi}, {Berry}, {Bersanetti},
  {Bertolini}, {Betzwieser}, {Bhandare}, {Bhandari}, {Bidler}, {Biggs},
  {Bilenko}, {Billingsley}, {Birney}, {Birnholtz}, {Biscans}, {Bischi},
  {Biscoveanu}, {Bisht}, {Bissenbayeva}, {Bitossi}, {Bizouard}, {Blackburn},
  {Blackman}, {Blair}, {Blair}, {Blair}, {Bobba}, {Bode}, {Boer}, {Boetzel},
  {Bogaert}, {Bondu}, {Bonilla}, {Bonnand}, {Booker}, {Boom}, {Bork}, {Boschi},
  {Bose}, {Bossilkov}, {Bosveld}, {Bouffanais}, {Bozzi}, {Bradaschia}, {Brady},
  {Bramley}, {Branchesi}, {Brau}, {Breschi}, {Briant}, {Briggs}, {Brighenti},
  {Brillet}, {Brinkmann}, {Brockill}, {Brooks}, {Brooks}, {Brown}, {Brunett},
  {Bruno}, {Bruntz}, {Buikema}, {Bulik}, {Bulten}, {Buonanno}, {Buscicchio},
  {Buskulic}, {Byer}, {Cabero}, {Cadonati}, {Cagnoli}, {Cahillane},
  {Calder{\'o}n Bustillo}, {Callaghan}, {Callister}, {Calloni}, {Camp},
  {Canepa}, {Cannon}, {Cao}, {Cao}, {Carapella}, {Carbognani}, {Caride},
  {Carney}, {Carullo}, {Casanueva Diaz}, {Casentini}, {Casta{\~n}eda},
  {Caudill}, {Cavagli{\`a}}, {Cavalier}, {Cavalieri}, {Cella},
  {Cerd{\'a}-Dur{\'a}n}, {Cesarini}, {Chaibi}, {Chakravarti}, {Chan}, {Chan},
  {Chandra}, {Chao}, {Charlton}, {Chase}, {Chassande-Mottin}, {Chatterjee},
  {Chaturvedi}, {Chatziioannou}, {Chen}, {Chen}, {Chen}, {Cheng}, {Cheong},
  {Chia}, {Chiadini}, {Chierici}, {Chincarini}, {Chiummo}, {Cho}, {Cho}, {Cho},
  {Christensen}, {Chu}, {Chua}, {Chung}, {Chung}, {Ciani}, {Ciecielag},
  {Cie{\'s}lar}, {Ciobanu}, {Ciolfi}, {Cipriano}, {Cirone}, {Clara}, {Clark},
  {Clearwater}, {Clesse}, {Cleva}, {Coccia}, {Cohadon}, {Cohen}, {Colleoni},
  {Collette}, {Collins}, {Colpi}, {Constancio}, {Conti}, {Cooper}, {Corban},
  {Corbitt}, {Cordero-Carri{\'o}n}, {Corezzi}, {Corley}, {Cornish}, {Corre},
  {Corsi}, {Cortese}, {Costa}, {Cotesta}, {Coughlin}, {Coughlin}, {Coulon},
  {Countryman}, {Couvares}, {Covas}, {Coward}, {Cowart}, {Coyne}, {Coyne},
  {Creighton}, {Creighton}, {Cripe}, {Croquette}, {Crowder}, {Cudell},
  {Cullen}, {Cumming}, {Cummings}, {Cunningham}, {Cuoco}, {Curylo}, {Canton},
  {D{\'a}lya}, {Dana}, {Daneshgaran-Bajastani}, {D'Angelo}, {Danilishin},
  {D'Antonio}, {Danzmann}, {Darsow-Fromm}, {Dasgupta}, {Datrier}, {Dattilo},
  {Dave}, {Davier}, {Davies}, {Davis}, {Daw}, {DeBra}, {Deenadayalan},
  {Degallaix}, {De Laurentis}, {Del{\'e}glise}, {Delfavero}, {De Lillo}, {Del
  Pozzo}, {DeMarchi}, {D'Emilio}, {Demos}, {Dent}, {De Pietri}, {De Rosa}, {De
  Rossi}, {DeSalvo}, {de Varona}, {Dhurandhar}, {D{\'\i}az}, {Diaz-Ortiz},
  {Dietrich}, {Di Fiore}, {Di Fronzo}, {Di Giorgio}, {Di Giovanni}, {Di
  Giovanni}, {Di Girolamo}, {Di Lieto}, {Ding}, {Di Pace}, {Di Palma}, {Di
  Renzo}, {Divakarla}, {Dmitriev}, {Doctor}, {Donovan}, {Dooley}, {Doravari},
  {Dorrington}, {Downes}, {Drago}, {Driggers}, {Du}, {Ducoin}, {Dupej},
  {Durante}, {D'Urso}, {Dwyer}, {Easter}, {Eddolls}, {Edelman}, {Edo}, {Edy},
  {Effler}, {Ehrens}, {Eichholz}, {Eikenberry}, {Eisenmann}, {Eisenstein},
  {Ejlli}, {Errico}, {Essick}, {Estelles}, {Estevez}, {Etienne}, {Etzel},
  {Evans}, {Evans}, {Ewing}, {Fafone}, {Fairhurst}, {Fan}, {Farinon}, {Farr},
  {Farr}, {Fauchon-Jones}, {Favata}, {Fays}, {Fazio}, {Feicht}, {Fejer},
  {Feng}, {Fenyvesi}, {Ferguson}, {Fernandez-Galiana}, {Ferrante}, {Ferreira},
  {Ferreira}, {Fidecaro}, {Fiori}, {Fiorucci}, {Fishbach}, {Fisher},
  {Fittipaldi}, {Fitz-Axen}, {Fiumara}, {Flaminio}, {Floden}, {Flynn}, {Fong},
  {Font}, {Forsyth}, {Fournier}, {Frasca}, {Frasconi}, {Frei}, {Freise},
  {Frey}, {Frey}, {Fritschel}, {Frolov}, {Fronz{\`e}}, {Fulda}, {Fyffe},
  {Gabbard}, {Gadre}, {Gaebel}, {Gair}, {Galaudage}, {Ganapathy}, {Ganguly},
  {Gaonkar}, {Garc{\'\i}a-Quir{\'o}s}, {Garufi}, {Gateley}, {Gaudio},
  {Gayathri}, {Gemme}, {Genin}, {Gennai}, {George}, {George}, {Gergely},
  {Ghonge}, {Ghosh}, {Ghosh}, {Ghosh}, {Giacomazzo}, {Giaime}, {Giardina},
  {Gibson}, {Gier}, {Gill}, {Glanzer}, {Gniesmer}, {Godwin}, {Goetz}, {Goetz},
  {Gohlke}, {Goncharov}, {Gonz{\'a}lez}, {Gopakumar}, {Gossan}, {Gosselin},
  {Gouaty}, {Grace}, {Grado}, {Granata}, {Grant}, {Gras}, {Grassia}, {Gray},
  {Gray}, {Greco}, {Green}, {Green}, {Gretarsson}, {Griggs}, {Grignani},
  {Grimaldi}, {Grimm}, {Grote}, {Grunewald}, {Gruning}, {Guidi}, {Guimaraes},
  {Guix{\'e}}, {Gulati}, {Guo}, {Gupta}, {Gupta}, {Gupta}, {Gustafson},
  {Gustafson}, {Haegel}, {Halim}, {Hall}, {Hamilton}, {Hammond}, {Haney},
  {Hanke}, {Hanks}, {Hanna}, {Hannam}, {Hannuksela}, {Hansen}, {Hanson},
  {Harder}, {Hardwick}, {Haris}, {Harms}, {Harry}, {Harry}, {Hasskew},
  {Haster}, {Haughian}, {Hayes}, {Healy}, {Heidmann}, {Heintze}, {Heinze},
  {Heitmann}, {Hellman}, {Hello}, {Hemming}, {Hendry}, {Heng}, {Hennes},
  {Hennig}, {Heurs}, {Hild}, {Hinderer}, {Hoback}, {Hochheim}, {Hofgard},
  {Hofman}, {Holgado}, {Holland}, {Holt}, {Holz}, {Hopkins}, {Horst}, {Hough},
  {Howell}, {Hoy}, {Huang}, {H{\"u}bner}, {Huerta}, {Huet}, {Hughey}, {Hui},
  {Husa}, {Huttner}, {Huxford}, {Huynh-Dinh}, {Idzkowski}, {Iess}, {Inchauspe},
  {Ingram}, {Intini}, {Isac}, {Isi}, {Iyer}, {Jacqmin}, {Jadhav}, {Jadhav},
  {James}, {Jani}, {Janthalur}, {Jaranowski}, {Jariwala}, {Jaume}, {Jenkins},
  {Jiang}, {Johns}, {Johnson-McDaniel}, {Jones}, {Jones}, {Jones}, {Jones},
  {Jones}, {Jonker}, {Ju}, {Junker}, {Kalaghatgi}, {Kalogera}, {Kamai},
  {Kandhasamy}, {Kang}, {Kanner}, {Kapadia}, {Karki}, {Kashyap}, {Kasprzack},
  {Kastaun}, {Katsanevas}, {Katsavounidis}, {Katzman}, {Kaufer}, {Kawabe},
  {K{\'e}f{\'e}lian}, {Keitel}, {Keivani}, {Kennedy}, {Key}, {Khadka},
  {Khalili}, {Khan}, {Khan}, {Khan}, {Khazanov}, {Khetan}, {Khursheed},
  {Kijbunchoo}, {Kim}, {Kim}, {Kim}, {Kim}, {Kim}, {Kim}, {Kim}, {Kimball},
  {King}, {Kinley-Hanlon}, {Kirchhoff}, {Kissel}, {Kleybolte}, {Klimenko},
  {Knowles}, {Knyazev}, {Koch}, {Koehlenbeck}, {Koekoek}, {Koley},
  {Kondrashov}, {Kontos}, {Koper}, {Korobko}, {Korth}, {Kovalam}, {Kozak},
  {Kringel}, {Krishnendu}, {Kr{\'o}lak}, {Krupinski}, {Kuehn}, {Kumar},
  {Kumar}, {Kumar}, {Kumar}, {Kumar}, {Kuo}, {Kutynia}, {Lackey}, {Laghi},
  {Lalande}, {Lam}, {Lamberts}, {Landry}, {Lane}, {Lang}, {Lange}, {Lantz},
  {Lanza}, {La Rosa}, {Lartaux-Vollard}, {Lasky}, {Laxen}, {Lazzarini},
  {Lazzaro}, {Leaci}, {Leavey}, {Lecoeuche}, {Lee}, {Lee}, {Lee}, {Lee}, {Lee},
  {Lehmann}, {Leroy}, {Letendre}, {Levin}, {Li}, {Li}, {li}, {Li}, {Li},
  {Linde}, {Linker}, {Linley}, {Littenberg}, {Liu}, {Liu},
  {Llorens-Monteagudo}, {Lo}, {Lockwood}, {London}, {Longo}, {Lorenzini},
  {Loriette}, {Lormand}, {Losurdo}, {Lough}, {Lousto}, {Lovelace}, {L{\"u}ck},
  {Lumaca}, {Lundgren}, {Ma}, {Macas}, {Macfoy}, {MacInnis}, {Macleod},
  {MacMillan}, {Macquet}, {Maga{\~n}a Hernandez}, {Maga{\~n}a-Sandoval},
  {Magee}, {Majorana}, {Maksimovic}, {Malik}, {Man}, {Mandic}, {Mangano},
  {Mansell}, {Manske}, {Mantovani}, {Mapelli}, {Marchesoni}, {Marion},
  {M{\'a}rka}, {M{\'a}rka}, {Markakis}, {Markosyan}, {Markowitz}, {Maros},
  {Marquina}, {Marsat}, {Martelli}, {Martin}, {Martin}, {Martinez}, {Martynov},
  {Masalehdan}, {Mason}, {Massera}, {Masserot}, {Massinger}, {Masso-Reid},
  {Mastrogiovanni}, {Matas}, {Matichard}, {Mavalvala}, {Maynard}, {McCann},
  {McCarthy}, {McClelland}, {McCormick}, {McCuller}, {McGuire}, {McIsaac},
  {McIver}, {McManus}, {McRae}, {McWilliams}, {Meacher}, {Meadors}, {Mehmet},
  {Mehta}, {Mejuto Villa}, {Melatos}, {Mendell}, {Mercer}, {Mereni}, {Merfeld},
  {Merilh}, {Merritt}, {Merzougui}, {Meshkov}, {Messenger}, {Messick},
  {Metzdorff}, {Meyers}, {Meylahn}, {Mhaske}, {Miani}, {Miao}, {Michaloliakos},
  {Michel}, {Middleton}, {Milano}, {Miller}, {Millhouse}, {Mills}, {Milotti},
  {Milovich-Goff}, {Minazzoli}, {Minenkov}, {Mishkin}, {Mishra}, {Mistry},
  {Mitra}, {Mitrofanov}, {Mitselmakher}, {Mittleman}, {Mo}, {Mogushi},
  {Mohapatra}, {Mohite}, {Molina-Ruiz}, {Mondin}, {Montani}, {Moore}, {Moraru},
  {Morawski}, {Moreno}, {Morisaki}, {Mours}, {Mow-Lowry}, {Mozzon},
  {Muciaccia}, {Mukherjee}, {Mukherjee}, {Mukherjee}, {Mukherjee}, {Mukund},
  {Mullavey}, {Munch}, {Mu{\~n}iz}, {Murray}, {Nagar}, {Nardecchia},
  {Naticchioni}, {Nayak}, {Neil}, {Neilson}, {Nelemans}, {Nelson}, {Nery},
  {Neunzert}, {Ng}, {Ng}, {Nguyen}, {Nguyen}, {Nichols}, {Nichols}, {Nissanke},
  {Nitz}, {Nocera}, {Noh}, {North}, {Nothard}, {Nuttall}, {Oberling},
  {O'Brien}, {Oganesyan}, {Ogin}, {Oh}, {Oh}, {Ohme}, {Ohta}, {Okada},
  {Oliver}, {Olivetto}, {Oppermann}, {Oram}, {O'Reilly}, {Ormiston}, {Ortega},
  {O'Shaughnessy}, {Ossokine}, {Osthelder}, {Ottaway}, {Overmier}, {Owen},
  {Pace}, {Pagano}, {Page}, {Pagliaroli}, {Pai}, {Pai}, {Palamos}, {Palashov},
  {Palomba}, {Pan}, {Panda}, {Pang}, {Pankow}, {Pannarale}, {Pant}, {Paoletti},
  {Paoli}, {Parida}, {Parker}, {Pascucci}, {Pasqualetti}, {Passaquieti},
  {Passuello}, {Patricelli}, {Payne}, {Pearlstone}, {Pechsiri}, {Pedersen},
  {Pedraza}, {Pele}, {Penn}, {Perego}, {Perez}, {P{\'e}rigois}, {Perreca},
  {Perri{\`e}s}, {Petermann}, {Pfeiffer}, {Phelps}, {Phukon}, {Piccinni},
  {Pichot}, {Piendibene}, {Piergiovanni}, {Pierro}, {Pillant}, {Pinard},
  {Pinto}, {Piotrzkowski}, {Pirello}, {Pitkin}, {Plastino}, {Poggiani}, {Pong},
  {Ponrathnam}, {Popolizio}, {Porter}, {Powell}, {Prajapati}, {Prasai},
  {Prasanna}, {Pratten}, {Prestegard}, {Principe}, {Prodi}, {Prokhorov},
  {Punturo}, {Puppo}, {P{\"u}rrer}, {Qi}, {Quetschke}, {Quinonez}, {Raab},
  {Raaijmakers}, {Radkins}, {Radulesco}, {Raffai}, {Rafferty}, {Raja}, {Rajan},
  {Rajbhandari}, {Rakhmanov}, {Ramirez}, {Ramos-Buades}, {Rana}, {Rao},
  {Rapagnani}, {Raymond}, {Razzano}, {Read}, {Regimbau}, {Rei}, {Reid},
  {Reitze}, {Rettegno}, {Ricci}, {Richardson}, {Richardson}, {Ricker},
  {Riemenschneider}, {Riles}, {Rizzo}, {Robertson}, {Robinet}, {Rocchi},
  {Rodriguez-Soto}, {Rolland}, {Rollins}, {Roma}, {Romanelli}, {Romano},
  {Romel}, {Romero-Shaw}, {Romie}, {Rose}, {Rose}, {Rose}, {Rosi{\'n}ska},
  {Rosofsky}, {Ross}, {Rowan}, {Rowlinson}, {Roy}, {Roy}, {Roy}, {Ruggi},
  {Rutins}, {Ryan}, {Sachdev}, {Sadecki}, {Sakellariadou}, {Salafia},
  {Salconi}, {Saleem}, {Salemi}, {Samajdar}, {Sanchez}, {Sanchez},
  {Sanchis-Gual}, {Sanders}, {Santiago}, {Santos}, {Sarin}, {Sassolas},
  {Sathyaprakash}, {Sauter}, {Savage}, {Savant}, {Sawant}, {Sayah}, {Schaetzl},
  {Schale}, {Scheel}, {Scheuer}, {Schmidt}, {Schnabel}, {Schofield},
  {Sch{\"o}nbeck}, {Schreiber}, {Schulte}, {Schutz}, {Schwarm}, {Schwartz},
  {Scott}, {Scott}, {Seidel}, {Sellers}, {Sengupta}, {Sennett}, {Sentenac},
  {Sequino}, {Sergeev}, {Setyawati}, {Shaddock}, {Shaffer}, {Sharifi},
  {Shahriar}, {Sharma}, {Sharma}, {Shawhan}, {Shen}, {Shikauchi}, {Shink},
  {Shoemaker}, {Shoemaker}, {Shukla}, {ShyamSundar}, {Siellez}, {Sieniawska},
  {Sigg}, {Singer}, {Singh}, {Singh}, {Singha}, {Singhal}, {Sintes}, {Sipala},
  {Skliris}, {Slagmolen}, {Slaven-Blair}, {Smetana}, {Smith}, {Smith},
  {Somala}, {Son}, {Soni}, {Sorazu}, {Sordini}, {Sorrentino}, {Souradeep},
  {Sowell}, {Spencer}, {Spera}, {Srivastava}, {Srivastava}, {Staats},
  {Stachie}, {Standke}, {Steer}, {Steinke}, {Steinlechner}, {Steinlechner},
  {Steinmeyer}, {Stevenson}, {Stocks}, {Stops}, {Stover}, {Strain}, {Stratta},
  {Strunk}, {Sturani}, {Stuver}, {Sudhagar}, {Sudhir}, {Summerscales}, {Sun},
  {Sunil}, {Sur}, {Suresh}, {Sutton}, {Swinkels}, {Szczepa{\'n}czyk}, {Tacca},
  {Tait}, {Talbot}, {Tanasijczuk}, {Tanner}, {Tao}, {T{\'a}pai}, {Tapia},
  {Tapia San Martin}, {Tasson}, {Taylor}, {Tenorio}, {Terkowski},
  {Thirugnanasambandam}, {Thomas}, {Thomas}, {Thompson}, {Thondapu}, {Thorne},
  {Thrane}, {Tinsman}, {Saravanan}, {Tiwari}, {Tiwari}, {Tiwari}, {Toland},
  {Tonelli}, {Tornasi}, {Torres-Forn{\'e}}, {Torrie}, {Tosta e Melo},
  {T{\"o}yr{\"a}}, {Travasso}, {Traylor}, {Tringali}, {Tripathee}, {Trovato},
  {Trudeau}, {Tsang}, {Tse}, {Tso}, {Tsukada}, {Tsuna}, {Tsutsui}, {Turconi},
  {Ubhi}, {Udall}, {Ueno}, {Ugolini}, {Unnikrishnan}, {Urban}, {Usman},
  {Utina}, {Vahlbruch}, {Vajente}, {Valdes}, {Valentini}, {van Bakel}, {van
  Beuzekom}, {van den Brand}, {Van Den Broeck}, {Vander-Hyde}, {van der
  Schaaf}, {Van Heijningen}, {van Veggel}, {Vardaro}, {Varma}, {Vass},
  {Vas{\'u}th}, {Vecchio}, {Vedovato}, {Veitch}, {Veitch}, {Venkateswara},
  {Venugopalan}, {Verkindt}, {Veske}, {Vetrano}, {Vicer{\'e}}, {Viets},
  {Vinciguerra}, {Vine}, {Vinet}, {Vitale}, {Vivanco}, {Vo}, {Vocca},
  {Vorvick}, {Vyatchanin}, {Wade}, {Wade}, {Wade}, {Walet}, {Walker},
  {Wallace}, {Wallace}, {Walsh}, {Wang}, {Wang}, {Wang}, {Ward}, {Warden},
  {Warner}, {Was}, {Watchi}, {Weaver}, {Wei}, {Weinert}, {Weinstein}, {Weiss},
  {Wellmann}, {Wen}, {We{\ss}els}, {Westhouse}, {Wette}, {Whelan}, {Whiting},
  {Whittle}, {Wilken}, {Williams}, {Willis}, {Willke}, {Winkler}, {Wipf},
  {Wittel}, {Woan}, {Woehler}, {Wofford}, {Wong}, {Wright}, {Wu}, {Wysocki},
  {Xiao}, {Yamamoto}, {Yang}, {Yang}, {Yang}, {Yap}, {Yazback}, {Yeeles}, {Yu},
  {Yu}, {Yuen}, {Zadro{\.Z}ny}, {Zadro{\.Z}ny}, {Zanolin}, {Zelenova},
  {Zendri}, {Zevin}, {Zhang}, {Zhang}, {Zhang}, {Zhao}, {Zhao}, {Zhou}, {Zhou},
  {Zhu}, {Zimmerman}, {Zucker}, {Zweizig}, {LIGO Scientific Collaboration}, \&
  {Virgo Collaboration}}]{2020PhRvL.125j1102A}
{Abbott}, R., {Abbott}, T.~D., {Abraham}, S., {et~al.} 2020{\natexlab{a}},
  \prl, 125, 101102

\bibitem[{{Abbott} {et~al.}(2020{\natexlab{b}}){Abbott}, {Abbott}, {Abraham},
  {Acernese}, {Ackley}, {Adams}, {Adhikari}, {Adya}, {Affeldt}, {Agathos},
  {Agatsuma}, {Aggarwal}, {Aguiar}, {Aich}, {Aiello}, {Ain}, {Ajith}, {Akcay},
  {Allen}, {Allocca}, {Altin}, {Amato}, {Anand}, {Ananyeva}, {Anderson},
  {Anderson}, {Angelova}, {Ansoldi}, {Antier}, {Appert}, {Arai}, {Araya},
  {Areeda}, {Ar{\`e}ne}, {Arnaud}, {Aronson}, {Arun}, {Asali}, {Ascenzi},
  {Ashton}, {Aston}, {Astone}, {Aubin}, {Aufmuth}, {AultONeal}, {Austin},
  {Avendano}, {Babak}, {Bacon}, {Badaracco}, {Bader}, {Bae}, {Baer}, {Baird},
  {Baldaccini}, {Ballardin}, {Ballmer}, {Bals}, {Balsamo}, {Baltus},
  {Banagiri}, {Bankar}, {Bankar}, {Barayoga}, {Barbieri}, {Barish}, {Barker},
  {Barkett}, {Barneo}, {Barone}, {Barr}, {Barsotti}, {Barsuglia}, {Barta},
  {Bartlett}, {Bartos}, {Bassiri}, {Basti}, {Bawaj}, {Bayley}, {Bazzan},
  {B{\'e}csy}, {Bejger}, {Belahcene}, {Bell}, {Beniwal}, {Benjamin}, {Bentley},
  {Bergamin}, {Berger}, {Bergmann}, {Bernuzzi}, {Berry}, {Bersanetti},
  {Bertolini}, {Betzwieser}, {Bhandare}, {Bhandari}, {Bidler}, {Biggs},
  {Bilenko}, {Billingsley}, {Birney}, {Birnholtz}, {Biscans}, {Bischi},
  {Biscoveanu}, {Bisht}, {Bissenbayeva}, {Bitossi}, {Bizouard}, {Blackburn},
  {Blackman}, {Blair}, {Blair}, {Blair}, {Bobba}, {Bode}, {Boer}, {Boetzel},
  {Bogaert}, {Bondu}, {Bonilla}, {Bonnand}, {Booker}, {Boom}, {Bork}, {Boschi},
  {Bose}, {Bossilkov}, {Bosveld}, {Bouffanais}, {Bozzi}, {Bradaschia}, {Brady},
  {Bramley}, {Branchesi}, {Brau}, {Breschi}, {Briant}, {Briggs}, {Brighenti},
  {Brillet}, {Brinkmann}, {Brockill}, {Brooks}, {Brooks}, {Brown}, {Brunett},
  {Bruno}, {Bruntz}, {Buikema}, {Bulik}, {Bulten}, {Buonanno}, {Buscicchio},
  {Buskulic}, {Byer}, {Cabero}, {Cadonati}, {Cagnoli}, {Cahillane}, {Bustillo},
  {Callaghan}, {Callister}, {Calloni}, {Camp}, {Canepa}, {Cannon}, {Cao},
  {Cao}, {Carapella}, {Carbognani}, {Caride}, {Carney}, {Carullo}, {Diaz},
  {Casentini}, {Casta{\~n}eda}, {Caudill}, {Cavagli{\`a}}, {Cavalier},
  {Cavalieri}, {Cella}, {Cerd{\'a}-Dur{\'a}n}, {Cesarini}, {Chaibi},
  {Chakravarti}, {Chan}, {Chan}, {Chao}, {Charlton}, {Chase},
  {Chassande-Mottin}, {Chatterjee}, {Chaturvedi}, {Chatziioannou}, {Chen},
  {Chen}, {Chen}, {Cheng}, {Cheong}, {Chia}, {Chiadini}, {Chierici},
  {Chincarini}, {Chiummo}, {Cho}, {Cho}, {Cho}, {Christensen}, {Chu}, {Chua},
  {Chung}, {Chung}, {Ciani}, {Ciecielag}, {Cie{\'s}lar}, {Ciobanu}, {Ciolfi},
  {Cipriano}, {Cirone}, {Clara}, {Clark}, {Clearwater}, {Clesse}, {Cleva},
  {Coccia}, {Cohadon}, {Cohen}, {Colleoni}, {Collette}, {Collins}, {Colpi},
  {Constancio}, {Conti}, {Cooper}, {Corban}, {Corbitt}, {Cordero-Carri{\'o}n},
  {Corezzi}, {Corley}, {Cornish}, {Corre}, {Corsi}, {Cortese}, {Costa},
  {Cotesta}, {Coughlin}, {Coughlin}, {Coulon}, {Countryman}, {Couvares},
  {Covas}, {Coward}, {Cowart}, {Coyne}, {Coyne}, {Creighton}, {Creighton},
  {Cripe}, {Croquette}, {Crowder}, {Cudell}, {Cullen}, {Cumming}, {Cummings},
  {Cunningham}, {Cuoco}, {Curylo}, {Canton}, {D{\'a}lya}, {Dana},
  {Daneshgaran-Bajastani}, {D'Angelo}, {Danilishin}, {D'Antonio}, {Danzmann},
  {Darsow-Fromm}, {Dasgupta}, {Datrier}, {Dattilo}, {Dave}, {Davier}, {Davies},
  {Davis}, {Daw}, {DeBra}, {Deenadayalan}, {Degallaix}, {De Laurentis},
  {Del{\'e}glise}, {Delfavero}, {De Lillo}, {Del Pozzo}, {DeMarchi},
  {D'Emilio}, {Demos}, {Dent}, {De Pietri}, {De Rosa}, {De Rossi}, {DeSalvo},
  {de Varona}, {Dhurandhar}, {D{\'\i}az}, {Diaz-Ortiz}, {Dietrich}, {Di Fiore},
  {Di Fronzo}, {Di Giorgio}, {Di Giovanni}, {Di Giovanni}, {Di Girolamo}, {Di
  Lieto}, {Ding}, {Di Pace}, {Di Palma}, {Di Renzo}, {Divakarla}, {Dmitriev},
  {Doctor}, {Donovan}, {Dooley}, {Doravari}, {Dorrington}, {Downes}, {Drago},
  {Driggers}, {Du}, {Ducoin}, {Dupej}, {Durante}, {D'Urso}, {Dwyer}, {Easter},
  {Eddolls}, {Edelman}, {Edo}, {Edy}, {Effler}, {Ehrens}, {Eichholz},
  {Eikenberry}, {Eisenmann}, {Eisenstein}, {Ejlli}, {Errico}, {Essick},
  {Estelles}, {Estevez}, {Etienne}, {Etzel}, {Evans}, {Evans}, {Ewing},
  {Fafone}, {Fairhurst}, {Fan}, {Farinon}, {Farr}, {Farr}, {Fauchon-Jones},
  {Favata}, {Fays}, {Fazio}, {Feicht}, {Fejer}, {Feng}, {Fenyvesi}, {Ferguson},
  {Fernandez-Galiana}, {Ferrante}, {Ferreira}, {Ferreira}, {Fidecaro}, {Fiori},
  {Fiorucci}, {Fishbach}, {Fisher}, {Fittipaldi}, {Fitz-Axen}, {Fiumara},
  {Flaminio}, {Floden}, {Flynn}, {Fong}, {Font}, {Forsyth}, {Fournier},
  {Frasca}, {Frasconi}, {Frei}, {Freise}, {Frey}, {Frey}, {Fritschel},
  {Frolov}, {Fronz{\`e}}, {Fulda}, {Fyffe}, {Gabbard}, {Gadre}, {Gaebel},
  {Gair}, {Galaudage}, {Ganapathy}, {Gaonkar}, {Garc{\'\i}a-Quir{\'o}s},
  {Garufi}, {Gateley}, {Gaudio}, {Gayathri}, {Gemme}, {Genin}, {Gennai},
  {George}, {George}, {Gergely}, {Ghonge}, {Ghosh}, {Ghosh}, {Ghosh},
  {Giacomazzo}, {Giaime}, {Giardina}, {Gibson}, {Gier}, {Gill}, {Glanzer},
  {Gniesmer}, {Godwin}, {Goetz}, {Goetz}, {Gohlke}, {Goncharov},
  {Gonz{\'a}lez}, {Gopakumar}, {Gossan}, {Gosselin}, {Gouaty}, {Grace},
  {Grado}, {Granata}, {Grant}, {Gras}, {Grassia}, {Gray}, {Gray}, {Greco},
  {Green}, {Green}, {Gretarsson}, {Griggs}, {Grignani}, {Grimaldi}, {Grimm},
  {Grote}, {Grunewald}, {Gruning}, {Guidi}, {Guimaraes}, {Guix{\'e}}, {Gulati},
  {Guo}, {Gupta}, {Gupta}, {Gupta}, {Gustafson}, {Gustafson}, {Haegel},
  {Halim}, {Hall}, {Hamilton}, {Hammond}, {Haney}, {Hanke}, {Hanks}, {Hanna},
  {Hannam}, {Hannuksela}, {Hansen}, {Hanson}, {Harder}, {Hardwick}, {Haris},
  {Harms}, {Harry}, {Harry}, {Hasskew}, {Haster}, {Haughian}, {Hayes}, {Healy},
  {Heidmann}, {Heintze}, {Heinze}, {Heitmann}, {Hellman}, {Hello}, {Hemming},
  {Hendry}, {Heng}, {Hennes}, {Hennig}, {Heurs}, {Hild}, {Hinderer}, {Hoback},
  {Hochheim}, {Hofgard}, {Hofman}, {Holgado}, {Holland}, {Holt}, {Holz},
  {Hopkins}, {Horst}, {Hough}, {Howell}, {Hoy}, {Huang}, {H{\"u}bner},
  {Huerta}, {Huet}, {Hughey}, {Hui}, {Husa}, {Huttner}, {Huxford},
  {Huynh-Dinh}, {Idzkowski}, {Iess}, {Inchauspe}, {Ingram}, {Intini}, {Isac},
  {Isi}, {Iyer}, {Jacqmin}, {Jadhav}, {Jadhav}, {James}, {Jani}, {Janthalur},
  {Jaranowski}, {Jariwala}, {Jaume}, {Jenkins}, {Jiang}, {Johns},
  {Johnson-McDaniel}, {Jones}, {Jones}, {Jones}, {Jones}, {Jones}, {Jonker},
  {Ju}, {Junker}, {Kalaghatgi}, {Kalogera}, {Kamai}, {Kandhasamy}, {Kang},
  {Kanner}, {Kapadia}, {Karki}, {Kashyap}, {Kasprzack}, {Kastaun},
  {Katsanevas}, {Katsavounidis}, {Katzman}, {Kaufer}, {Kawabe},
  {K{\'e}f{\'e}lian}, {Keitel}, {Keivani}, {Kennedy}, {Key}, {Khadka},
  {Khalili}, {Khan}, {Khan}, {Khan}, {Khazanov}, {Khetan}, {Khursheed},
  {Kijbunchoo}, {Kim}, {Kim}, {Kim}, {Kim}, {Kim}, {Kim}, {Kim}, {Kimball},
  {King}, {Kinley-Hanlon}, {Kirchhoff}, {Kissel}, {Kleybolte}, {Klimenko},
  {Knowles}, {Knyazev}, {Koch}, {Koehlenbeck}, {Koekoek}, {Koley},
  {Kondrashov}, {Kontos}, {Koper}, {Korobko}, {Korth}, {Kovalam}, {Kozak},
  {Kringel}, {Krishnendu}, {Kr{\'o}lak}, {Krupinski}, {Kuehn}, {Kumar},
  {Kumar}, {Kumar}, {Kumar}, {Kumar}, {Kuo}, {Kutynia}, {Lackey}, {Laghi},
  {Lalande}, {Lam}, {Lamberts}, {Landry}, {Lane}, {Lang}, {Lange}, {Lantz},
  {Lanza}, {La Rosa}, {Lartaux-Vollard}, {Lasky}, {Laxen}, {Lazzarini},
  {Lazzaro}, {Leaci}, {Leavey}, {Lecoeuche}, {Lee}, {Lee}, {Lee}, {Lee}, {Lee},
  {Lehmann}, {Leroy}, {Letendre}, {Levin}, {Li}, {Li}, {li}, {Li}, {Li},
  {Linde}, {Linker}, {Linley}, {Littenberg}, {Liu}, {Liu},
  {Llorens-Monteagudo}, {Lo}, {Lockwood}, {London}, {Longo}, {Lorenzini},
  {Loriette}, {Lormand}, {Losurdo}, {Lough}, {Lousto}, {Lovelace}, {L{\"u}ck},
  {Lumaca}, {Lundgren}, {Ma}, {Macas}, {Macfoy}, {MacInnis}, {Macleod},
  {MacMillan}, {Macquet}, {Hernandez}, {Maga{\~n}a-Sandoval}, {Magee},
  {Majorana}, {Maksimovic}, {Malik}, {Man}, {Mandic}, {Mangano}, {Mansell},
  {Manske}, {Mantovani}, {Mapelli}, {Marchesoni}, {Marion}, {M{\'a}rka},
  {M{\'a}rka}, {Markakis}, {Markosyan}, {Markowitz}, {Maros}, {Marquina},
  {Marsat}, {Martelli}, {Martin}, {Martin}, {Martinez}, {Martynov},
  {Masalehdan}, {Mason}, {Massera}, {Masserot}, {Massinger}, {Masso-Reid},
  {Mastrogiovanni}, {Matas}, {Matichard}, {Mavalvala}, {Maynard}, {McCann},
  {McCarthy}, {McClelland}, {McCormick}, {McCuller}, {McGuire}, {McIsaac},
  {McIver}, {McManus}, {McRae}, {McWilliams}, {Meacher}, {Meadors}, {Mehmet},
  {Mehta}, {Villa}, {Melatos}, {Mendell}, {Mercer}, {Mereni}, {Merfeld},
  {Merilh}, {Merritt}, {Merzougui}, {Meshkov}, {Messenger}, {Messick},
  {Metzdorff}, {Meyers}, {Meylahn}, {Mhaske}, {Miani}, {Miao}, {Michaloliakos},
  {Michel}, {Middleton}, {Milano}, {Miller}, {Millhouse}, {Mills}, {Milotti},
  {Milovich-Goff}, {Minazzoli}, {Minenkov}, {Mishkin}, {Mishra}, {Mistry},
  {Mitra}, {Mitrofanov}, {Mitselmakher}, {Mittleman}, {Mo}, {Mogushi},
  {Mohapatra}, {Mohite}, {Molina-Ruiz}, {Mondin}, {Montani}, {Moore}, {Moraru},
  {Morawski}, {Moreno}, {Morisaki}, {Mours}, {Mow-Lowry}, {Mozzon},
  {Muciaccia}, {Mukherjee}, {Mukherjee}, {Mukherjee}, {Mukherjee}, {Mukund},
  {Mullavey}, {Munch}, {Mu{\~n}iz}, {Murray}, {Nagar}, {Nardecchia},
  {Naticchioni}, {Nayak}, {Neil}, {Neilson}, {Nelemans}, {Nelson}, {Nery},
  {Neunzert}, {Ng}, {Ng}, {Nguyen}, {Nguyen}, {Nichols}, {Nichols}, {Nissanke},
  {Nocera}, {Noh}, {North}, {Nothard}, {Nuttall}, {Oberling}, {O'Brien},
  {Oganesyan}, {Ogin}, {Oh}, {Oh}, {Ohme}, {Ohta}, {Okada}, {Oliver},
  {Olivetto}, {Oppermann}, {Oram}, {O'Reilly}, {Ormiston}, {Ortega},
  {O'Shaughnessy}, {Ossokine}, {Osthelder}, {Ottaway}, {Overmier}, {Owen},
  {Pace}, {Pagano}, {Page}, {Pagliaroli}, {Pai}, {Pai}, {Palamos}, {Palashov},
  {Palomba}, {Pan}, {Panda}, {Pang}, {Pankow}, {Pannarale}, {Pant}, {Paoletti},
  {Paoli}, {Parida}, {Parker}, {Pascucci}, {Pasqualetti}, {Passaquieti},
  {Passuello}, {Patricelli}, {Payne}, {Pearlstone}, {Pechsiri}, {Pedersen},
  {Pedraza}, {Pele}, {Penn}, {Perego}, {Perez}, {P{\'e}rigois}, {Perreca},
  {Perri{\`e}s}, {Petermann}, {Pfeiffer}, {Phelps}, {Phukon}, {Piccinni},
  {Pichot}, {Piendibene}, {Piergiovanni}, {Pierro}, {Pillant}, {Pinard},
  {Pinto}, {Piotrzkowski}, {Pirello}, {Pitkin}, {Plastino}, {Poggiani}, {Pong},
  {Ponrathnam}, {Popolizio}, {Porter}, {Powell}, {Prajapati}, {Prasai},
  {Prasanna}, {Pratten}, {Prestegard}, {Principe}, {Prodi}, {Prokhorov},
  {Punturo}, {Puppo}, {P{\"u}rrer}, {Qi}, {Quetschke}, {Quinonez}, {Raab},
  {Raaijmakers}, {Radkins}, {Radulesco}, {Raffai}, {Rafferty}, {Raja}, {Rajan},
  {Rajbhandari}, {Rakhmanov}, {Ramirez}, {Ramos-Buades}, {Rana}, {Rao},
  {Rapagnani}, {Raymond}, {Razzano}, {Read}, {Regimbau}, {Rei}, {Reid},
  {Reitze}, {Rettegno}, {Ricci}, {Richardson}, {Richardson}, {Ricker},
  {Riemenschneider}, {Riles}, {Rizzo}, {Robertson}, {Robinet}, {Rocchi},
  {Rodriguez-Soto}, {Rolland}, {Rollins}, {Roma}, {Romanelli}, {Romano},
  {Romel}, {Romero-Shaw}, {Romie}, {Rose}, {Rose}, {Rose}, {Rosi{\'n}ska},
  {Rosofsky}, {Ross}, {Rowan}, {Rowlinson}, {Roy}, {Roy}, {Roy}, {Ruggi},
  {Rutins}, {Ryan}, {Sachdev}, {Sadecki}, {Sakellariadou}, {Salafia},
  {Salconi}, {Saleem}, {Samajdar}, {Sanchez}, {Sanchez}, {Sanchis-Gual},
  {Sanders}, {Santiago}, {Santos}, {Sarin}, {Sassolas}, {Sathyaprakash},
  {Sauter}, {Savage}, {Savant}, {Sawant}, {Sayah}, {Schaetzl}, {Schale},
  {Scheel}, {Scheuer}, {Schmidt}, {Schnabel}, {Schofield}, {Sch{\"o}nbeck},
  {Schreiber}, {Schulte}, {Schutz}, {Schwarm}, {Schwartz}, {Scott}, {Scott},
  {Seidel}, {Sellers}, {Sengupta}, {Sennett}, {Sentenac}, {Sequino}, {Sergeev},
  {Setyawati}, {Shaddock}, {Shaffer}, {Shahriar}, {Sharifi}, {Sharma},
  {Sharma}, {Shawhan}, {Shen}, {Shikauchi}, {Shink}, {Shoemaker}, {Shoemaker},
  {Shukla}, {ShyamSundar}, {Siellez}, {Sieniawska}, {Sigg}, {Singer}, {Singh},
  {Singh}, {Singha}, {Singhal}, {Sintes}, {Sipala}, {Skliris}, {Slagmolen},
  {Slaven-Blair}, {Smetana}, {Smith}, {Smith}, {Somala}, {Son}, {Soni},
  {Sorazu}, {Sordini}, {Sorrentino}, {Souradeep}, {Sowell}, {Spencer}, {Spera},
  {Srivastava}, {Srivastava}, {Staats}, {Stachie}, {Standke}, {Steer},
  {Steinke}, {Steinlechner}, {Steinlechner}, {Steinmeyer}, {Stevenson},
  {Stocks}, {Stops}, {Stover}, {Strain}, {Stratta}, {Strunk}, {Sturani},
  {Stuver}, {Sudhagar}, {Sudhir}, {Summerscales}, {Sun}, {Sunil}, {Sur},
  {Suresh}, {Sutton}, {Swinkels}, {Szczepa{\'n}czyk}, {Tacca}, {Tait},
  {Talbot}, {Tanasijczuk}, {Tanner}, {Tao}, {T{\'a}pai}, {Tapia}, {San Martin},
  {Tasson}, {Taylor}, {Tenorio}, {Terkowski}, {Thirugnanasambandam}, {Thomas},
  {Thomas}, {Thompson}, {Thondapu}, {Thorne}, {Thrane}, {Tinsman}, {Saravanan},
  {Tiwari}, {Tiwari}, {Tiwari}, {Toland}, {Tonelli}, {Tornasi},
  {Torres-Forn{\'e}}, {Torrie}, {Tosta e Melo}, {T{\"o}yr{\"a}}, {Trail},
  {Travasso}, {Traylor}, {Tringali}, {Tripathee}, {Trovato}, {Trudeau},
  {Tsang}, {Tse}, {Tso}, {Tsukada}, {Tsuna}, {Tsutsui}, {Turconi}, {Ubhi},
  {Udall}, {Ueno}, {Ugolini}, {Unnikrishnan}, {Urban}, {Usman}, {Utina},
  {Vahlbruch}, {Vajente}, {Valdes}, {Valentini}, {van Bakel}, {van Beuzekom},
  {van den Brand}, {Van Den Broeck}, {Vander-Hyde}, {van der Schaaf}, {Van
  Heijningen}, {van Veggel}, {Vardaro}, {Varma}, {Vass}, {Vas{\'u}th},
  {Vecchio}, {Vedovato}, {Veitch}, {Veitch}, {Venkateswara}, {Venugopalan},
  {Verkindt}, {Veske}, {Vetrano}, {Vicer{\'e}}, {Viets}, {Vinciguerra}, {Vine},
  {Vinet}, {Vitale}, {Vivanco}, {Vo}, {Vocca}, {Vorvick}, {Vyatchanin}, {Wade},
  {Wade}, {Wade}, {Walet}, {Walker}, {Wallace}, {Wallace}, {Walsh}, {Wang},
  {Wang}, {Wang}, {Ward}, {Warden}, {Warner}, {Was}, {Watchi}, {Weaver}, {Wei},
  {Weinert}, {Weinstein}, {Weiss}, {Wellmann}, {Wen}, {We{\ss}els},
  {Westhouse}, {Wette}, {Whelan}, {Whiting}, {Whittle}, {Wilken}, {Williams},
  {Willis}, {Willke}, {Winkler}, {Wipf}, {Wittel}, {Woan}, {Woehler},
  {Wofford}, {Wong}, {Wright}, {Wu}, {Wysocki}, {Xiao}, {Yamamoto}, {Yang},
  {Yang}, {Yang}, {Yap}, {Yazback}, {Yeeles}, {Yu}, {Yu}, {Yuen},
  {Zadro{\.z}ny}, {Zadro{\.z}ny}, {Zanolin}, {Zelenova}, {Zendri}, {Zevin},
  {Zhang}, {Zhang}, {Zhang}, {Zhao}, {Zhao}, {Zhou}, {Zhou}, {Zhu},
  {Zimmerman}, {Zlochower}, {Zucker}, {Zweizig}, {LIGO Scientific
  Collaboration}, \& {Virgo Collaboration}}]{2020ApJ...900L..13A}
{Abbott}, R., {Abbott}, T.~D., {Abraham}, S., {et~al.} 2020{\natexlab{b}},
  \apjl, 900, L13

\bibitem[{{Adams} {et~al.}(2017{\natexlab{a}}){Adams}, {Kochanek}, {Gerke}, \&
  {Stanek}}]{2017MNRAS.469.1445A}
{Adams}, S.~M., {Kochanek}, C.~S., {Gerke}, J.~R., \& {Stanek}, K.~Z.
  2017{\natexlab{a}}, \mnras, 469, 1445

\bibitem[{{Adams} {et~al.}(2017{\natexlab{b}}){Adams}, {Kochanek}, {Gerke},
  {Stanek}, \& {Dai}}]{2017MNRAS.468.4968A}
{Adams}, S.~M., {Kochanek}, C.~S., {Gerke}, J.~R., {Stanek}, K.~Z., \& {Dai},
  X. 2017{\natexlab{b}}, \mnras, 468, 4968

\bibitem[{{Aguilera-Dena} {et~al.}(2023){Aguilera-Dena}, {M{\"u}ller},
  {Antoniadis}, {Langer}, {Dessart}, {Vigna-G{\'o}mez}, \&
  {Yoon}}]{2023A&A...671A.134A}
{Aguilera-Dena}, D.~R., {M{\"u}ller}, B., {Antoniadis}, J., {et~al.} 2023,
  \aap, 671, A134

\bibitem[{{Anderson}(2019)}]{2019A&A...628A...7A}
{Anderson}, J.~P. 2019, \aap, 628, A7

\bibitem[{{Arnett} {et~al.}(1989){Arnett}, {Bahcall}, {Kirshner}, \&
  {Woosley}}]{1989ARA&A..27..629A}
{Arnett}, W.~D., {Bahcall}, J.~N., {Kirshner}, R.~P., \& {Woosley}, S.~E. 1989,
  \araa, 27, 629

\bibitem[{{Asplund} {et~al.}(2009){Asplund}, {Grevesse}, {Sauval}, \&
  {Scott}}]{2009ARA&A..47..481A}
{Asplund}, M., {Grevesse}, N., {Sauval}, A.~J., \& {Scott}, P. 2009, \araa, 47,
  481

\bibitem[{{Ballone} {et~al.}(2023){Ballone}, {Costa}, {Mapelli}, {MacLeod},
  {Torniamenti}, \& {Pacheco-Arias}}]{2023MNRAS.519.5191B}
{Ballone}, A., {Costa}, G., {Mapelli}, M., {et~al.} 2023, \mnras, 519, 5191

\bibitem[{{Barkat} \& {Wheeler}(1988)}]{1988ApJ...332..247B}
{Barkat}, Z. \& {Wheeler}, J.~C. 1988, \apj, 332, 247

\bibitem[{{Baron} \& {Cooperstein}(1990)}]{1990ApJ...353..597B}
{Baron}, E. \& {Cooperstein}, J. 1990, \apj, 353, 597

\bibitem[{{Basinger} {et~al.}(2021){Basinger}, {Kochanek}, {Adams}, {Dai}, \&
  {Stanek}}]{2021MNRAS.508.1156B}
{Basinger}, C.~M., {Kochanek}, C.~S., {Adams}, S.~M., {Dai}, X., \& {Stanek},
  K.~Z. 2021, \mnras, 508, 1156

\bibitem[{{Beasor} {et~al.}(2023){Beasor}, {Hosseinzadeh}, {Smith}, {Davies},
  {Jencson}, {Pearson}, \& {Sand}}]{2023arXiv230916121B}
{Beasor}, E.~R., {Hosseinzadeh}, G., {Smith}, N., {et~al.} 2023, arXiv
  e-prints, arXiv:2309.16121

\bibitem[{{Bernini-Peron} {et~al.}(2023){Bernini-Peron}, {Marcolino}, {Sander},
  {Bouret}, {Ramachandran}, {Saling}, {Schneider}, {Oskinova}, \&
  {Najarro}}]{2023A&A...677A..50B}
{Bernini-Peron}, M., {Marcolino}, W.~L.~F., {Sander}, A.~A.~C., {et~al.} 2023,
  \aap, 677, A50

\bibitem[{{Braun} \& {Langer}(1995)}]{1995A&A...297..483B}
{Braun}, H. \& {Langer}, N. 1995, \aap, 297, 483

\bibitem[{{Brown} {et~al.}(2001){Brown}, {Heger}, {Langer}, {Lee}, {Wellstein},
  \& {Bethe}}]{2001NewA....6..457B}
{Brown}, G.~E., {Heger}, A., {Langer}, N., {et~al.} 2001, \na, 6, 457

\bibitem[{{Burrows} {et~al.}(2020){Burrows}, {Radice}, {Vartanyan}, {Nagakura},
  {Skinner}, \& {Dolence}}]{2020MNRAS.491.2715B}
{Burrows}, A., {Radice}, D., {Vartanyan}, D., {et~al.} 2020, \mnras, 491, 2715

\bibitem[{{Burrows} {et~al.}(1995){Burrows}, {Krist}, {Hester}, {Sahai},
  {Trauger}, {Stapelfeldt}, {Gallagher}, {Ballester}, {Casertano}, {Clarke},
  {Crisp}, {Evans}, {Griffiths}, {Hoessel}, {Holtzman}, {Mould}, {Scowen},
  {Watson}, \& {Westphal}}]{1995ApJ...452..680B}
{Burrows}, C.~J., {Krist}, J., {Hester}, J.~J., {et~al.} 1995, \apj, 452, 680

\bibitem[{{Chan} {et~al.}(2020){Chan}, {M{\"u}ller}, \&
  {Heger}}]{2020MNRAS.495.3751C}
{Chan}, C., {M{\"u}ller}, B., \& {Heger}, A. 2020, \mnras, 495, 3751

\bibitem[{{Chan} {et~al.}(2018){Chan}, {M{\"u}ller}, {Heger}, {Pakmor}, \&
  {Springel}}]{2018ApJ...852L..19C}
{Chan}, C., {M{\"u}ller}, B., {Heger}, A., {Pakmor}, R., \& {Springel}, V.
  2018, \apjl, 852, L19

\bibitem[{{Chieffi} \& {Limongi}(2020)}]{2020ApJ...890...43C}
{Chieffi}, A. \& {Limongi}, M. 2020, \apj, 890, 43

\bibitem[{{Chini} {et~al.}(2012){Chini}, {Hoffmeister}, {Nasseri}, {Stahl}, \&
  {Zinnecker}}]{2012MNRAS.424.1925C}
{Chini}, R., {Hoffmeister}, V.~H., {Nasseri}, A., {Stahl}, O., \& {Zinnecker},
  H. 2012, \mnras, 424, 1925

\bibitem[{{Claeys} {et~al.}(2011){Claeys}, {de Mink}, {Pols}, {Eldridge}, \&
  {Baes}}]{2011A&A...528A.131C}
{Claeys}, J.~S.~W., {de Mink}, S.~E., {Pols}, O.~R., {Eldridge}, J.~J., \&
  {Baes}, M. 2011, \aap, 528, A131

\bibitem[{{Costa} {et~al.}(2022){Costa}, {Ballone}, {Mapelli}, \&
  {Bressan}}]{2022MNRAS.516.1072C}
{Costa}, G., {Ballone}, A., {Mapelli}, M., \& {Bressan}, A. 2022, \mnras, 516,
  1072

\bibitem[{{Couch} {et~al.}(2020){Couch}, {Warren}, \&
  {O'Connor}}]{2020ApJ...890..127C}
{Couch}, S.~M., {Warren}, M.~L., \& {O'Connor}, E.~P. 2020, \apj, 890, 127

\bibitem[{{Cyburt} {et~al.}(2010){Cyburt}, {Amthor}, {Ferguson}, {Meisel},
  {Smith}, {Warren}, {Heger}, {Hoffman}, {Rauscher}, {Sakharuk}, {Schatz},
  {Thielemann}, \& {Wiescher}}]{2010ApJS..189..240C}
{Cyburt}, R.~H., {Amthor}, A.~M., {Ferguson}, R., {et~al.} 2010, \apjs, 189,
  240

\bibitem[{{Davies} \& {Beasor}(2020)}]{2020MNRAS.493..468D}
{Davies}, B. \& {Beasor}, E.~R. 2020, \mnras, 493, 468

\bibitem[{{Davies} {et~al.}(2018){Davies}, {Crowther}, \&
  {Beasor}}]{2018MNRAS.478.3138D}
{Davies}, B., {Crowther}, P.~A., \& {Beasor}, E.~R. 2018, \mnras, 478, 3138

\bibitem[{{de Mink} {et~al.}(2013){de Mink}, {Langer}, {Izzard}, {Sana}, \& {de
  Koter}}]{2013ApJ...764..166D}
{de Mink}, S.~E., {Langer}, N., {Izzard}, R.~G., {Sana}, H., \& {de Koter}, A.
  2013, \apj, 764, 166

\bibitem[{{de Mink} {et~al.}(2007){de Mink}, {Pols}, \&
  {Hilditch}}]{2007A&A...467.1181D}
{de Mink}, S.~E., {Pols}, O.~R., \& {Hilditch}, R.~W. 2007, \aap, 467, 1181

\bibitem[{{Di Carlo} {et~al.}(2019){Di Carlo}, {Giacobbo}, {Mapelli},
  {Pasquato}, {Spera}, {Wang}, \& {Haardt}}]{2019MNRAS.487.2947D}
{Di Carlo}, U.~N., {Giacobbo}, N., {Mapelli}, M., {et~al.} 2019, \mnras, 487,
  2947

\bibitem[{{Di Carlo} {et~al.}(2020){Di Carlo}, {Mapelli}, {Bouffanais},
  {Giacobbo}, {Santoliquido}, {Bressan}, {Spera}, \&
  {Haardt}}]{2020MNRAS.497.1043D}
{Di Carlo}, U.~N., {Mapelli}, M., {Bouffanais}, Y., {et~al.} 2020, \mnras, 497,
  1043

\bibitem[{{Donati} \& {Landstreet}(2009)}]{2009ARA&A..47..333D}
{Donati}, J.-F. \& {Landstreet}, J.~D. 2009, \araa, 47, 333

\bibitem[{{Dray} \& {Tout}(2007)}]{2007MNRAS.376...61D}
{Dray}, L.~M. \& {Tout}, C.~A. 2007, \mnras, 376, 61

\bibitem[{{Dufton} {et~al.}(2013){Dufton}, {Langer}, {Dunstall}, {Evans},
  {Brott}, {de Mink}, {Howarth}, {Kennedy}, {McEvoy}, {Potter},
  {Ram{\'{\i}}rez-Agudelo}, {Sana}, {Sim{\'o}n-D{\'{\i}}az}, {Taylor}, \&
  {Vink}}]{2013A&A...550A.109D}
{Dufton}, P.~L., {Langer}, N., {Dunstall}, P.~R., {et~al.} 2013, \aap, 550,
  A109

\bibitem[{{Edelman} {et~al.}(2023){Edelman}, {Farr}, \&
  {Doctor}}]{2023ApJ...946...16E}
{Edelman}, B., {Farr}, B., \& {Doctor}, Z. 2023, \apj, 946, 16

\bibitem[{{Ertl} {et~al.}(2016){Ertl}, {Janka}, {Woosley}, {Sukhbold}, \&
  {Ugliano}}]{2016ApJ...818..124E}
{Ertl}, T., {Janka}, H.-T., {Woosley}, S.~E., {Sukhbold}, T., \& {Ugliano}, M.
  2016, \apj, 818, 124

\bibitem[{{Ertl} {et~al.}(2020){Ertl}, {Woosley}, {Sukhbold}, \&
  {Janka}}]{2020ApJ...890...51E}
{Ertl}, T., {Woosley}, S.~E., {Sukhbold}, T., \& {Janka}, H.~T. 2020, \apj,
  890, 51

\bibitem[{{Farag} {et~al.}(2022){Farag}, {Renzo}, {Farmer}, {Chidester}, \&
  {Timmes}}]{2022ApJ...937..112F}
{Farag}, E., {Renzo}, M., {Farmer}, R., {Chidester}, M.~T., \& {Timmes}, F.~X.
  2022, \apj, 937, 112

\bibitem[{{Farmer} {et~al.}(2021){Farmer}, {Laplace}, {de Mink}, \&
  {Justham}}]{2021ApJ...923..214F}
{Farmer}, R., {Laplace}, E., {de Mink}, S.~E., \& {Justham}, S. 2021, \apj,
  923, 214

\bibitem[{{Farmer} {et~al.}(2023){Farmer}, {Laplace}, {Ma}, {de Mink}, \&
  {Justham}}]{2023ApJ...948..111F}
{Farmer}, R., {Laplace}, E., {Ma}, J.-z., {de Mink}, S.~E., \& {Justham}, S.
  2023, \apj, 948, 111

\bibitem[{{Farmer} {et~al.}(2020){Farmer}, {Renzo}, {de Mink}, {Fishbach}, \&
  {Justham}}]{2020ApJ...902L..36F}
{Farmer}, R., {Renzo}, M., {de Mink}, S.~E., {Fishbach}, M., \& {Justham}, S.
  2020, \apjl, 902, L36

\bibitem[{{Farmer} {et~al.}(2019){Farmer}, {Renzo}, {de Mink}, {Marchant}, \&
  {Justham}}]{2019ApJ...887...53F}
{Farmer}, R., {Renzo}, M., {de Mink}, S.~E., {Marchant}, P., \& {Justham}, S.
  2019, \apj, 887, 53

\bibitem[{{Farrell} {et~al.}(2022){Farrell}, {Groh}, {Meynet}, \&
  {Eldridge}}]{2022MNRAS.512.4116F}
{Farrell}, E., {Groh}, J.~H., {Meynet}, G., \& {Eldridge}, J.~J. 2022, \mnras,
  512, 4116

\bibitem[{{Ferrario} {et~al.}(2015){Ferrario}, {Melatos}, \&
  {Zrake}}]{2015SSRv..191...77F}
{Ferrario}, L., {Melatos}, A., \& {Zrake}, J. 2015, \ssr, 191, 77

\bibitem[{{Ferraro} {et~al.}(2023){Ferraro}, {Mucciarelli}, {Lanzoni},
  {Pallanca}, {Cadelano}, {Billi}, {Sills}, {Vesperini}, {Dalessandro},
  {Beccari}, {Monaco}, \& {Mateo}}]{2023NatCo..14.2584F}
{Ferraro}, F.~R., {Mucciarelli}, A., {Lanzoni}, B., {et~al.} 2023, Nature
  Communications, 14, 2584

\bibitem[{{Fossati} {et~al.}(2015){Fossati}, {Castro}, {Sch{\"o}ller},
  {Hubrig}, {Langer}, {Morel}, {Briquet}, {Herrero}, {Przybilla}, {Sana},
  {Schneider}, {de Koter}, \& {BOB Collaboration}}]{2015A&A...582A..45F}
{Fossati}, L., {Castro}, N., {Sch{\"o}ller}, M., {et~al.} 2015, \aap, 582, A45

\bibitem[{{Freitag} \& {Benz}(2005)}]{2005MNRAS.358.1133F}
{Freitag}, M. \& {Benz}, W. 2005, \mnras, 358, 1133

\bibitem[{{Fryer}(2014)}]{2014frap.confE...4F}
{Fryer}, C. 2014, in Proceedings of Frontier Research in Astrophysics
  (FRAPWS2014) held 26-31 May, 4

\bibitem[{{Gerke} {et~al.}(2015){Gerke}, {Kochanek}, \&
  {Stanek}}]{2015MNRAS.450.3289G}
{Gerke}, J.~R., {Kochanek}, C.~S., \& {Stanek}, K.~Z. 2015, \mnras, 450, 3289

\bibitem[{{Gieles} {et~al.}(2018){Gieles}, {Charbonnel}, {Krause},
  {H{\'e}nault-Brunet}, {Agertz}, {Lamers}, {Bastian}, {Gualandris}, {Zocchi},
  \& {Petts}}]{2018MNRAS.478.2461G}
{Gieles}, M., {Charbonnel}, C., {Krause}, M. G.~H., {et~al.} 2018, \mnras, 478,
  2461

\bibitem[{{Glebbeek} {et~al.}(2013){Glebbeek}, {Gaburov}, {Portegies Zwart}, \&
  {Pols}}]{2013MNRAS.434.3497G}
{Glebbeek}, E., {Gaburov}, E., {Portegies Zwart}, S., \& {Pols}, O.~R. 2013,
  \mnras, 434, 3497

\bibitem[{{Grassitelli} {et~al.}(2021){Grassitelli}, {Langer}, {Mackey},
  {Gr{\"a}fener}, {Grin}, {Sander}, \& {Vink}}]{2021A&A...647A..99G}
{Grassitelli}, L., {Langer}, N., {Mackey}, J., {et~al.} 2021, \aap, 647, A99

\bibitem[{{Grunhut} {et~al.}(2017){Grunhut}, {Wade}, {Neiner}, {Oksala},
  {Petit}, {Alecian}, {Bohlender}, {Bouret}, {Henrichs}, {Hussain},
  {Kochukhov}, \& {MiMeS Collaboration}}]{2017MNRAS.465.2432G}
{Grunhut}, J.~H., {Wade}, G.~A., {Neiner}, C., {et~al.} 2017, \mnras, 465, 2432

\bibitem[{{Hachinger} {et~al.}(2012){Hachinger}, {Mazzali}, {Taubenberger},
  {Hillebrandt}, {Nomoto}, \& {Sauer}}]{2012MNRAS.422...70H}
{Hachinger}, S., {Mazzali}, P.~A., {Taubenberger}, S., {et~al.} 2012, \mnras,
  422, 70

\bibitem[{{Hamuy}(2003)}]{2003ApJ...582..905H}
{Hamuy}, M. 2003, \apj, 582, 905

\bibitem[{{Hayashi} \& {Hoshi}(1961)}]{1961PASJ...13..442H}
{Hayashi}, C. \& {Hoshi}, R. 1961, \pasj, 13, 442

\bibitem[{{Heger} {et~al.}(2023){Heger}, {M{\"u}ller}, \&
  {Mandel}}]{heger2023a}
{Heger}, A., {M{\"u}ller}, B., \& {Mandel}, I. 2023, Black Holes as the End
  State of Stellar Evolution: Theory and Simulations, ed. Z.~{Haima}, 61--111

\bibitem[{{Hellings}(1983)}]{1983Ap&SS..96...37H}
{Hellings}, P. 1983, \apss, 96, 37

\bibitem[{{Hellings}(1984)}]{1984Ap&SS.104...83H}
{Hellings}, P. 1984, \apss, 104, 83

\bibitem[{{Henneco} {et~al.}(2024){Henneco}, {Schneider}, \&
  {Laplace}}]{2023arXiv231112124H}
{Henneco}, J., {Schneider}, F. R.~N., \& {Laplace}, E. 2024, \aap, 682, A169

\bibitem[{{Henyey} {et~al.}(1965){Henyey}, {Vardya}, \&
  {Bodenheimer}}]{1965ApJ...142..841H}
{Henyey}, L., {Vardya}, M.~S., \& {Bodenheimer}, P. 1965, \apj, 142, 841

\bibitem[{{Hillebrandt} \& {Meyer}(1989)}]{1989A&A...219L...3H}
{Hillebrandt}, W. \& {Meyer}, F. 1989, \aap, 219, L3

\bibitem[{{Hirai} {et~al.}(2021){Hirai}, {Podsiadlowski}, {Owocki},
  {Schneider}, \& {Smith}}]{2021MNRAS.503.4276H}
{Hirai}, R., {Podsiadlowski}, P., {Owocki}, S.~P., {Schneider}, F. R.~N., \&
  {Smith}, N. 2021, \mnras, 503, 4276

\bibitem[{{Hobbs} {et~al.}(2005){Hobbs}, {Lorimer}, {Lyne}, \&
  {Kramer}}]{2005MNRAS.360..974H}
{Hobbs}, G., {Lorimer}, D.~R., {Lyne}, A.~G., \& {Kramer}, M. 2005, \mnras,
  360, 974

\bibitem[{{Humphreys} \& {Davidson}(1979)}]{1979ApJ...232..409H}
{Humphreys}, R.~M. \& {Davidson}, K. 1979, \apj, 232, 409

\bibitem[{{Humphreys} \& {Davidson}(1994)}]{1994PASP..106.1025H}
{Humphreys}, R.~M. \& {Davidson}, K. 1994, \pasp, 106, 1025

\bibitem[{Hunter(2007)}]{hunter2007matplotlib}
Hunter, J.~D. 2007, Computing in Science \& Engineering, 9, 90

\bibitem[{{Ivanova}(2002)}]{2002PhDT........25I}
{Ivanova}, N. 2002, PhD thesis, University of Oxford, UK

\bibitem[{{Ivanova} \& {Podsiadlowski}(2002)}]{2002Ap&SS.281..191I}
{Ivanova}, N. \& {Podsiadlowski}, P. 2002, \apss, 281, 191

\bibitem[{{Ivanova} {et~al.}(2002){Ivanova}, {Podsiadlowski}, \&
  {Spruit}}]{2002MNRAS.334..819I}
{Ivanova}, N., {Podsiadlowski}, P., \& {Spruit}, H. 2002, \mnras, 334, 819

\bibitem[{{Jencson} {et~al.}(2023){Jencson}, {Pearson}, {Beasor}, {Lau},
  {Andrews}, {Bostroem}, {Dong}, {Engesser}, {Gomez}, {Guolo}, {Hoang},
  {Hosseinzadeh}, {Jha}, {Karambelkar}, {Kasliwal}, {Lundquist}, {Meza
  Retamal}, {Rest}, {Sand}, {Shahbandeh}, {Shrestha}, {Smith}, {Strader},
  {Valenti}, {Wang}, \& {Zenati}}]{2023ApJ...952L..30J}
{Jencson}, J.~E., {Pearson}, J., {Beasor}, E.~R., {et~al.} 2023, \apjl, 952,
  L30

\bibitem[{{Jiang} {et~al.}(2018){Jiang}, {Cantiello}, {Bildsten}, {Quataert},
  {Blaes}, \& {Stone}}]{2018Natur.561..498J}
{Jiang}, Y.-F., {Cantiello}, M., {Bildsten}, L., {et~al.} 2018, \nat, 561, 498

\bibitem[{{Justham} {et~al.}(2014){Justham}, {Podsiadlowski}, \&
  {Vink}}]{2014ApJ...796..121J}
{Justham}, S., {Podsiadlowski}, P., \& {Vink}, J.~S. 2014, \apj, 796, 121

\bibitem[{{Keszthelyi} {et~al.}(2019){Keszthelyi}, {Meynet}, {Georgy}, {Wade},
  {Petit}, \& {David-Uraz}}]{2019MNRAS.485.5843K}
{Keszthelyi}, Z., {Meynet}, G., {Georgy}, C., {et~al.} 2019, \mnras, 485, 5843

\bibitem[{{Kilpatrick} {et~al.}(2023){Kilpatrick}, {Foley},
  {Jacobson-Gal{\'a}n}, {Piro}, {Smartt}, {Drout}, {Gagliano}, {Gall},
  {Hjorth}, {Jones}, {Mandel}, {Margutti}, {Ramirez-Ruiz}, {Ransome}, {Villar},
  {Coulter}, {Gao}, {Matthews}, {Taggart}, \& {Zenati}}]{2023ApJ...952L..23K}
{Kilpatrick}, C.~D., {Foley}, R.~J., {Jacobson-Gal{\'a}n}, W.~V., {et~al.}
  2023, \apjl, 952, L23

\bibitem[{{Kippenhahn} \& {Weigert}(1994)}]{1994sse..book.....K}
{Kippenhahn}, R. \& {Weigert}, A. 1994, {Stellar Structure and Evolution}
  (Springer-Verlag Berlin Heidelberg New York)

\bibitem[{Kluyver {et~al.}(2016)Kluyver, Ragan-Kelley, P{\'e}rez, Granger,
  Bussonnier, Frederic, Kelley, Hamrick, Grout, Corlay, Ivanov, Avila, Abdalla,
  Willing, \& et~al.}]{kluyver2016jupyternotebook}
Kluyver, T., Ragan-Kelley, B., P{\'e}rez, F., {et~al.} 2016, in Positioning and
  Power in Academic Publishing: Players, Agents and Agendas, ed. F.~Loizides \&
  B.~Schmidt (IOS Press), 87--90

\bibitem[{{Kobulnicky} \& {Fryer}(2007)}]{2007ApJ...670..747K}
{Kobulnicky}, H.~A. \& {Fryer}, C.~L. 2007, \apj, 670, 747

\bibitem[{{Kobulnicky} {et~al.}(2014){Kobulnicky}, {Kiminki}, {Lundquist},
  {Burke}, {Chapman}, {Keller}, {Lester}, {Rolen}, {Topel}, {Bhattacharjee},
  {Smullen}, {Vargas {\'A}lvarez}, {Runnoe}, {Dale}, \&
  {Brotherton}}]{2014ApJS..213...34K}
{Kobulnicky}, H.~A., {Kiminki}, D.~C., {Lundquist}, M.~J., {et~al.} 2014,
  \apjs, 213, 34

\bibitem[{{Kochanek} {et~al.}(2024){Kochanek}, {Neustadt}, \&
  {Stanek}}]{2024ApJ...962..145K}
{Kochanek}, C.~S., {Neustadt}, J. M.~M., \& {Stanek}, K.~Z. 2024, \apj, 962,
  145

\bibitem[{{Kuroda} {et~al.}(2018){Kuroda}, {Kotake}, {Takiwaki}, \&
  {Thielemann}}]{2018MNRAS.477L..80K}
{Kuroda}, T., {Kotake}, K., {Takiwaki}, T., \& {Thielemann}, F.-K. 2018,
  \mnras, 477, L80

\bibitem[{{Langer}(1989)}]{1989A&A...210...93L}
{Langer}, N. 1989, \aap, 210, 93

\bibitem[{{Langer}(2012)}]{2012ARA&A..50..107L}
{Langer}, N. 2012, \araa, 50, 107

\bibitem[{{Langer}(2014)}]{2014IAUS..302....1L}
{Langer}, N. 2014, in IAU Symposium, Vol. 302, IAU Symposium, 1--9

\bibitem[{{Laplace}(2022)}]{2022A&C....3800516L}
{Laplace}, E. 2022, Astronomy and Computing, 38, 100516

\bibitem[{{Laplace} {et~al.}(2021){Laplace}, {Justham}, {Renzo}, {G{\"o}tberg},
  {Farmer}, {Vartanyan}, \& {de Mink}}]{2021A&A...656A..58L}
{Laplace}, E., {Justham}, S., {Renzo}, M., {et~al.} 2021, \aap, 656, A58

\bibitem[{{Laplace} {et~al.}(2023){Laplace}, {Schneider}, {Temaj}, \&
  {Podsiadlowski}}]{Laplace+2023}
{Laplace}, E., {Schneider}, F.~R.~N., {Temaj}, D., \& {Podsiadlowski}, P. 2023,
  in prep.

\bibitem[{{Leonard} \& {Livio}(1995)}]{1995ApJ...447L.121L}
{Leonard}, P. J.~T. \& {Livio}, M. 1995, \apjl, 447, L121

\bibitem[{{Lombardi} {et~al.}(2002){Lombardi}, {Warren}, {Rasio}, {Sills}, \&
  {Warren}}]{2002ApJ...568..939L}
{Lombardi}, Jr., J.~C., {Warren}, J.~S., {Rasio}, F.~A., {Sills}, A., \&
  {Warren}, A.~R. 2002, \apj, 568, 939

\bibitem[{{Lundqvist} \& {Fransson}(1996)}]{1996ApJ...464..924L}
{Lundqvist}, P. \& {Fransson}, C. 1996, \apj, 464, 924

\bibitem[{{Mapelli}(2016)}]{2016MNRAS.459.3432M}
{Mapelli}, M. 2016, \mnras, 459, 3432

\bibitem[{{Marchant} \& {Moriya}(2020)}]{2020A&A...640L..18M}
{Marchant}, P. \& {Moriya}, T.~J. 2020, \aap, 640, L18

\bibitem[{{Mason} {et~al.}(2009){Mason}, {Hartkopf}, {Gies}, {Henry}, \&
  {Helsel}}]{2009AJ....137.3358M}
{Mason}, B.~D., {Hartkopf}, W.~I., {Gies}, D.~R., {Henry}, T.~J., \& {Helsel},
  J.~W. 2009, \aj, 137, 3358

\bibitem[{{Mauron} \& {Josselin}(2011)}]{2011A&A...526A.156M}
{Mauron}, N. \& {Josselin}, E. 2011, \aap, 526, A156

\bibitem[{{Menon} {et~al.}(2023){Menon}, {Ercolino}, {Urbaneja}, {Lennon},
  {Herrero}, {Hirai}, {Langer}, {Schootemeijer}, {Chatzopoulos}, {Frank}, \&
  {Shiber}}]{2023arXiv231105581M}
{Menon}, A., {Ercolino}, A., {Urbaneja}, M.~A., {et~al.} 2023, arXiv e-prints,
  arXiv:2311.05581

\bibitem[{{Menon} \& {Heger}(2017)}]{2017MNRAS.469.4649M}
{Menon}, A. \& {Heger}, A. 2017, \mnras, 469, 4649

\bibitem[{{Menon} {et~al.}(2019){Menon}, {Utrobin}, \&
  {Heger}}]{2019MNRAS.482..438M}
{Menon}, A., {Utrobin}, V., \& {Heger}, A. 2019, \mnras, 482, 438

\bibitem[{{Meynet} {et~al.}(2011){Meynet}, {Eggenberger}, \&
  {Maeder}}]{2011A&A...525L..11M}
{Meynet}, G., {Eggenberger}, P., \& {Maeder}, A. 2011, \aap, 525, L11

\bibitem[{{Meza} \& {Anderson}(2020)}]{2020A&A...641A.177M}
{Meza}, N. \& {Anderson}, J.~P. 2020, \aap, 641, A177

\bibitem[{{Moe} \& {Di Stefano}(2017)}]{2017ApJS..230...15M}
{Moe}, M. \& {Di Stefano}, R. 2017, \apjs, 230, 15

\bibitem[{{Moreno} {et~al.}(2022){Moreno}, {Schneider}, {R{\"o}pke}, {Ohlmann},
  {Pakmor}, {Podsiadlowski}, \& {Sand}}]{2022A&A...667A..72M}
{Moreno}, M.~M., {Schneider}, F. R.~N., {R{\"o}pke}, F.~K., {et~al.} 2022,
  \aap, 667, A72

\bibitem[{{Morris} \& {Podsiadlowski}(2007)}]{2007Sci...315.1103M}
{Morris}, T. \& {Podsiadlowski}, P. 2007, Science, 315, 1103

\bibitem[{{Morris} \& {Podsiadlowski}(2009)}]{2009MNRAS.399..515M}
{Morris}, T. \& {Podsiadlowski}, P. 2009, \mnras, 399, 515

\bibitem[{{M{\"u}ller} {et~al.}(2016){M{\"u}ller}, {Heger}, {Liptai}, \&
  {Cameron}}]{2016MNRAS.460..742M}
{M{\"u}ller}, B., {Heger}, A., {Liptai}, D., \& {Cameron}, J.~B. 2016, \mnras,
  460, 742

\bibitem[{{Neiner} {et~al.}(2015){Neiner}, {Mathis}, {Alecian}, {Emeriau},
  {Grunhut}, {BinaMIcS}, \& {MiMeS Collaborations}}]{2015IAUS..305...61N}
{Neiner}, C., {Mathis}, S., {Alecian}, E., {et~al.} 2015, in Polarimetry, ed.
  K.~N. {Nagendra}, S.~{Bagnulo}, R.~{Centeno}, \& M.~{Jes{\'u}s Mart{\'\i}nez
  Gonz{\'a}lez}, Vol. 305, 61--66

\bibitem[{{Nieuwenhuijzen} \& {de Jager}(1990)}]{1990A&A...231..134N}
{Nieuwenhuijzen}, H. \& {de Jager}, C. 1990, \aap, 231, 134

\bibitem[{{Nugis} \& {Lamers}(2000)}]{2000A&A...360..227N}
{Nugis}, T. \& {Lamers}, H.~J.~G.~L.~M. 2000, \aap, 360, 227

\bibitem[{{O'Connor} \& {Ott}(2011)}]{2011ApJ...730...70O}
{O'Connor}, E. \& {Ott}, C.~D. 2011, \apj, 730, 70

\bibitem[{{Offner} {et~al.}(2023){Offner}, {Moe}, {Kratter}, {Sadavoy},
  {Jensen}, \& {Tobin}}]{2023ASPC..534..275O}
{Offner}, S.~S.~R., {Moe}, M., {Kratter}, K.~M., {et~al.} 2023, in Astronomical
  Society of the Pacific Conference Series, Vol. 534, Protostars and Planets
  VII, ed. S.~{Inutsuka}, Y.~{Aikawa}, T.~{Muto}, K.~{Tomida}, \& M.~{Tamura},
  275

\bibitem[{Oliphant(2006)}]{oliphant2006numpy}
Oliphant, T.~E. 2006, A guide to NumPy, Vol.~1 (Trelgol Publishing USA)

\bibitem[{{Ott} {et~al.}(2018){Ott}, {Roberts}, {da Silva Schneider}, {Fedrow},
  {Haas}, \& {Schnetter}}]{2018ApJ...855L...3O}
{Ott}, C.~D., {Roberts}, L.~F., {da Silva Schneider}, A., {et~al.} 2018, \apjl,
  855, L3

\bibitem[{{Patton} \& {Sukhbold}(2020)}]{2020MNRAS.499.2803P}
{Patton}, R.~A. \& {Sukhbold}, T. 2020, \mnras, 499, 2803

\bibitem[{{Paxton} {et~al.}(2011){Paxton}, {Bildsten}, {Dotter}, {Herwig},
  {Lesaffre}, \& {Timmes}}]{2011ApJS..192....3P}
{Paxton}, B., {Bildsten}, L., {Dotter}, A., {et~al.} 2011, \apjs, 192, 3

\bibitem[{{Paxton} {et~al.}(2013){Paxton}, {Cantiello}, {Arras}, {Bildsten},
  {Brown}, {Dotter}, {Mankovich}, {Montgomery}, {Stello}, {Timmes}, \&
  {Townsend}}]{2013ApJS..208....4P}
{Paxton}, B., {Cantiello}, M., {Arras}, P., {et~al.} 2013, \apjs, 208, 4

\bibitem[{{Paxton} {et~al.}(2015){Paxton}, {Marchant}, {Schwab}, {Bauer},
  {Bildsten}, {Cantiello}, {Dessart}, {Farmer}, {Hu}, {Langer}, {Townsend},
  {Townsley}, \& {Timmes}}]{2015ApJS..220...15P}
{Paxton}, B., {Marchant}, P., {Schwab}, J., {et~al.} 2015, \apjs, 220, 15

\bibitem[{{Paxton} {et~al.}(2018){Paxton}, {Schwab}, {Bauer}, {Bildsten},
  {Blinnikov}, {Duffell}, {Farmer}, {Goldberg}, {Marchant}, {Sorokina},
  {Thoul}, {Townsend}, \& {Timmes}}]{2018ApJS..234...34P}
{Paxton}, B., {Schwab}, J., {Bauer}, E.~B., {et~al.} 2018, \apjs, 234, 34

\bibitem[{{Paxton} {et~al.}(2019){Paxton}, {Smolec}, {Schwab}, {Gautschy},
  {Bildsten}, {Cantiello}, {Dotter}, {Farmer}, {Goldberg}, {Jermyn}, {Kanbur},
  {Marchant}, {Thoul}, {Townsend}, {Wolf}, {Zhang}, \&
  {Timmes}}]{2019ApJS..243...10P}
{Paxton}, B., {Smolec}, R., {Schwab}, J., {et~al.} 2019, \apjs, 243, 10

\bibitem[{{Petermann} {et~al.}(2015){Petermann}, {Langer}, {Castro}, \&
  {Fossati}}]{2015A&A...584A..54P}
{Petermann}, I., {Langer}, N., {Castro}, N., \& {Fossati}, L. 2015, \aap, 584,
  A54

\bibitem[{{Petit} {et~al.}(2017){Petit}, {Keszthelyi}, {MacInnis}, {Cohen},
  {Townsend}, {Wade}, {Thomas}, {Owocki}, {Puls}, \&
  {ud-Doula}}]{2017MNRAS.466.1052P}
{Petit}, V., {Keszthelyi}, Z., {MacInnis}, R., {et~al.} 2017, \mnras, 466, 1052

\bibitem[{{Podsiadlowski}(1992)}]{1992PASP..104..717P}
{Podsiadlowski}, P. 1992, \pasp, 104, 717

\bibitem[{{Podsiadlowski}(2017{\natexlab{a}})}]{2017arXiv170203973P}
{Podsiadlowski}, P. 2017{\natexlab{a}}, arXiv e-prints, arXiv:1702.03973

\bibitem[{{Podsiadlowski}(2017{\natexlab{b}})}]{2017hsn..book..635P}
{Podsiadlowski}, P. 2017{\natexlab{b}}, in Handbook of Supernovae, ed. A.~W.
  {Alsabti} \& P.~{Murdin}, 635

\bibitem[{{Podsiadlowski} \& {Joss}(1989)}]{1989Natur.338..401P}
{Podsiadlowski}, P. \& {Joss}, P.~C. 1989, \nat, 338, 401

\bibitem[{{Podsiadlowski} {et~al.}(1992){Podsiadlowski}, {Joss}, \&
  {Hsu}}]{1992ApJ...391..246P}
{Podsiadlowski}, P., {Joss}, P.~C., \& {Hsu}, J.~J.~L. 1992, \apj, 391, 246

\bibitem[{{Podsiadlowski} {et~al.}(1990){Podsiadlowski}, {Joss}, \&
  {Rappaport}}]{1990A&A...227L...9P}
{Podsiadlowski}, P., {Joss}, P.~C., \& {Rappaport}, S. 1990, \aap, 227, L9

\bibitem[{{Podsiadlowski} {et~al.}(2004){Podsiadlowski}, {Langer},
  {Poelarends}, {Rappaport}, {Heger}, \& {Pfahl}}]{2004ApJ...612.1044P}
{Podsiadlowski}, P., {Langer}, N., {Poelarends}, A.~J.~T., {et~al.} 2004, \apj,
  612, 1044

\bibitem[{{Podsiadlowski} {et~al.}(2007){Podsiadlowski}, {Morris}, \&
  {Ivanova}}]{2007AIPC..937..125P}
{Podsiadlowski}, P., {Morris}, T.~S., \& {Ivanova}, N. 2007, in American
  Institute of Physics Conference Series, Vol. 937, Supernova 1987A: 20 Years
  After: Supernovae and Gamma-Ray Bursters, ed. S.~{Immler}, K.~{Weiler}, \&
  R.~{McCray}, 125--133

\bibitem[{{Pols}(1994)}]{1994A&A...290..119P}
{Pols}, O.~R. 1994, \aap, 290, 119

\bibitem[{{Portegies Zwart} {et~al.}(2004){Portegies Zwart}, {Baumgardt},
  {Hut}, {Makino}, \& {McMillan}}]{2004Natur.428..724P}
{Portegies Zwart}, S.~F., {Baumgardt}, H., {Hut}, P., {Makino}, J., \&
  {McMillan}, S. L.~W. 2004, \nat, 428, 724

\bibitem[{{Portegies Zwart} \& {McMillan}(2002)}]{2002ApJ...576..899P}
{Portegies Zwart}, S.~F. \& {McMillan}, S. L.~W. 2002, \apj, 576, 899

\bibitem[{{Potter} {et~al.}(2012){Potter}, {Chitre}, \&
  {Tout}}]{2012MNRAS.424.2358P}
{Potter}, A.~T., {Chitre}, S.~M., \& {Tout}, C.~A. 2012, \mnras, 424, 2358

\bibitem[{{Powell} \& {M{\"u}ller}(2020)}]{2020MNRAS.494.4665P}
{Powell}, J. \& {M{\"u}ller}, B. 2020, \mnras, 494, 4665

\bibitem[{{Pumo} {et~al.}(2023){Pumo}, {Cosentino}, {Pastorello}, {Benetti},
  {Cherubini}, {Manic{\`o}}, \& {Zampieri}}]{2023MNRAS.tmp..837P}
{Pumo}, M.~L., {Cosentino}, S.~P., {Pastorello}, A., {et~al.} 2023, \mnras

\bibitem[{{Quentin} \& {Tout}(2018)}]{2018MNRAS.477.2298Q}
{Quentin}, L.~G. \& {Tout}, C.~A. 2018, \mnras, 477, 2298

\bibitem[{{Ram{\'{\i}}rez-Agudelo} {et~al.}(2015){Ram{\'{\i}}rez-Agudelo},
  {Sana}, {de Mink}, {H{\'e}nault-Brunet}, {de Koter}, {Langer}, {Tramper},
  {Gr{\"a}fener}, {Evans}, {Vink}, {Dufton}, \& {Taylor}}]{2015A&A...580A..92R}
{Ram{\'{\i}}rez-Agudelo}, O.~H., {Sana}, H., {de Mink}, S.~E., {et~al.} 2015,
  \aap, 580, A92

\bibitem[{{Ram{\'{\i}}rez-Agudelo} {et~al.}(2013){Ram{\'{\i}}rez-Agudelo},
  {Sim{\'o}n-D{\'{\i}}az}, {Sana}, {de Koter}, {Sab{\'{\i}}n-Sanjul{\'{\i}}an},
  {de Mink}, {Dufton}, {Gr{\"a}fener}, {Evans}, {Herrero}, {Langer}, {Lennon},
  {Ma{\'{\i}}z Apell{\'a}niz}, {Markova}, {Najarro}, {Puls}, {Taylor}, \&
  {Vink}}]{2013A&A...560A..29R}
{Ram{\'{\i}}rez-Agudelo}, O.~H., {Sim{\'o}n-D{\'{\i}}az}, S., {Sana}, H.,
  {et~al.} 2013, \aap, 560, A29

\bibitem[{{Renzo} {et~al.}(2020{\natexlab{a}}){Renzo}, {Cantiello}, {Metzger},
  \& {Jiang}}]{2020ApJ...904L..13R}
{Renzo}, M., {Cantiello}, M., {Metzger}, B.~D., \& {Jiang}, Y.~F.
  2020{\natexlab{a}}, \apjl, 904, L13

\bibitem[{{Renzo} {et~al.}(2020{\natexlab{b}}){Renzo}, {Farmer}, {Justham}, {de
  Mink}, {G{\"o}tberg}, \& {Marchant}}]{2020MNRAS.493.4333R}
{Renzo}, M., {Farmer}, R.~J., {Justham}, S., {et~al.} 2020{\natexlab{b}},
  \mnras, 493, 4333

\bibitem[{{Renzo} {et~al.}(2023){Renzo}, {Zapartas}, {Justham}, {Breivik},
  {Lau}, {Farmer}, {Cantiello}, \& {Metzger}}]{2023ApJ...942L..32R}
{Renzo}, M., {Zapartas}, E., {Justham}, S., {et~al.} 2023, \apjl, 942, L32

\bibitem[{{Saio} {et~al.}(1988){Saio}, {Kato}, \&
  {Nomoto}}]{1988ApJ...331..388S}
{Saio}, H., {Kato}, M., \& {Nomoto}, K. 1988, \apj, 331, 388

\bibitem[{{Sana} {et~al.}(2013){Sana}, {de Koter}, {de Mink}, {Dunstall},
  {Evans}, {H{\'e}nault-Brunet}, {Ma{\'{\i}}z Apell{\'a}niz},
  {Ram{\'{\i}}rez-Agudelo}, {Taylor}, {Walborn}, {Clark}, {Crowther},
  {Herrero}, {Gieles}, {Langer}, {Lennon}, \& {Vink}}]{2013A&A...550A.107S}
{Sana}, H., {de Koter}, A., {de Mink}, S.~E., {et~al.} 2013, \aap, 550, A107

\bibitem[{{Sana} {et~al.}(2012){Sana}, {de Mink}, {de Koter}, {Langer},
  {Evans}, {Gieles}, {Gosset}, {Izzard}, {Le Bouquin}, \&
  {Schneider}}]{2012Sci...337..444S}
{Sana}, H., {de Mink}, S.~E., {de Koter}, A., {et~al.} 2012, Science, 337, 444

\bibitem[{{Sana} {et~al.}(2014){Sana}, {Le Bouquin}, {Lacour}, {Berger},
  {Duvert}, {Gauchet}, {Norris}, {Olofsson}, {Pickel}, {Zins}, {Absil}, {de
  Koter}, {Kratter}, {Schnurr}, \& {Zinnecker}}]{2014ApJS..215...15S}
{Sana}, H., {Le Bouquin}, J.~B., {Lacour}, S., {et~al.} 2014, \apjs, 215, 15

\bibitem[{{Schneider} {et~al.}(2015){Schneider}, {Izzard}, {Langer}, \& {de
  Mink}}]{2015ApJ...805...20S}
{Schneider}, F.~R.~N., {Izzard}, R.~G., {Langer}, N., \& {de Mink}, S.~E. 2015,
  \apj, 805, 20

\bibitem[{{Schneider} {et~al.}(2024){Schneider}, {Laplace}, \&
  {Podsiadlowski}}]{Schneider+2024b}
{Schneider}, F.~R.~N., {Laplace}, E., \& {Podsiadlowski}, P. 2024, in prep.

\bibitem[{{Schneider} {et~al.}(2020){Schneider}, {Ohlmann}, {Podsiadlowski},
  {R{\"o}pke}, {Balbus}, \& {Pakmor}}]{2020MNRAS.495.2796S}
{Schneider}, F.~R.~N., {Ohlmann}, S.~T., {Podsiadlowski}, P., {et~al.} 2020,
  \mnras, 495, 2796

\bibitem[{{Schneider} {et~al.}(2019){Schneider}, {Ohlmann}, {Podsiadlowski},
  {R{\"o}pke}, {Balbus}, {Pakmor}, \& {Springel}}]{2019Natur.574..211S}
{Schneider}, F.~R.~N., {Ohlmann}, S.~T., {Podsiadlowski}, P., {et~al.} 2019,
  \nat, 574, 211

\bibitem[{{Schneider} {et~al.}(2016){Schneider}, {Podsiadlowski}, {Langer},
  {Castro}, \& {Fossati}}]{2016MNRAS.457.2355S}
{Schneider}, F.~R.~N., {Podsiadlowski}, P., {Langer}, N., {Castro}, N., \&
  {Fossati}, L. 2016, \mnras, 457, 2355

\bibitem[{{Schneider} {et~al.}(2023){Schneider}, {Podsiadlowski}, \&
  {Laplace}}]{2023ApJ...950L...9S}
{Schneider}, F. R.~N., {Podsiadlowski}, P., \& {Laplace}, E. 2023, \apjl, 950,
  L9

\bibitem[{{Schneider} {et~al.}(2021){Schneider}, {Podsiadlowski}, \&
  {M{\"u}ller}}]{2021A&A...645A...5S}
{Schneider}, F.~R.~N., {Podsiadlowski}, P., \& {M{\"u}ller}, B. 2021, \aap,
  645, A5

\bibitem[{{Sch{\"o}ller} {et~al.}(2017){Sch{\"o}ller}, {Hubrig}, {Fossati},
  {Carroll}, {Briquet}, {Oskinova}, {J{\"a}rvinen}, {Ilyin}, {Castro}, {Morel},
  {Langer}, {Przybilla}, {Nieva}, {Kholtygin}, {Sana}, {Herrero}, {Barb{\'a}},
  {de Koter}, \& {BOB Collaboration}}]{2017A&A...599A..66S}
{Sch{\"o}ller}, M., {Hubrig}, S., {Fossati}, L., {et~al.} 2017, \aap, 599, A66

\bibitem[{{Schootemeijer} {et~al.}(2019){Schootemeijer}, {Langer}, {Grin}, \&
  {Wang}}]{2019A&A...625A.132S}
{Schootemeijer}, A., {Langer}, N., {Grin}, N.~J., \& {Wang}, C. 2019, \aap,
  625, A132

\bibitem[{{Sibony} {et~al.}(2023){Sibony}, {Georgy}, {Ekstr{\"o}m}, \&
  {Meynet}}]{2023A&A...680A.101S}
{Sibony}, Y., {Georgy}, C., {Ekstr{\"o}m}, S., \& {Meynet}, G. 2023, \aap, 680,
  A101

\bibitem[{{Sills} {et~al.}(2005){Sills}, {Adams}, \&
  {Davies}}]{2005MNRAS.358..716S}
{Sills}, A., {Adams}, T., \& {Davies}, M.~B. 2005, \mnras, 358, 716

\bibitem[{{Smartt}(2015)}]{2015PASA...32...16S}
{Smartt}, S.~J. 2015, \pasa, 32, e016

\bibitem[{{Smith}(2014)}]{2014ARA&A..52..487S}
{Smith}, N. 2014, \araa, 52, 487

\bibitem[{{Smith} {et~al.}(2004){Smith}, {Vink}, \& {de
  Koter}}]{2004ApJ...615..475S}
{Smith}, N., {Vink}, J.~S., \& {de Koter}, A. 2004, \apj, 615, 475

\bibitem[{{Sravan} {et~al.}(2019){Sravan}, {Marchant}, \&
  {Kalogera}}]{2019ApJ...885..130S}
{Sravan}, N., {Marchant}, P., \& {Kalogera}, V. 2019, \apj, 885, 130

\bibitem[{{Stegmann} {et~al.}(2022){Stegmann}, {Antonini}, \&
  {Moe}}]{2022MNRAS.516.1406S}
{Stegmann}, J., {Antonini}, F., \& {Moe}, M. 2022, \mnras, 516, 1406

\bibitem[{{Sukhbold} \& {Adams}(2020)}]{2020MNRAS.492.2578S}
{Sukhbold}, T. \& {Adams}, S. 2020, \mnras, 492, 2578

\bibitem[{{Sukhbold} {et~al.}(2016){Sukhbold}, {Ertl}, {Woosley}, {Brown}, \&
  {Janka}}]{2016ApJ...821...38S}
{Sukhbold}, T., {Ertl}, T., {Woosley}, S.~E., {Brown}, J.~M., \& {Janka}, H.-T.
  2016, \apj, 821, 38

\bibitem[{{Sukhbold} \& {Woosley}(2014)}]{2014ApJ...783...10S}
{Sukhbold}, T. \& {Woosley}, S.~E. 2014, \apj, 783, 10

\bibitem[{{Taddia} {et~al.}(2016){Taddia}, {Sollerman}, {Fremling}, {Migotto},
  {Gal-Yam}, {Armen}, {Duggan}, {Ergon}, {Filippenko}, {Fransson},
  {Hosseinzadeh}, {Kasliwal}, {Laher}, {Leloudas}, {Leonard}, {Lunnan},
  {Masci}, {Moon}, {Silverman}, \& {Wozniak}}]{2016A&A...588A...5T}
{Taddia}, F., {Sollerman}, J., {Fremling}, C., {et~al.} 2016, \aap, 588, A5

\bibitem[{{Takahashi} {et~al.}(2023){Takahashi}, {Takiwaki}, \&
  {Yoshida}}]{2023ApJ...945...19T}
{Takahashi}, K., {Takiwaki}, T., \& {Yoshida}, T. 2023, \apj, 945, 19

\bibitem[{{Temaj} {et~al.}(2024){Temaj}, {Schneider}, {Laplace}, {Wei}, \&
  {Podsiadlowski}}]{2024A&A...682A.123T}
{Temaj}, D., {Schneider}, F.~R.~N., {Laplace}, E., {Wei}, D., \&
  {Podsiadlowski}, P. 2024, \aap, 682, A123

\bibitem[{{The LIGO Scientific Collaboration} {et~al.}(2021{\natexlab{a}}){The
  LIGO Scientific Collaboration}, {the Virgo Collaboration}, {the KAGRA
  Collaboration}, {Abbott}, {Abbott}, {Acernese}, {Ackley}, {Adams},
  {Adhikari}, {Adhikari}, {Adya}, {Affeldt}, {Agarwal}, {Agathos}, {Agatsuma},
  {Aggarwal}, {Aguiar}, {Aiello}, {Ain}, {Ajith}, {Akcay}, {Akutsu},
  {Albanesi}, {Allocca}, {Altin}, {Amato}, {Anand}, {Anand}, {Ananyeva},
  {Anderson}, {Anderson}, {Ando}, {Andrade}, {Andres}, {Andri{\'c}},
  {Angelova}, {Ansoldi}, {Antelis}, {Antier}, {Appert}, {Arai}, {Arai}, {Arai},
  {Araki}, {Araya}, {Araya}, {Areeda}, {Ar{\`e}ne}, {Aritomi}, {Arnaud},
  {Arogeti}, {Aronson}, {Arun}, {Asada}, {Asali}, {Ashton}, {Aso}, {Assiduo},
  {Aston}, {Astone}, {Aubin}, {Austin}, {Babak}, {Badaracco}, {Bader},
  {Badger}, {Bae}, {Bae}, {Baer}, {Bagnasco}, {Bai}, {Baiotti}, {Baird},
  {Bajpai}, {Ball}, {Ballardin}, {Ballmer}, {Balsamo}, {Baltus}, {Banagiri},
  {Bankar}, {Barayoga}, {Barbieri}, {Barish}, {Barker}, {Barneo}, {Barone},
  {Barr}, {Barsotti}, {Barsuglia}, {Barta}, {Bartlett}, {Barton}, {Bartos},
  {Bassiri}, {Basti}, {Bawaj}, {Bayley}, {Baylor}, {Bazzan}, {B{\'e}csy},
  {Bedakihale}, {Bejger}, {Belahcene}, {Benedetto}, {Beniwal}, {Bennett},
  {Bentley}, {BenYaala}, {Bergamin}, {Berger}, {Bernuzzi}, {Berry},
  {Bersanetti}, {Bertolini}, {Betzwieser}, {Beveridge}, {Bhandare}, {Bhardwaj},
  {Bhattacharjee}, {Bhaumik}, {Bilenko}, {Billingsley}, {Bini}, {Birney},
  {Birnholtz}, {Biscans}, {Bischi}, {Biscoveanu}, {Bisht}, {Biswas}, {Bitossi},
  {Bizouard}, {Blackburn}, {Blair}, {Blair}, {Blair}, {Bobba}, {Bode}, {Boer},
  {Bogaert}, {Boldrini}, {Bonavena}, {Bondu}, {Bonilla}, {Bonnand}, {Booker},
  {Boom}, {Bork}, {Boschi}, {Bose}, {Bose}, {Bossilkov}, {Boudart},
  {Bouffanais}, {Bozzi}, {Bradaschia}, {Brady}, {Bramley}, {Branch},
  {Branchesi}, {Brandt}, {Brau}, {Breschi}, {Briant}, {Briggs}, {Brillet},
  {Brinkmann}, {Brockill}, {Brooks}, {Brooks}, {Brown}, {Brunett}, {Bruno},
  {Bruntz}, {Bryant}, {Bulik}, {Bulten}, {Buonanno}, {Buscicchio}, {Buskulic},
  {Buy}, {Byer}, {Cabourn Davies}, {Cadonati}, {Cagnoli}, {Cahillane},
  {Calder{\'o}n Bustillo}, {Callaghan}, {Callister}, {Calloni}, {Cameron},
  {Camp}, {Canepa}, {Canevarolo}, {Cannavacciuolo}, {Cannon}, {Cao}, {Cao},
  {Capocasa}, {Capote}, {Carapella}, {Carbognani}, {Carlin}, {Carney},
  {Carpinelli}, {Carrillo}, {Carullo}, {Carver}, {Casanueva Diaz}, {Casentini},
  {Castaldi}, {Caudill}, {Cavagli{\`a}}, {Cavalier}, {Cavalieri}, {Ceasar},
  {Cella}, {Cerd{\'a}-Dur{\'a}n}, {Cesarini}, {Chaibi}, {Chakravarti},
  {Chalathadka Subrahmanya}, {Champion}, {Chan}, {Chan}, {Chan}, {Chan},
  {Chan}, {Chandra}, {Chanial}, {Chao}, {Chapman-Bird}, {Charlton}, {Chase},
  {Chassande-Mottin}, {Chatterjee}, {Chatterjee}, {Chatterjee}, {Chaturvedi},
  {Chaty}, {Chatziioannou}, {Chen}, {Chen}, {Chen}, {Chen}, {Chen}, {Chen},
  {Chen}, {Chen}, {Cheng}, {Cheong}, {Cheung}, {Chia}, {Chiadini}, {Chiang},
  {Chiarini}, {Chierici}, {Chincarini}, {Chiofalo}, {Chiummo}, {Cho}, {Cho},
  {Choudhary}, {Choudhary}, {Christensen}, {Chu}, {Chu}, {Chu}, {Chua},
  {Chung}, {Ciani}, {Ciecielag}, {Cie{\'s}lar}, {Cifaldi}, {Ciobanu}, {Ciolfi},
  {Cipriano}, {Cirone}, {Clara}, {Clark}, {Clark}, {Clarke}, {Clearwater},
  {Clesse}, {Cleva}, {Coccia}, {Codazzo}, {Cohadon}, {Cohen}, {Cohen},
  {Colleoni}, {Collette}, {Colombo}, {Colpi}, {Compton}, {Constancio}, {Conti},
  {Cooper}, {Corban}, {Corbitt}, {Cordero-Carri{\'o}n}, {Corezzi}, {Corley},
  {Cornish}, {Corre}, {Corsi}, {Cortese}, {Costa}, {Cotesta}, {Coughlin},
  {Coulon}, {Countryman}, {Cousins}, {Couvares}, {Coward}, {Cowart}, {Coyne},
  {Coyne}, {Creighton}, {Creighton}, {Criswell}, {Croquette}, {Crowder},
  {Cudell}, {Cullen}, {Cumming}, {Cummings}, {Cunningham}, {Cuoco},
  {Cury{\l}o}, {Dabadie}, {Dal Canton}, {Dall'Osso}, {D{\'a}lya}, {Dana},
  {DaneshgaranBajastani}, {D'Angelo}, {Danila}, {Danilishin}, {D'Antonio},
  {Danzmann}, {Darsow-Fromm}, {Dasgupta}, {Datrier}, {Datta}, {Dattilo},
  {Dave}, {Davier}, {Davis}, {Davis}, {Daw}, {de Alarc{\'o}n}, {Dean}, {DeBra},
  {Deenadayalan}, {Degallaix}, {De Laurentis}, {Del{\'e}glise}, {Del Favero},
  {De Lillo}, {De Lillo}, {Del Pozzo}, {DeMarchi}, {De Matteis}, {D'Emilio},
  {Demos}, {Dent}, {Depasse}, {De Pietri}, {De Rosa}, {De Rossi}, {DeSalvo},
  {De Simone}, {Dhurandhar}, {D{\'\i}az}, {Diaz-Ortiz}, {Didio}, {Dietrich},
  {Di Fiore}, {Di Fronzo}, {Di Giorgio}, {Di Giovanni}, {Di Giovanni}, {Di
  Girolamo}, {Di Lieto}, {Ding}, {Di Pace}, {Di Palma}, {Di Renzo},
  {Divakarla}, {Dmitriev}, {Doctor}, {D'Onofrio}, {Donovan}, {Dooley},
  {Doravari}, {Dorrington}, {Drago}, {Driggers}, {Drori}, {Ducoin}, {Dupej},
  {Durante}, {D'Urso}, {Duverne}, {Dwyer}, {Eassa}, {Easter}, {Ebersold},
  {Eckhardt}, {Eddolls}, {Edelman}, {Edo}, {Edy}, {Effler}, {Eguchi},
  {Eichholz}, {Eikenberry}, {Eisenmann}, {Eisenstein}, {Ejlli}, {Engelby},
  {Enomoto}, {Errico}, {Essick}, {Estell{\'e}s}, {Estevez}, {Etienne}, {Etzel},
  {Evans}, {Evans}, {Ewing}, {Fafone}, {Fair}, {Fairhurst}, {Farah}, {Farinon},
  {Farr}, {Farr}, {Farrow}, {Fauchon-Jones}, {Favaro}, {Favata}, {Fays},
  {Fazio}, {Feicht}, {Fejer}, {Fenyvesi}, {Ferguson}, {Fernandez-Galiana},
  {Ferrante}, {Ferreira}, {Fidecaro}, {Figura}, {Fiori}, {Fishbach}, {Fisher},
  {Fittipaldi}, {Fiumara}, {Flaminio}, {Floden}, {Fong}, {Font}, {Fornal},
  {Forsyth}, {Franke}, {Frasca}, {Frasconi}, {Frederick}, {Freed}, {Frei},
  {Freise}, {Frey}, {Fritschel}, {Frolov}, {Fronz{\'e}}, {Fujii}, {Fujikawa},
  {Fukunaga}, {Fukushima}, {Fulda}, {Fyffe}, {Gabbard}, {Gabella}, {Gadre},
  {Gair}, {Gais}, {Galaudage}, {Gamba}, {Ganapathy}, {Ganguly}, {Gao},
  {Gaonkar}, {Garaventa}, {Garc{\'\i}a}, {Garc{\'\i}a-N{\'u}{\~n}ez},
  {Garc{\'\i}a-Quir{\'o}s}, {Garufi}, {Gateley}, {Gaudio}, {Gayathri}, {Ge},
  {Gemme}, {Gennai}, {George}, {George}, {Gerberding}, {Gergely}, {Gewecke},
  {Ghonge}, {Ghosh}, {Ghosh}, {Ghosh}, {Ghosh}, {Giacomazzo}, {Giacoppo},
  {Giaime}, {Giardina}, {Gibson}, {Gier}, {Giesler}, {Giri}, {Gissi},
  {Glanzer}, {Gleckl}, {Godwin}, {Goetz}, {Goetz}, {Gohlke}, {Golomb},
  {Goncharov}, {Gonz{\'a}lez}, {Gopakumar}, {Gosselin}, {Gouaty}, {Gould},
  {Grace}, {Grado}, {Granata}, {Granata}, {Grant}, {Gras}, {Grassia}, {Gray},
  {Gray}, {Greco}, {Green}, {Green}, {Gretarsson}, {Gretarsson}, {Griffith},
  {Griffiths}, {Griggs}, {Grignani}, {Grimaldi}, {Grimm}, {Grote}, {Grunewald},
  {Gruning}, {Guerra}, {Guidi}, {Guimaraes}, {Guix{\'e}}, {Gulati}, {Guo},
  {Guo}, {Gupta}, {Gupta}, {Gupta}, {Gustafson}, {Gustafson}, {Guzman}, {Ha},
  {Haegel}, {Hagiwara}, {Haino}, {Halim}, {Hall}, {Hamilton}, {Hammond}, {Han},
  {Haney}, {Hanks}, {Hanna}, {Hannam}, {Hannuksela}, {Hansen}, {Hansen},
  {Hanson}, {Harder}, {Hardwick}, {Haris}, {Harms}, {Harry}, {Harry},
  {Hartwig}, {Hasegawa}, {Haskell}, {Hasskew}, {Haster}, {Hattori}, {Haughian},
  {Hayakawa}, {Hayama}, {Hayes}, {Healy}, {Heidmann}, {Heidt}, {Heintze},
  {Heinze}, {Heinzel}, {Heitmann}, {Hellman}, {Hello}, {Helmling-Cornell},
  {Hemming}, {Hendry}, {Heng}, {Hennes}, {Hennig}, {Hennig}, {Hernandez},
  {Hernandez Vivanco}, {Heurs}, {Hild}, {Hill}, {Himemoto}, {Hines},
  {Hiranuma}, {Hirata}, {Hirose}, {Hochheim}, {Hofman}, {Hohmann}, {Holcomb},
  {Holland}, {Holley-Bockelmann}, {Hollows}, {Holmes}, {Holt}, {Holz}, {Hong},
  {Hopkins}, {Hough}, {Hourihane}, {Howell}, {Hoy}, {Hoyland}, {Hreibi},
  {Hsieh}, {Hsu}, {Huang}, {Huang}, {Huang}, {Huang}, {Huang}, {Huang},
  {H{\"u}bner}, {Huddart}, {Hughey}, {Hui}, {Hui}, {Husa}, {Huttner},
  {Huxford}, {Huynh-Dinh}, {Ide}, {Idzkowski}, {Iess}, {Ikenoue}, {Imam},
  {Inayoshi}, {Ingram}, {Inoue}, {Ioka}, {Isi}, {Isleif}, {Ito}, {Itoh},
  {Iyer}, {Izumi}, {JaberianHamedan}, {Jacqmin}, {Jadhav}, {Jadhav}, {James},
  {Jan}, {Jani}, {Janquart}, {Janssens}, {Janthalur}, {Jaranowski}, {Jariwala},
  {Jaume}, {Jenkins}, {Jenner}, {Jeon}, {Jeunon}, {Jia}, {Jin}, {Johns},
  {Johnson-McDaniel}, {Jones}, {Jones}, {Jones}, {Jones}, {Jones}, {Jonker},
  {Ju}, {Jung}, {Jung}, {Junker}, {Juste}, {Kaihotsu}, {Kajita}, {Kakizaki},
  {Kalaghatgi}, {Kalogera}, {Kamai}, {Kamiizumi}, {Kanda}, {Kandhasamy},
  {Kang}, {Kanner}, {Kao}, {Kapadia}, {Kapasi}, {Karat}, {Karathanasis},
  {Karki}, {Kashyap}, {Kasprzack}, {Kastaun}, {Katsanevas}, {Katsavounidis},
  {Katzman}, {Kaur}, {Kawabe}, {Kawaguchi}, {Kawai}, {Kawasaki},
  {K{\'e}f{\'e}lian}, {Keitel}, {Key}, {Khadka}, {Khalili}, {Khan}, {Khazanov},
  {Khetan}, {Khursheed}, {Kijbunchoo}, {Kim}, {Kim}, {Kim}, {Kim}, {Kim},
  {Kim}, {Kimball}, {Kimura}, {Kinley-Hanlon}, {Kirchhoff}, {Kissel}, {Kita},
  {Kitazawa}, {Kleybolte}, {Klimenko}, {Knee}, {Knowles}, {Knyazev}, {Koch},
  {Koekoek}, {Kojima}, {Kokeyama}, {Koley}, {Kolitsidou}, {Kolstein}, {Komori},
  {Kondrashov}, {Kong}, {Kontos}, {Koper}, {Korobko}, {Kotake}, {Kovalam},
  {Kozak}, {Kozakai}, {Kozu}, {Kringel}, {Krishnendu}, {Kr{\'o}lak}, {Kuehn},
  {Kuei}, {Kuijer}, {Kulkarni}, {Kumar}, {Kumar}, {Kumar}, {Kumar}, {Kume},
  {Kuns}, {Kuo}, {Kuo}, {Kuromiya}, {Kuroyanagi}, {Kusayanagi}, {Kuwahara},
  {Kwak}, {Lagabbe}, {Laghi}, {Lalande}, {Lam}, {Lamberts}, {Landry}, {Lane},
  {Lang}, {Lange}, {Lantz}, {La Rosa}, {Lartaux-Vollard}, {Lasky}, {Laxen},
  {Lazzarini}, {Lazzaro}, {Leaci}, {Leavey}, {Lecoeuche}, {Lee}, {Lee}, {Lee},
  {Lee}, {Lee}, {Lee}, {Lehmann}, {Lema{\^\i}tre}, {Leonardi}, {Leroy},
  {Letendre}, {Levesque}, {Levin}, {Leviton}, {Leyde}, {Li}, {Li}, {Li}, {Li},
  {Li}, {Li}, {Lin}, {Lin}, {Lin}, {Lin}, {Lin}, {Linde}, {Linker}, {Linley},
  {Littenberg}, {Liu}, {Liu}, {Liu}, {Liu}, {Llamas}, {Llorens-Monteagudo},
  {Lo}, {Lockwood}, {Loh}, {London}, {Longo}, {Lopez}, {Lopez Portilla},
  {Lorenzini}, {Loriette}, {Lormand}, {Losurdo}, {Lott}, {Lough}, {Lousto},
  {Lovelace}, {Lucaccioni}, {L{\"u}ck}, {Lumaca}, {Lundgren}, {Luo}, {Lynam},
  {Macas}, {MacInnis}, {Macleod}, {MacMillan}, {Macquet}, {Maga{\~n}a
  Hernandez}, {Magazz{\`u}}, {Magee}, {Maggiore}, {Magnozzi}, {Mahesh},
  {Majorana}, {Makarem}, {Maksimovic}, {Maliakal}, {Malik}, {Man}, {Mandic},
  {Mangano}, {Mango}, {Mansell}, {Manske}, {Mantovani}, {Mapelli},
  {Marchesoni}, {Marchio}, {Marion}, {Mark}, {M{\'a}rka}, {M{\'a}rka},
  {Markakis}, {Markosyan}, {Markowitz}, {Maros}, {Marquina}, {Marsat},
  {Martelli}, {Martin}, {Martin}, {Martinez}, {Martinez}, {Martinez},
  {Martinovic}, {Martynov}, {Marx}, {Masalehdan}, {Mason}, {Massera},
  {Masserot}, {Massinger}, {Masso-Reid}, {Mastrogiovanni}, {Matas},
  {Mateu-Lucena}, {Matichard}, {Matiushechkina}, {Mavalvala}, {McCann},
  {McCarthy}, {McClelland}, {McClincy}, {McCormick}, {McCuller}, {McGhee},
  {McGuire}, {McIsaac}, {McIver}, {McRae}, {McWilliams}, {Meacher}, {Mehmet},
  {Mehta}, {Meijer}, {Melatos}, {Melchor}, {Mendell}, {Menendez-Vazquez},
  {Menoni}, {Mercer}, {Mereni}, {Merfeld}, {Merilh}, {Merritt}, {Merzougui},
  {Meshkov}, {Messenger}, {Messick}, {Meyers}, {Meylahn}, {Mhaske}, {Miani},
  {Miao}, {Michaloliakos}, {Michel}, {Michimura}, {Middleton}, {Milano},
  {Miller}, {Miller}, {Miller}, {Millhouse}, {Mills}, {Milotti}, {Minazzoli},
  {Minenkov}, {Mio}, {Mir}, {Miravet-Ten{\'e}s}, {Mishra}, {Mishra}, {Mistry},
  {Mitra}, {Mitrofanov}, {Mitselmakher}, {Mittleman}, {Miyakawa}, {Miyamoto},
  {Miyazaki}, {Miyo}, {Miyoki}, {Mo}, {Modafferi}, {Moguel}, {Mogushi},
  {Mohapatra}, {Mohite}, {Molina}, {Molina-Ruiz}, {Mondin}, {Montani}, {Moore},
  {Moraru}, {Morawski}, {More}, {Moreno}, {Moreno}, {Mori}, {Morisaki},
  {Moriwaki}, {Morr{\'a}s}, {Mours}, {Mow-Lowry}, {Mozzon}, {Muciaccia},
  {Mukherjee}, {Mukherjee}, {Mukherjee}, {Mukherjee}, {Mukherjee}, {Mukund},
  {Mullavey}, {Munch}, {Mu{\~n}iz}, {Murray}, {Musenich}, {Muusse}, {Nadji},
  {Nagano}, {Nagano}, {Nagar}, {Nakamura}, {Nakano}, {Nakano}, {Nakashima},
  {Nakayama}, {Napolano}, {Nardecchia}, {Narikawa}, {Naticchioni}, {Nayak},
  {Nayak}, {Negishi}, {Neil}, {Neilson}, {Nelemans}, {Nelson}, {Nery},
  {Neubauer}, {Neunzert}, {Ng}, {Ng}, {Nguyen}, {Nguyen}, {Nguyen}, {Nguyen
  Quynh}, {Ni}, {Nichols}, {Nishizawa}, {Nissanke}, {Nitoglia}, {Nocera},
  {Norman}, {North}, {Nozaki}, {Nu{\~n}o Siles}, {Nuttall}, {Oberling},
  {O'Brien}, {Obuchi}, {O'Dell}, {Oelker}, {Ogaki}, {Oganesyan}, {Oh}, {Oh},
  {Oh}, {Ohashi}, {Ohishi}, {Ohkawa}, {Ohme}, {Ohta}, {Okada}, {Okutani},
  {Okutomi}, {Olivetto}, {Oohara}, {Ooi}, {Oram}, {O'Reilly}, {Ormiston},
  {Ormsby}, {Ortega}, {O'Shaughnessy}, {O'Shea}, {Oshino}, {Ossokine},
  {Osthelder}, {Otabe}, {Ottaway}, {Overmier}, {Pace}, {Pagano}, {Page},
  {Pagliaroli}, {Pai}, {Pai}, {Palamos}, {Palashov}, {Palomba}, {Pan}, {Pan},
  {Panda}, {Pang}, {Pang}, {Pankow}, {Pannarale}, {Pant}, {Panther},
  {Paoletti}, {Paoli}, {Paolone}, {Parisi}, {Park}, {Park}, {Parker},
  {Pascucci}, {Pasqualetti}, {Passaquieti}, {Passuello}, {Patel}, {Pathak},
  {Patricelli}, {Patron}, {Paul}, {Payne}, {Pedraza}, {Pegoraro}, {Pele},
  {Pe{\~n}a Arellano}, {Penn}, {Perego}, {Pereira}, {Pereira}, {Perez},
  {P{\'e}rigois}, {Perkins}, {Perreca}, {Perri{\`e}s}, {Petermann},
  {Petterson}, {Pfeiffer}, {Pham}, {Phukon}, {Piccinni}, {Pichot},
  {Piendibene}, {Piergiovanni}, {Pierini}, {Pierro}, {Pillant}, {Pillas},
  {Pilo}, {Pinard}, {Pinto}, {Pinto}, {Piotrzkowski}, {Piotrzkowski},
  {Pirello}, {Pitkin}, {Placidi}, {Planas}, {Plastino}, {Pluchar}, {Poggiani},
  {Polini}, {Pong}, {Ponrathnam}, {Popolizio}, {Porter}, {Poulton}, {Powell},
  {Pracchia}, {Pradier}, {Prajapati}, {Prasai}, {Prasanna}, {Pratten},
  {Principe}, {Prodi}, {Prokhorov}, {Prosposito}, {Prudenzi}, {Puecher},
  {Punturo}, {Puosi}, {Puppo}, {P{\"u}rrer}, {Qi}, {Quetschke},
  {Quitzow-James}, {Qutob}, {Raab}, {Raaijmakers}, {Radkins}, {Radulesco},
  {Raffai}, {Rail}, {Raja}, {Rajan}, {Ramirez}, {Ramirez}, {Ramos-Buades},
  {Rana}, {Rapagnani}, {Rapol}, {Ray}, {Raymond}, {Raza}, {Razzano}, {Read},
  {Rees}, {Regimbau}, {Rei}, {Reid}, {Reid}, {Reitze}, {Relton}, {Renzini},
  {Rettegno}, {Reza}, {Rezac}, {Ricci}, {Richards}, {Richardson}, {Richardson},
  {Riemenschneider}, {Riles}, {Rinaldi}, {Rink}, {Rizzo}, {Robertson}, {Robie},
  {Robinet}, {Rocchi}, {Rodriguez}, {Rolland}, {Rollins}, {Romanelli},
  {Romano}, {Romel}, {Romero-Rodr{\'\i}guez}, {Romero-Shaw}, {Romie},
  {Ronchini}, {Rosa}, {Rose}, {Rosi{\'n}ska}, {Ross}, {Rowan}, {Rowlinson},
  {Roy}, {Roy}, {Roy}, {Rozza}, {Ruggi}, {Ruiz-Rocha}, {Ryan}, {Sachdev},
  {Sadecki}, {Sadiq}, {Sago}, {Saito}, {Saito}, {Sakai}, {Sakai},
  {Sakellariadou}, {Sakuno}, {Salafia}, {Salconi}, {Saleem}, {Salemi},
  {Samajdar}, {Sanchez}, {Sanchez}, {Sanchez}, {Sanchis-Gual}, {Sanders},
  {Sanuy}, {Saravanan}, {Sarin}, {Sassolas}, {Satari}, {Sathyaprakash}, {Sato},
  {Sato}, {Sauter}, {Savage}, {Sawada}, {Sawant}, {Sawant}, {Sayah},
  {Schaetzl}, {Scheel}, {Scheuer}, {Schiworski}, {Schmidt}, {Schmidt},
  {Schnabel}, {Schneewind}, {Schofield}, {Sch{\"o}nbeck}, {Schulte}, {Schutz},
  {Schwartz}, {Scott}, {Scott}, {Seglar-Arroyo}, {Sekiguchi}, {Sekiguchi},
  {Sellers}, {Sengupta}, {Sentenac}, {Seo}, {Sequino}, {Sergeev}, {Setyawati},
  {Shaffer}, {Shahriar}, {Shams}, {Shao}, {Sharma}, {Sharma}, {Shawhan},
  {Shcheblanov}, {Shibagaki}, {Shikauchi}, {Shimizu}, {Shimoda}, {Shimode},
  {Shinkai}, {Shishido}, {Shoda}, {Shoemaker}, {Shoemaker}, {ShyamSundar},
  {Sieniawska}, {Sigg}, {Singer}, {Singh}, {Singh}, {Singha}, {Sintes},
  {Sipala}, {Skliris}, {Slagmolen}, {Slaven-Blair}, {Smetana}, {Smith},
  {Smith}, {Soldateschi}, {Somala}, {Somiya}, {Son}, {Soni}, {Soni}, {Sordini},
  {Sorrentino}, {Sorrentino}, {Sotani}, {Soulard}, {Souradeep}, {Sowell},
  {Spagnuolo}, {Spencer}, {Spera}, {Srinivasan}, {Srivastava}, {Srivastava},
  {Staats}, {Stachie}, {Steer}, {Steinhoff}, {Steinlechner}, {Steinlechner},
  {Stevenson}, {Stops}, {Stover}, {Strain}, {Strang}, {Stratta}, {Strunk},
  {Sturani}, {Stuver}, {Sudhagar}, {Sudhir}, {Sugimoto}, {Suh}, {Sullivan},
  {Sullivan}, {Summerscales}, {Sun}, {Sun}, {Sunil}, {Sur}, {Suresh}, {Sutton},
  {Suzuki}, {Suzuki}, {Swinkels}, {Szczepa{\'n}czyk}, {Szewczyk}, {Tacca},
  {Tagoshi}, {Tait}, {Takahashi}, {Takahashi}, {Takamori}, {Takano}, {Takeda},
  {Takeda}, {Talbot}, {Talbot}, {Tanaka}, {Tanaka}, {Tanaka}, {Tanaka},
  {Tanaka}, {Tanasijczuk}, {Tanioka}, {Tanner}, {Tao}, {Tao}, {Tapia San
  Mart{\'\i}n}, {Taranto}, {Tasson}, {Telada}, {Tenorio}, {Terhune},
  {Terkowski}, {Thirugnanasambandam}, {Thomas}, {Thomas}, {Thomas}, {Thompson},
  {Thondapu}, {Thorne}, {Thrane}, {Tiwari}, {Tiwari}, {Tiwari}, {Toivonen},
  {Toland}, {Tolley}, {Tomaru}, {Tomigami}, {Tomura}, {Tonelli},
  {Torres-Forn{\'e}}, {Torrie}, {Tosta e Melo}, {T{\"o}yr{\"a}}, {Trapananti},
  {Travasso}, {Traylor}, {Trevor}, {Tringali}, {Tripathee}, {Troiano},
  {Trovato}, {Trozzo}, {Trudeau}, {Tsai}, {Tsai}, {Tsang}, {Tsang}, {Tsao},
  {Tse}, {Tso}, {Tsubono}, {Tsuchida}, {Tsukada}, {Tsuna}, {Tsutsui},
  {Tsuzuki}, {Turbang}, {Turconi}, {Tuyenbayev}, {Ubhi}, {Uchikata},
  {Uchiyama}, {Udall}, {Ueda}, {Uehara}, {Ueno}, {Ueshima}, {Unnikrishnan},
  {Uraguchi}, {Urban}, {Ushiba}, {Utina}, {Vahlbruch}, {Vajente}, {Vajpeyi},
  {Valdes}, {Valentini}, {Valsan}, {van Bakel}, {van Beuzekom}, {van den
  Brand}, {Van Den Broeck}, {Vander-Hyde}, {van der Schaaf}, {van Heijningen},
  {Vanosky}, {van Putten}, {van Remortel}, {Vardaro}, {Vargas}, {Varma},
  {Vas{\'u}th}, {Vecchio}, {Vedovato}, {Veitch}, {Veitch}, {Venneberg},
  {Venugopalan}, {Verkindt}, {Verma}, {Verma}, {Veske}, {Vetrano},
  {Vicer{\'e}}, {Vidyant}, {Viets}, {Vijaykumar}, {Villa-Ortega}, {Vinet},
  {Virtuoso}, {Vitale}, {Vo}, {Vocca}, {von Reis}, {von Wrangel}, {Vorvick},
  {Vyatchanin}, {Wade}, {Wade}, {Wagner}, {Walet}, {Walker}, {Wallace},
  {Wallace}, {Walsh}, {Wang}, {Wang}, {Wang}, {Ward}, {Warner}, {Was},
  {Washimi}, {Washington}, {Watchi}, {Weaver}, {Webster}, {Weinert},
  {Weinstein}, {Weiss}, {Weller}, {Weller}, {Wellmann}, {Wen}, {We{\ss}els},
  {Wette}, {Whelan}, {White}, {Whiting}, {Whittle}, {Wilken}, {Williams},
  {Williams}, {Williams}, {Williamson}, {Willis}, {Willke}, {Wilson},
  {Winkler}, {Wipf}, {Wlodarczyk}, {Woan}, {Woehler}, {Wofford}, {Wong}, {Wu},
  {Wu}, {Wu}, {Wu}, {Wysocki}, {Xiao}, {Xu}, {Yamada}, {Yamamoto}, {Yamamoto},
  {Yamamoto}, {Yamamoto}, {Yamashita}, {Yamazaki}, {Yang}, {Yang}, {Yang},
  {Yang}, {Yang}, {Yap}, {Yeeles}, {Yelikar}, {Ying}, {Yokogawa}, {Yokoyama},
  {Yokozawa}, {Yoo}, {Yoshioka}, {Yu}, {Yu}, {Yuzurihara}, {Zadro{\.z}ny},
  {Zanolin}, {Zeidler}, {Zelenova}, {Zendri}, {Zevin}, {Zhan}, {Zhang},
  {Zhang}, {Zhang}, {Zhang}, {Zhang}, {Zhao}, {Zhao}, {Zhao}, {Zhao}, {Zheng},
  {Zhou}, {Zhou}, {Zhu}, {Zhu}, {Zimmerman}, {Zlochower}, {Zucker}, \&
  {Zweizig}}]{2021arXiv211103606T}
{The LIGO Scientific Collaboration}, {the Virgo Collaboration}, {the KAGRA
  Collaboration}, {et~al.} 2021{\natexlab{a}}, arXiv e-prints, arXiv:2111.03606

\bibitem[{{The LIGO Scientific Collaboration} {et~al.}(2021{\natexlab{b}}){The
  LIGO Scientific Collaboration}, {the Virgo Collaboration}, {the KAGRA
  Collaboration}, {Abbott}, {Abbott}, {Acernese}, {Ackley}, {Adams},
  {Adhikari}, {Adhikari}, {Adya}, {Affeldt}, {Agarwal}, {Agathos}, {Agatsuma},
  {Aggarwal}, {Aguiar}, {Aiello}, {Ain}, {Ajith}, {Akutsu}, {Albanesi},
  {Allocca}, {Altin}, {Amato}, {Anand}, {Anand}, {Ananyeva}, {Anderson},
  {Anderson}, {Ando}, {Andrade}, {Andres}, {Andri{\'c}}, {Angelova}, {Ansoldi},
  {Antelis}, {Antier}, {Antonini}, {Appert}, {Arai}, {Arai}, {Arai}, {Araki},
  {Araya}, {Araya}, {Areeda}, {Ar{\`e}ne}, {Aritomi}, {Arnaud}, {Aronson},
  {Arun}, {Asada}, {Asali}, {Ashton}, {Aso}, {Assiduo}, {Aston}, {Astone},
  {Aubin}, {Austin}, {Babak}, {Badaracco}, {Bader}, {Badger}, {Bae}, {Bae},
  {Baer}, {Bagnasco}, {Bai}, {Baiotti}, {Baird}, {Bajpai}, {Ball}, {Ballardin},
  {Ballmer}, {Balsamo}, {Baltus}, {Banagiri}, {Bankar}, {Barayoga}, {Barbieri},
  {Barish}, {Barker}, {Barneo}, {Barone}, {Barr}, {Barsotti}, {Barsuglia},
  {Barta}, {Bartlett}, {Barton}, {Bartos}, {Bassiri}, {Basti}, {Bawaj},
  {Bayley}, {Baylor}, {Bazzan}, {B{\'e}csy}, {Bedakihale}, {Bejger},
  {Belahcene}, {Benedetto}, {Beniwal}, {Bennett}, {Bentley}, {BenYaala},
  {Bergamin}, {Berger}, {Bernuzzi}, {Berry}, {Bersanetti}, {Bertolini},
  {Betzwieser}, {Beveridge}, {Bhandare}, {Bhardwaj}, {Bhattacharjee},
  {Bhaumik}, {Bilenko}, {Billingsley}, {Bini}, {Birney}, {Birnholtz},
  {Biscans}, {Bischi}, {Biscoveanu}, {Bisht}, {Biswas}, {Bitossi}, {Bizouard},
  {Blackburn}, {Blair}, {Blair}, {Blair}, {Bobba}, {Bode}, {Boer}, {Bogaert},
  {Boldrini}, {Bonavena}, {Bondu}, {Bonilla}, {Bonnand}, {Booker}, {Boom},
  {Bork}, {Boschi}, {Bose}, {Bose}, {Bossilkov}, {Boudart}, {Bouffanais},
  {Bozzi}, {Bradaschia}, {Brady}, {Bramley}, {Branch}, {Branchesi}, {Brau},
  {Breschi}, {Briant}, {Briggs}, {Brillet}, {Brinkmann}, {Brockill}, {Brooks},
  {Brooks}, {Brown}, {Brunett}, {Bruno}, {Bruntz}, {Bryant}, {Bulik}, {Bulten},
  {Buonanno}, {Buscicchio}, {Buskulic}, {Buy}, {Byer}, {Cadonati}, {Cagnoli},
  {Cahillane}, {Calder{\'o}n Bustillo}, {Callaghan}, {Callister}, {Calloni},
  {Cameron}, {Camp}, {Canepa}, {Canevarolo}, {Cannavacciuolo}, {Cannon}, {Cao},
  {Cao}, {Capocasa}, {Capote}, {Carapella}, {Carbognani}, {Carlin}, {Carney},
  {Carpinelli}, {Carrillo}, {Carullo}, {Carver}, {Casanueva Diaz}, {Casentini},
  {Castaldi}, {Caudill}, {Cavagli{\`a}}, {Cavalier}, {Cavalieri}, {Ceasar},
  {Cella}, {Cerd{\'a}-Dur{\'a}n}, {Cesarini}, {Chaibi}, {Chakravarti},
  {Chalathadka Subrahmanya}, {Champion}, {Chan}, {Chan}, {Chan}, {Chan},
  {Chan}, {Chandra}, {Chanial}, {Chao}, {Charlton}, {Chase},
  {Chassande-Mottin}, {Chatterjee}, {Chatterjee}, {Chatterjee}, {Chaturvedi},
  {Chaty}, {Chatziioannou}, {Chen}, {Chen}, {Chen}, {Chen}, {Chen}, {Chen},
  {Chen}, {Chen}, {Cheng}, {Cheong}, {Cheung}, {Chia}, {Chiadini}, {Chiang},
  {Chiarini}, {Chierici}, {Chincarini}, {Chiofalo}, {Chiummo}, {Cho}, {Cho},
  {Choudhary}, {Choudhary}, {Christensen}, {Chu}, {Chu}, {Chu}, {Chua},
  {Chung}, {Ciani}, {Ciecielag}, {Cie{\'s}lar}, {Cifaldi}, {Ciobanu}, {Ciolfi},
  {Cipriano}, {Cirone}, {Clara}, {Clark}, {Clark}, {Clarke}, {Clearwater},
  {Clesse}, {Cleva}, {Coccia}, {Codazzo}, {Cohadon}, {Cohen}, {Cohen},
  {Colleoni}, {Collette}, {Colombo}, {Colpi}, {Compton}, {Constancio}, {Conti},
  {Cooper}, {Corban}, {Corbitt}, {Cordero-Carri{\'o}n}, {Corezzi}, {Corley},
  {Cornish}, {Corre}, {Corsi}, {Cortese}, {Costa}, {Cotesta}, {Coughlin},
  {Coulon}, {Countryman}, {Cousins}, {Couvares}, {Coward}, {Cowart}, {Coyne},
  {Coyne}, {Creighton}, {Creighton}, {Criswell}, {Croquette}, {Crowder},
  {Cudell}, {Cullen}, {Cumming}, {Cummings}, {Cunningham}, {Cuoco},
  {Cury{\l}o}, {Dabadie}, {Dal Canton}, {Dall'Osso}, {D{\'a}lya}, {Dana},
  {DaneshgaranBajastani}, {D'Angelo}, {Danilishin}, {D'Antonio}, {Danzmann},
  {Darsow-Fromm}, {Dasgupta}, {Datrier}, {Datta}, {Dattilo}, {Dave}, {Davier},
  {Davies}, {Davis}, {Davis}, {Daw}, {Dean}, {DeBra}, {Deenadayalan},
  {Degallaix}, {De Laurentis}, {Del{\'e}glise}, {Del Favero}, {De Lillo}, {De
  Lillo}, {Del Pozzo}, {DeMarchi}, {De Matteis}, {D'Emilio}, {Demos}, {Dent},
  {Depasse}, {De Pietri}, {De Rosa}, {De Rossi}, {DeSalvo}, {De Simone},
  {Dhurandhar}, {D{\'\i}az}, {Diaz-Ortiz}, {Didio}, {Dietrich}, {Di Fiore}, {Di
  Fronzo}, {Di Giorgio}, {Di Giovanni}, {Di Giovanni}, {Di Girolamo}, {Di
  Lieto}, {Ding}, {Di Pace}, {Di Palma}, {Di Renzo}, {Divakarla}, {Dmitriev},
  {Doctor}, {D'Onofrio}, {Donovan}, {Dooley}, {Doravari}, {Dorrington},
  {Drago}, {Driggers}, {Drori}, {Ducoin}, {Dupej}, {Durante}, {D'Urso},
  {Duverne}, {Dwyer}, {Eassa}, {Easter}, {Ebersold}, {Eckhardt}, {Eddolls},
  {Edelman}, {Edo}, {Edy}, {Effler}, {Eguchi}, {Eichholz}, {Eikenberry},
  {Eisenmann}, {Eisenstein}, {Ejlli}, {Engelby}, {Enomoto}, {Errico}, {Essick},
  {Estell{\'e}s}, {Estevez}, {Etienne}, {Etzel}, {Evans}, {Evans}, {Ewing},
  {Fafone}, {Fair}, {Fairhurst}, {Farah}, {Farinon}, {Farr}, {Farr}, {Farrow},
  {Fauchon-Jones}, {Favaro}, {Favata}, {Fays}, {Fazio}, {Feicht}, {Fejer},
  {Fenyvesi}, {Ferguson}, {Fernandez-Galiana}, {Ferrante}, {Ferreira},
  {Fidecaro}, {Figura}, {Fiori}, {Fishbach}, {Fisher}, {Fittipaldi}, {Fiumara},
  {Flaminio}, {Floden}, {Fong}, {Font}, {Fornal}, {Forsyth}, {Franke},
  {Frasca}, {Frasconi}, {Frederick}, {Freed}, {Frei}, {Freise}, {Frey},
  {Fritschel}, {Frolov}, {Fronz{\'e}}, {Fujii}, {Fujikawa}, {Fukunaga},
  {Fukushima}, {Fulda}, {Fyffe}, {Gabbard}, {Gadre}, {Gair}, {Gais},
  {Galaudage}, {Gamba}, {Ganapathy}, {Ganguly}, {Gao}, {Gaonkar}, {Garaventa},
  {Garc{\'\i}a-N{\'u}{\~n}ez}, {Garc{\'\i}a-Quir{\'o}s}, {Garufi}, {Gateley},
  {Gaudio}, {Gayathri}, {Ge}, {Gemme}, {Gennai}, {George}, {Gerberding},
  {Gergely}, {Gewecke}, {Ghonge}, {Ghosh}, {Ghosh}, {Ghosh}, {Ghosh},
  {Giacomazzo}, {Giacoppo}, {Giaime}, {Giardina}, {Gibson}, {Gier}, {Giesler},
  {Giri}, {Gissi}, {Glanzer}, {Gleckl}, {Godwin}, {Goetz}, {Goetz}, {Gohlke},
  {Golomb}, {Goncharov}, {Gonz{\'a}lez}, {Gopakumar}, {Gosselin}, {Gouaty},
  {Gould}, {Grace}, {Grado}, {Granata}, {Granata}, {Grant}, {Gras}, {Grassia},
  {Gray}, {Gray}, {Greco}, {Green}, {Green}, {Gretarsson}, {Gretarsson},
  {Griffith}, {Griffiths}, {Griggs}, {Grignani}, {Grimaldi}, {Grimm}, {Grote},
  {Grunewald}, {Gruning}, {Guerra}, {Guidi}, {Guimaraes}, {Guix{\'e}},
  {Gulati}, {Guo}, {Guo}, {Gupta}, {Gupta}, {Gupta}, {Gustafson}, {Gustafson},
  {Guzman}, {Ha}, {Haegel}, {Hagiwara}, {Haino}, {Halim}, {Hall}, {Hamilton},
  {Hammond}, {Han}, {Haney}, {Hanks}, {Hanna}, {Hannam}, {Hannuksela},
  {Hansen}, {Hansen}, {Hanson}, {Harder}, {Hardwick}, {Haris}, {Harms},
  {Harry}, {Harry}, {Hartwig}, {Hasegawa}, {Haskell}, {Hasskew}, {Haster},
  {Hattori}, {Haughian}, {Hayakawa}, {Hayama}, {Hayes}, {Healy}, {Heidmann},
  {Heidt}, {Heintze}, {Heinze}, {Heinzel}, {Heitmann}, {Hellman}, {Hello},
  {Helmling-Cornell}, {Hemming}, {Hendry}, {Heng}, {Hennes}, {Hennig},
  {Hennig}, {Hernandez}, {Hernandez Vivanco}, {Heurs}, {Hild}, {Hill},
  {Himemoto}, {Hines}, {Hiranuma}, {Hirata}, {Hirose}, {Hochheim}, {Hofman},
  {Hohmann}, {Holcomb}, {Holland}, {Hollows}, {Holmes}, {Holt}, {Holz}, {Hong},
  {Hopkins}, {Hough}, {Hourihane}, {Howell}, {Hoy}, {Hoyland}, {Hreibi},
  {Hsieh}, {Hsu}, {Huang}, {Huang}, {Huang}, {Huang}, {Huang}, {Huang},
  {H{\"u}bner}, {Huddart}, {Hughey}, {Hui}, {Hui}, {Husa}, {Huttner},
  {Huxford}, {Huynh-Dinh}, {Ide}, {Idzkowski}, {Iess}, {Ikenoue}, {Imam},
  {Inayoshi}, {Ingram}, {Inoue}, {Ioka}, {Isi}, {Isleif}, {Ito}, {Itoh},
  {Iyer}, {Izumi}, {JaberianHamedan}, {Jacqmin}, {Jadhav}, {Jadhav}, {James},
  {Jan}, {Jani}, {Janquart}, {Janssens}, {Janthalur}, {Jaranowski}, {Jariwala},
  {Jaume}, {Jenkins}, {Jenner}, {Jeon}, {Jeunon}, {Jia}, {Jin}, {Johns},
  {Jones}, {Jones}, {Jones}, {Jones}, {Jones}, {Jonker}, {Ju}, {Jung}, {Jung},
  {Junker}, {Juste}, {Kaihotsu}, {Kajita}, {Kakizaki}, {Kalaghatgi},
  {Kalogera}, {Kamai}, {Kamiizumi}, {Kanda}, {Kandhasamy}, {Kang}, {Kanner},
  {Kao}, {Kapadia}, {Kapasi}, {Karat}, {Karathanasis}, {Karki}, {Kashyap},
  {Kasprzack}, {Kastaun}, {Katsanevas}, {Katsavounidis}, {Katzman}, {Kaur},
  {Kawabe}, {Kawaguchi}, {Kawai}, {Kawasaki}, {K{\'e}f{\'e}lian}, {Keitel},
  {Key}, {Khadka}, {Khalili}, {Khan}, {Khazanov}, {Khetan}, {Khursheed},
  {Kijbunchoo}, {Kim}, {Kim}, {Kim}, {Kim}, {Kim}, {Kim}, {Kimball}, {Kimura},
  {Kinley-Hanlon}, {Kirchhoff}, {Kissel}, {Kita}, {Kitazawa}, {Kleybolte},
  {Klimenko}, {Knee}, {Knowles}, {Knyazev}, {Koch}, {Koekoek}, {Kojima},
  {Kokeyama}, {Koley}, {Kolitsidou}, {Kolstein}, {Komori}, {Kondrashov},
  {Kong}, {Kontos}, {Koper}, {Korobko}, {Kotake}, {Kovalam}, {Kozak},
  {Kozakai}, {Kozu}, {Kringel}, {Krishnendu}, {Kr{\'o}lak}, {Kuehn}, {Kuei},
  {Kuijer}, {Kumar}, {Kumar}, {Kumar}, {Kumar}, {Kume}, {Kuns}, {Kuo}, {Kuo},
  {Kuromiya}, {Kuroyanagi}, {Kusayanagi}, {Kuwahara}, {Kwak}, {Lagabbe},
  {Laghi}, {Lalande}, {Lam}, {Lamberts}, {Landry}, {Landry}, {Lane}, {Lang},
  {Lange}, {Lantz}, {La Rosa}, {Lartaux-Vollard}, {Lasky}, {Laxen},
  {Lazzarini}, {Lazzaro}, {Leaci}, {Leavey}, {Lecoeuche}, {Lee}, {Lee}, {Lee},
  {Lee}, {Lee}, {Lee}, {Lehmann}, {Lema{\^\i}tre}, {Leonardi}, {Leroy},
  {Letendre}, {Levesque}, {Levin}, {Leviton}, {Leyde}, {Li}, {Li}, {Li}, {Li},
  {Li}, {Li}, {Lin}, {Lin}, {Lin}, {Lin}, {Lin}, {Linde}, {Linker}, {Linley},
  {Littenberg}, {Liu}, {Liu}, {Liu}, {Liu}, {Llamas}, {Llorens-Monteagudo},
  {Lo}, {Lockwood}, {London}, {Longo}, {Lopez}, {Lopez Portilla}, {Lorenzini},
  {Loriette}, {Lormand}, {Losurdo}, {Lott}, {Lough}, {Lousto}, {Lovelace},
  {Lucaccioni}, {L{\"u}ck}, {Lumaca}, {Lundgren}, {Luo}, {Lynam}, {Macas},
  {MacInnis}, {Macleod}, {MacMillan}, {Macquet}, {Maga{\~n}a Hernandez},
  {Magazz{\`u}}, {Magee}, {Maggiore}, {Magnozzi}, {Mahesh}, {Majorana},
  {Makarem}, {Maksimovic}, {Maliakal}, {Malik}, {Man}, {Mandic}, {Mangano},
  {Mango}, {Mansell}, {Manske}, {Mantovani}, {Mapelli}, {Marchesoni},
  {Marchio}, {Marion}, {Mark}, {M{\'a}rka}, {M{\'a}rka}, {Markakis},
  {Markosyan}, {Markowitz}, {Maros}, {Marquina}, {Marsat}, {Martelli},
  {Martin}, {Martin}, {Martinez}, {Martinez}, {Martinez}, {Martinovic},
  {Martynov}, {Marx}, {Masalehdan}, {Mason}, {Massera}, {Masserot},
  {Massinger}, {Masso-Reid}, {Mastrogiovanni}, {Matas}, {Mateu-Lucena},
  {Matichard}, {Matiushechkina}, {Mavalvala}, {McCann}, {McCarthy},
  {McClelland}, {McClincy}, {McCormick}, {McCuller}, {McGhee}, {McGuire},
  {McIsaac}, {McIver}, {McRae}, {McWilliams}, {Meacher}, {Mehmet}, {Mehta},
  {Meijer}, {Melatos}, {Melchor}, {Mendell}, {Menendez-Vazquez}, {Menoni},
  {Mercer}, {Mereni}, {Merfeld}, {Merilh}, {Merritt}, {Merzougui}, {Meshkov},
  {Messenger}, {Messick}, {Meyers}, {Meylahn}, {Mhaske}, {Miani}, {Miao},
  {Michaloliakos}, {Michel}, {Michimura}, {Middleton}, {Milano}, {Miller},
  {Miller}, {Miller}, {Miller}, {Millhouse}, {Mills}, {Milotti}, {Minazzoli},
  {Minenkov}, {Mio}, {Mir}, {Miravet-Ten{\'e}s}, {Mishra}, {Mishra}, {Mistry},
  {Mitra}, {Mitrofanov}, {Mitselmakher}, {Mittleman}, {Miyakawa}, {Miyamoto},
  {Miyazaki}, {Miyo}, {Miyoki}, {Mo}, {Moguel}, {Mogushi}, {Mohapatra},
  {Mohite}, {Molina}, {Molina-Ruiz}, {Mondin}, {Montani}, {Moore}, {Moraru},
  {Morawski}, {More}, {Moreno}, {Moreno}, {Mori}, {Morisaki}, {Moriwaki},
  {Mours}, {Mow-Lowry}, {Mozzon}, {Muciaccia}, {Mukherjee}, {Mukherjee},
  {Mukherjee}, {Mukherjee}, {Mukherjee}, {Mukund}, {Mullavey}, {Munch},
  {Mu{\~n}iz}, {Murray}, {Musenich}, {Muusse}, {Nadji}, {Nagano}, {Nagano},
  {Nagar}, {Nakamura}, {Nakano}, {Nakano}, {Nakashima}, {Nakayama}, {Napolano},
  {Nardecchia}, {Narikawa}, {Naticchioni}, {Nayak}, {Nayak}, {Negishi}, {Neil},
  {Neilson}, {Nelemans}, {Nelson}, {Nery}, {Neubauer}, {Neunzert}, {Ng}, {Ng},
  {Nguyen}, {Nguyen}, {Nguyen}, {Nguyen Quynh}, {Ni}, {Nichols}, {Nishizawa},
  {Nissanke}, {Nitoglia}, {Nocera}, {Norman}, {North}, {Nozaki}, {Nuttall},
  {Oberling}, {O'Brien}, {Obuchi}, {O'Dell}, {Oelker}, {Ogaki}, {Oganesyan},
  {Oh}, {Oh}, {Oh}, {Ohashi}, {Ohishi}, {Ohkawa}, {Ohme}, {Ohta}, {Okada},
  {Okutani}, {Okutomi}, {Olivetto}, {Oohara}, {Ooi}, {Oram}, {O'Reilly},
  {Ormiston}, {Ormsby}, {Ortega}, {O'Shaughnessy}, {O'Shea}, {Oshino},
  {Ossokine}, {Osthelder}, {Otabe}, {Ottaway}, {Overmier}, {Pace}, {Pagano},
  {Page}, {Pagliaroli}, {Pai}, {Pai}, {Palamos}, {Palashov}, {Palomba}, {Pan},
  {Pan}, {Panda}, {Pang}, {Pang}, {Pankow}, {Pannarale}, {Pant}, {Panther},
  {Paoletti}, {Paoli}, {Paolone}, {Parisi}, {Park}, {Park}, {Parker},
  {Pascucci}, {Pasqualetti}, {Passaquieti}, {Passuello}, {Patel}, {Pathak},
  {Patricelli}, {Patron}, {Paul}, {Payne}, {Pedraza}, {Pegoraro}, {Pele},
  {Pe{\~n}a Arellano}, {Penn}, {Perego}, {Pereira}, {Pereira}, {Perez},
  {P{\'e}rigois}, {Perkins}, {Perreca}, {Perri{\`e}s}, {Petermann},
  {Petterson}, {Pfeiffer}, {Pham}, {Phukon}, {Piccinni}, {Pichot},
  {Piendibene}, {Piergiovanni}, {Pierini}, {Pierro}, {Pillant}, {Pillas},
  {Pilo}, {Pinard}, {Pinto}, {Pinto}, {Piotrzkowski}, {Pirello}, {Pitkin},
  {Placidi}, {Planas}, {Plastino}, {Pluchar}, {Poggiani}, {Polini}, {Pong},
  {Ponrathnam}, {Popolizio}, {Porter}, {Poulton}, {Powell}, {Pracchia},
  {Pradier}, {Prajapati}, {Prasai}, {Prasanna}, {Pratten}, {Principe}, {Prodi},
  {Prokhorov}, {Prosposito}, {Prudenzi}, {Puecher}, {Punturo}, {Puosi},
  {Puppo}, {P{\"u}rrer}, {Qi}, {Quetschke}, {Quitzow-James}, {Raab},
  {Raaijmakers}, {Radkins}, {Radulesco}, {Raffai}, {Rail}, {Raja}, {Rajan},
  {Ramirez}, {Ramirez}, {Ramos-Buades}, {Rana}, {Rapagnani}, {Rapol}, {Ray},
  {Raymond}, {Raza}, {Razzano}, {Read}, {Rees}, {Regimbau}, {Rei}, {Reid},
  {Reid}, {Reitze}, {Relton}, {Renzini}, {Rettegno}, {Rezac}, {Ricci},
  {Richards}, {Richardson}, {Richardson}, {Riemenschneider}, {Riles},
  {Rinaldi}, {Rink}, {Rizzo}, {Robertson}, {Robie}, {Robinet}, {Rocchi},
  {Rodriguez}, {Rolland}, {Rollins}, {Romanelli}, {Romano}, {Romel},
  {Romero-Rodr{\'\i}guez}, {Romero-Shaw}, {Romie}, {Ronchini}, {Rosa}, {Rose},
  {Rosi{\'n}ska}, {Ross}, {Rowan}, {Rowlinson}, {Roy}, {Roy}, {Roy}, {Rozza},
  {Ruggi}, {Ryan}, {Sachdev}, {Sadecki}, {Sadiq}, {Sago}, {Saito}, {Saito},
  {Sakai}, {Sakai}, {Sakellariadou}, {Sakuno}, {Salafia}, {Salconi}, {Saleem},
  {Salemi}, {Samajdar}, {Sanchez}, {Sanchez}, {Sanchez}, {Sanchis-Gual},
  {Sanders}, {Sanuy}, {Saravanan}, {Sarin}, {Sassolas}, {Satari},
  {Sathyaprakash}, {Sato}, {Sato}, {Sauter}, {Savage}, {Sawada}, {Sawant},
  {Sawant}, {Sayah}, {Schaetzl}, {Scheel}, {Scheuer}, {Schiworski}, {Schmidt},
  {Schmidt}, {Schnabel}, {Schneewind}, {Schofield}, {Sch{\"o}nbeck}, {Schulte},
  {Schutz}, {Schwartz}, {Scott}, {Scott}, {Seglar-Arroyo}, {Sekiguchi},
  {Sekiguchi}, {Sellers}, {Sengupta}, {Sentenac}, {Seo}, {Sequino}, {Sergeev},
  {Setyawati}, {Shaffer}, {Shahriar}, {Shams}, {Shao}, {Sharma}, {Sharma},
  {Shawhan}, {Shcheblanov}, {Shibagaki}, {Shikauchi}, {Shimizu}, {Shimoda},
  {Shimode}, {Shinkai}, {Shishido}, {Shoda}, {Shoemaker}, {Shoemaker},
  {ShyamSundar}, {Sieniawska}, {Sigg}, {Singer}, {Singh}, {Singh}, {Singha},
  {Sintes}, {Sipala}, {Skliris}, {Slagmolen}, {Slaven-Blair}, {Smetana},
  {Smith}, {Smith}, {Soldateschi}, {Somala}, {Somiya}, {Son}, {Soni}, {Soni},
  {Sordini}, {Sorrentino}, {Sorrentino}, {Sotani}, {Soulard}, {Souradeep},
  {Sowell}, {Spagnuolo}, {Spencer}, {Spera}, {Srinivasan}, {Srivastava},
  {Srivastava}, {Staats}, {Stachie}, {Steer}, {Steinlechner}, {Steinlechner},
  {Stops}, {Stover}, {Strain}, {Strang}, {Stratta}, {Strunk}, {Sturani},
  {Stuver}, {Sudhagar}, {Sudhir}, {Sugimoto}, {Suh}, {Summerscales}, {Sun},
  {Sun}, {Sunil}, {Sur}, {Suresh}, {Sutton}, {Suzuki}, {Suzuki}, {Swinkels},
  {Szczepa{\'n}czyk}, {Szewczyk}, {Tacca}, {Tagoshi}, {Tait}, {Takahashi},
  {Takahashi}, {Takamori}, {Takano}, {Takeda}, {Takeda}, {Talbot}, {Talbot},
  {Tanaka}, {Tanaka}, {Tanaka}, {Tanaka}, {Tanaka}, {Tanasijczuk}, {Tanioka},
  {Tanner}, {Tao}, {Tao}, {Tapia San Mart{\'\i}n}, {Taranto}, {Tasson},
  {Telada}, {Tenorio}, {Terhune}, {Terkowski}, {Thirugnanasambandam}, {Thomas},
  {Thomas}, {Thompson}, {Thondapu}, {Thorne}, {Thrane}, {Tiwari}, {Tiwari},
  {Tiwari}, {Toivonen}, {Toland}, {Tolley}, {Tomaru}, {Tomigami}, {Tomura},
  {Tonelli}, {Torres-Forn{\'e}}, {Torrie}, {Tosta e Melo}, {T{\"o}yr{\"a}},
  {Trapananti}, {Travasso}, {Traylor}, {Trevor}, {Tringali}, {Tripathee},
  {Troiano}, {Trovato}, {Trozzo}, {Trudeau}, {Tsai}, {Tsai}, {Tsang}, {Tsang},
  {Tsao}, {Tse}, {Tso}, {Tsubono}, {Tsuchida}, {Tsukada}, {Tsuna}, {Tsutsui},
  {Tsuzuki}, {Turbang}, {Turconi}, {Tuyenbayev}, {Ubhi}, {Uchikata},
  {Uchiyama}, {Udall}, {Ueda}, {Uehara}, {Ueno}, {Ueshima}, {Unnikrishnan},
  {Uraguchi}, {Urban}, {Ushiba}, {Utina}, {Vahlbruch}, {Vajente}, {Vajpeyi},
  {Valdes}, {Valentini}, {Valsan}, {van Bakel}, {van Beuzekom}, {van den
  Brand}, {Van Den Broeck}, {Vander-Hyde}, {van der Schaaf}, {van Heijningen},
  {Vanosky}, {van Putten}, {van Remortel}, {Vardaro}, {Vargas}, {Varma},
  {Vas{\'u}th}, {Vecchio}, {Vedovato}, {Veitch}, {Veitch}, {Venneberg},
  {Venugopalan}, {Verkindt}, {Verma}, {Verma}, {Veske}, {Vetrano},
  {Vicer{\'e}}, {Vidyant}, {Viets}, {Vijaykumar}, {Villa-Ortega}, {Vinet},
  {Virtuoso}, {Vitale}, {Vo}, {Vocca}, {von Reis}, {von Wrangel}, {Vorvick},
  {Vyatchanin}, {Wade}, {Wade}, {Wagner}, {Walet}, {Walker}, {Wallace},
  {Wallace}, {Walsh}, {Wang}, {Wang}, {Wang}, {Ward}, {Warner}, {Was},
  {Washimi}, {Washington}, {Watchi}, {Weaver}, {Webster}, {Weinert},
  {Weinstein}, {Weiss}, {Weller}, {Wellmann}, {Wen}, {We{\ss}els}, {Wette},
  {Whelan}, {White}, {Whiting}, {Whittle}, {Wilken}, {Williams}, {Williams},
  {Williamson}, {Willis}, {Willke}, {Wilson}, {Winkler}, {Wipf}, {Wlodarczyk},
  {Woan}, {Woehler}, {Wofford}, {Wong}, {Wu}, {Wu}, {Wu}, {Wu}, {Wysocki},
  {Xiao}, {Xu}, {Yamada}, {Yamamoto}, {Yamamoto}, {Yamamoto}, {Yamamoto},
  {Yamashita}, {Yamazaki}, {Yang}, {Yang}, {Yang}, {Yang}, {Yang}, {Yap},
  {Yeeles}, {Yelikar}, {Ying}, {Yokogawa}, {Yokoyama}, {Yokozawa}, {Yoo},
  {Yoshioka}, {Yu}, {Yu}, {Yuzurihara}, {Zadro{\.z}ny}, {Zanolin}, {Zeidler},
  {Zelenova}, {Zendri}, {Zevin}, {Zhan}, {Zhang}, {Zhang}, {Zhang}, {Zhang},
  {Zhang}, {Zhao}, {Zhao}, {Zhao}, {Zhao}, {Zhou}, {Zhou}, {Zhu}, {Zhu},
  {Zimmerman}, {Zlochower}, {Zucker}, \& {Zweizig}}]{2021arXiv211103634T}
{The LIGO Scientific Collaboration}, {the Virgo Collaboration}, {the KAGRA
  Collaboration}, {et~al.} 2021{\natexlab{b}}, arXiv e-prints, arXiv:2111.03634

\bibitem[{{Timmes} {et~al.}(1996){Timmes}, {Woosley}, \&
  {Weaver}}]{1996ApJ...457..834T}
{Timmes}, F.~X., {Woosley}, S.~E., \& {Weaver}, T.~A. 1996, \apj, 457, 834

\bibitem[{{Tiwari}(2022)}]{2022ApJ...928..155T}
{Tiwari}, V. 2022, \apj, 928, 155

\bibitem[{{Tiwari} \& {Fairhurst}(2021)}]{2021ApJ...913L..19T}
{Tiwari}, V. \& {Fairhurst}, S. 2021, \apjl, 913, L19

\bibitem[{{Ugliano} {et~al.}(2012){Ugliano}, {Janka}, {Marek}, \&
  {Arcones}}]{2012ApJ...757...69U}
{Ugliano}, M., {Janka}, H.-T., {Marek}, A., \& {Arcones}, A. 2012, \apj, 757,
  69

\bibitem[{{Urushibata} {et~al.}(2018){Urushibata}, {Takahashi}, {Umeda}, \&
  {Yoshida}}]{2018MNRAS.473L.101U}
{Urushibata}, T., {Takahashi}, K., {Umeda}, H., \& {Yoshida}, T. 2018, \mnras,
  473, L101

\bibitem[{{Vanbeveren} {et~al.}(2013){Vanbeveren}, {Mennekens}, {Van
  Rensbergen}, \& {De Loore}}]{2013A&A...552A.105V}
{Vanbeveren}, D., {Mennekens}, N., {Van Rensbergen}, W., \& {De Loore}, C.
  2013, \aap, 552, A105

\bibitem[{{Vigna-G{\'o}mez} {et~al.}(2019){Vigna-G{\'o}mez}, {Justham},
  {Mandel}, {de Mink}, \& {Podsiadlowski}}]{2019ApJ...876L..29V}
{Vigna-G{\'o}mez}, A., {Justham}, S., {Mandel}, I., {de Mink}, S.~E., \&
  {Podsiadlowski}, P. 2019, \apjl, 876, L29

\bibitem[{{Vigna-G{\'o}mez} {et~al.}(2021){Vigna-G{\'o}mez}, {Toonen},
  {Ramirez-Ruiz}, {Leigh}, {Riley}, \& {Haster}}]{2021ApJ...907L..19V}
{Vigna-G{\'o}mez}, A., {Toonen}, S., {Ramirez-Ruiz}, E., {et~al.} 2021, \apjl,
  907, L19

\bibitem[{{Vink} \& {de Koter}(2005)}]{2005A&A...442..587V}
{Vink}, J.~S. \& {de Koter}, A. 2005, \aap, 442, 587

\bibitem[{{Vink} {et~al.}(2000){Vink}, {de Koter}, \&
  {Lamers}}]{2000A&A...362..295V}
{Vink}, J.~S., {de Koter}, A., \& {Lamers}, H.~J.~G.~L.~M. 2000, \aap, 362, 295

\bibitem[{{Vink} {et~al.}(2001){Vink}, {de Koter}, \&
  {Lamers}}]{2001A&A...369..574V}
{Vink}, J.~S., {de Koter}, A., \& {Lamers}, H.~J.~G.~L.~M. 2001, \aap, 369, 574

\bibitem[{{Vink} {et~al.}(2021){Vink}, {Higgins}, {Sander}, \&
  {Sabhahit}}]{2021MNRAS.504..146V}
{Vink}, J.~S., {Higgins}, E.~R., {Sander}, A. A.~C., \& {Sabhahit}, G.~N. 2021,
  \mnras, 504, 146

\bibitem[{{Virtanen} {et~al.}(2020){Virtanen}, {Gommers}, {Oliphant},
  {Haberland}, {Reddy}, {Cournapeau}, {Burovski}, {Peterson}, {Weckesser},
  {Bright}, {van der Walt}, {Brett}, {Wilson}, {Millman}, {Mayorov}, {Nelson},
  {Jones}, {Kern}, {Larson}, {Carey}, {Polat}, {Feng}, {Moore}, {Vand erPlas},
  {Laxalde}, {Perktold}, {Cimrman}, {Henriksen}, {Quintero}, {Harris},
  {Archibald}, {Ribeiro}, {Pedregosa}, {van Mulbregt}, \& {SciPy 1. 0
  Contributors}}]{2020NatMe..17..261V}
{Virtanen}, P., {Gommers}, R., {Oliphant}, T.~E., {et~al.} 2020, Nature
  Methods, 17, 261

\bibitem[{{Wade} {et~al.}(2014){Wade}, {Grunhut}, {Alecian}, {Neiner},
  {Auri{\`e}re}, {Bohlender}, {David-Uraz}, {Folsom}, {Henrichs}, {Kochukhov},
  {Mathis}, {Owocki}, {Petit}, \& {Petit}}]{2014IAUS..302..265W}
{Wade}, G.~A., {Grunhut}, J., {Alecian}, E., {et~al.} 2014, in IAU Symposium,
  Vol. 302, IAU Symposium, 265--269

\bibitem[{{Wang} {et~al.}(2020){Wang}, {Langer}, {Schootemeijer}, {Castro},
  {Adscheid}, {Marchant}, \& {Hastings}}]{2020ApJ...888L..12W}
{Wang}, C., {Langer}, N., {Schootemeijer}, A., {et~al.} 2020, \apjl, 888, L12

\bibitem[{{Weis} \& {Bomans}(2020)}]{2020Galax...8...20W}
{Weis}, K. \& {Bomans}, D.~J. 2020, Galaxies, 8, 20

\bibitem[{{Wellstein} \& {Langer}(1999)}]{1999A&A...350..148W}
{Wellstein}, S. \& {Langer}, N. 1999, \aap, 350, 148

\bibitem[{{Wellstein} {et~al.}(2001){Wellstein}, {Langer}, \&
  {Braun}}]{2001A&A...369..939W}
{Wellstein}, S., {Langer}, N., \& {Braun}, H. 2001, \aap, 369, 939

\bibitem[{{Woosley}(1988)}]{1988ApJ...330..218W}
{Woosley}, S.~E. 1988, \apj, 330, 218

\bibitem[{{Woosley}(2017)}]{2017ApJ...836..244W}
{Woosley}, S.~E. 2017, \apj, 836, 244

\bibitem[{{Woosley}(2019)}]{2019ApJ...878...49W}
{Woosley}, S.~E. 2019, \apj, 878, 49

\bibitem[{{Woosley} {et~al.}(2002){Woosley}, {Heger}, \&
  {Weaver}}]{2002RvMP...74.1015W}
{Woosley}, S.~E., {Heger}, A., \& {Weaver}, T.~A. 2002, Reviews of Modern
  Physics, 74, 1015

\bibitem[{{Woosley} {et~al.}(1993){Woosley}, {Langer}, \&
  {Weaver}}]{1993ApJ...411..823W}
{Woosley}, S.~E., {Langer}, N., \& {Weaver}, T.~A. 1993, \apj, 411, 823

\bibitem[{{Xu} {et~al.}(2013){Xu}, {Takahashi}, {Goriely}, {Arnould}, {Ohta},
  \& {Utsunomiya}}]{2013NuPhA.918...61X}
{Xu}, Y., {Takahashi}, K., {Goriely}, S., {et~al.} 2013, \nphysa, 918, 61

\bibitem[{{Zapartas} {et~al.}(2019){Zapartas}, {de Mink}, {Justham}, {Smith},
  {de Koter}, {Renzo}, {Arcavi}, {Farmer}, {G{\"o}tberg}, \&
  {Toonen}}]{2019A&A...631A...5Z}
{Zapartas}, E., {de Mink}, S.~E., {Justham}, S., {et~al.} 2019, \aap, 631, A5

\bibitem[{{Zapartas} {et~al.}(2021){Zapartas}, {de Mink}, {Justham}, {Smith},
  {Renzo}, \& {de Koter}}]{2021A&A...645A...6Z}
{Zapartas}, E., {de Mink}, S.~E., {Justham}, S., {et~al.} 2021, \aap, 645, A6

\end{thebibliography}
 \newcommand{\noop}[1]{}

\clearpage
\begin{appendix}

\section{Supplemental information}

\begin{figure}
    \centering
    \includegraphics[width=\linewidth]{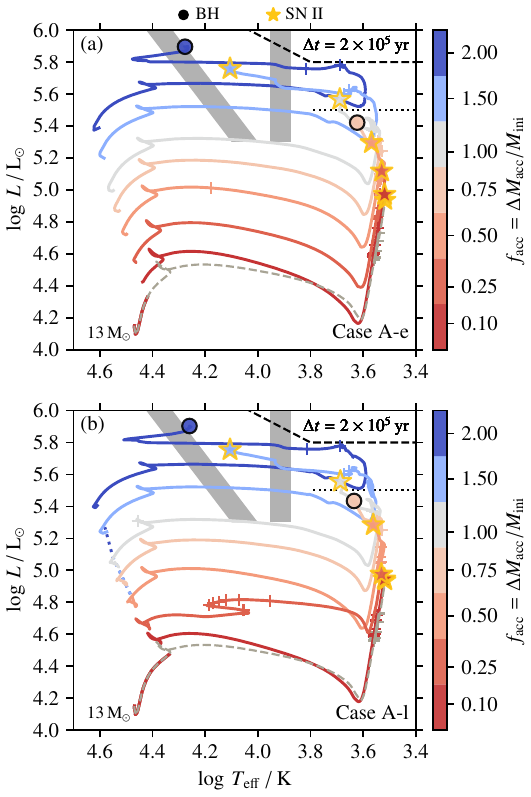}
    \caption{Same as Fig.~\ref{fig:hrd-case-be-m13} but for early (panel a) and late (panel b) \casea accretors with initially $13\,\msun$.}
    \label{fig:hrd-case-ae-al-m13}
\end{figure}

\begin{figure}
    \centering
    \includegraphics[width=\linewidth]{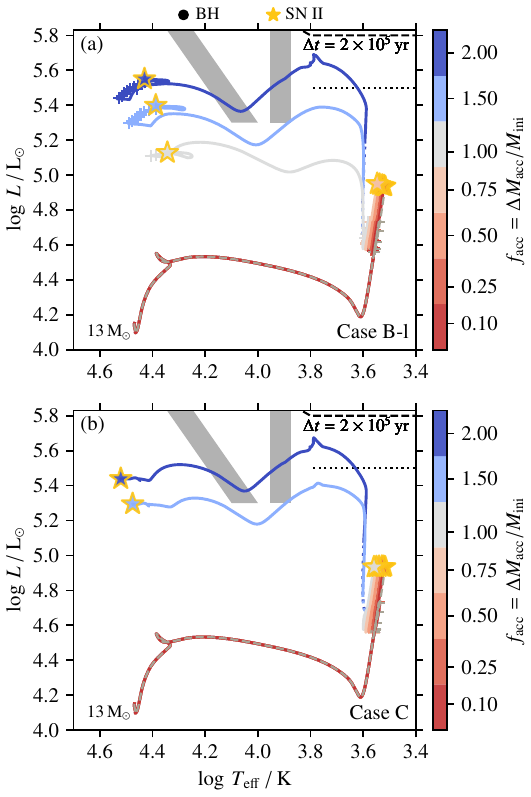}
    \caption{Same as Fig.~\ref{fig:hrd-case-be-m13} but for late \caseb (panel a) and \casec (panel b) accretors with initially $13\,\msun$.}
    \label{fig:hrd-case-bl-c-m13}
\end{figure}

\begin{figure*}
    \centering
    \includegraphics[width=\linewidth]{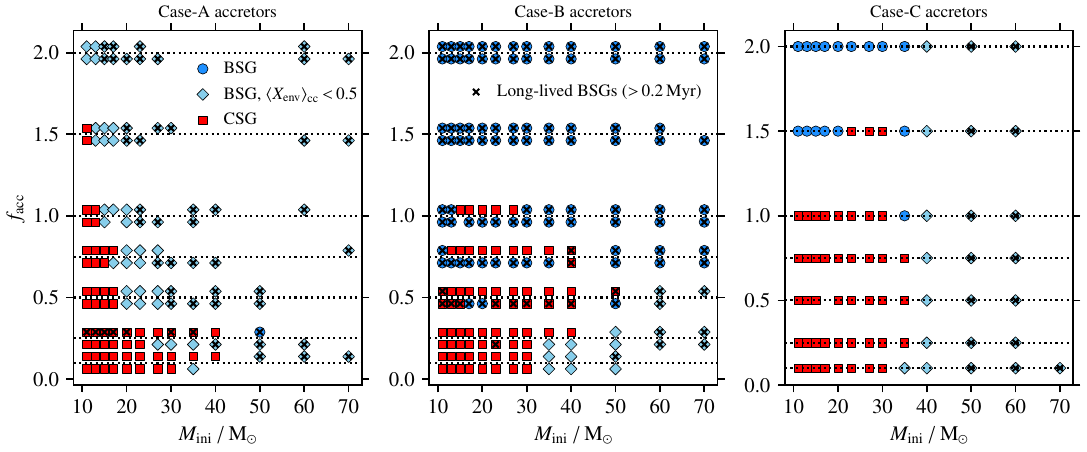}
    \caption{Endstages of the accretor models as BSGs and CSGs in the grid spanned by the initial mass $M_\mathrm{ini}$ and the accretion fraction $f_\mathrm{acc}$. Supergiants that are hot (blue; BSGs) because of a high surface helium mass fraction ($\left<X_\mathrm{env}\right>_\mathrm{cc}<0.5$) are shown separately. Cross symbols indicate models that burn helium in their cores for more than $0.2\myr$ at effective temperatures of $\log T_\mathrm{eff}/K<3.9$, \ie as BSGs, and the three panels are for \casea, B and~C accretors. Early \casea and-B accretors are offset to lower values along their respective $f_\mathrm{acc}$ indicated by the black dotted lines while late \casea and-B models are offset to slightly higher values. This overview of the model grid visually underlines two key results presented in Sect.~\ref{sec:pre-sn-evolution-and-structure}. Firstly, there are models that burn helium in their cores as BSGs but explode as CSGs while other models also explode as BSGs and never become CSGs because of the accretion. Secondly, the need for larger $f_\mathrm{acc}$ in more evolved pre-accretors to turn them into long-lived BSGs is illustrated clearly. The threshold $f_\mathrm{acc}$ values to turn stars into long-lived BSGs is significantly lower in merger models that also account for mixing of core material into the envelope such that the values here should be understood as upper limits.}
    \label{fig:mini-facc-overview}
\end{figure*}

\begin{figure*}
    \centering
    \includegraphics[width=0.9\textwidth]{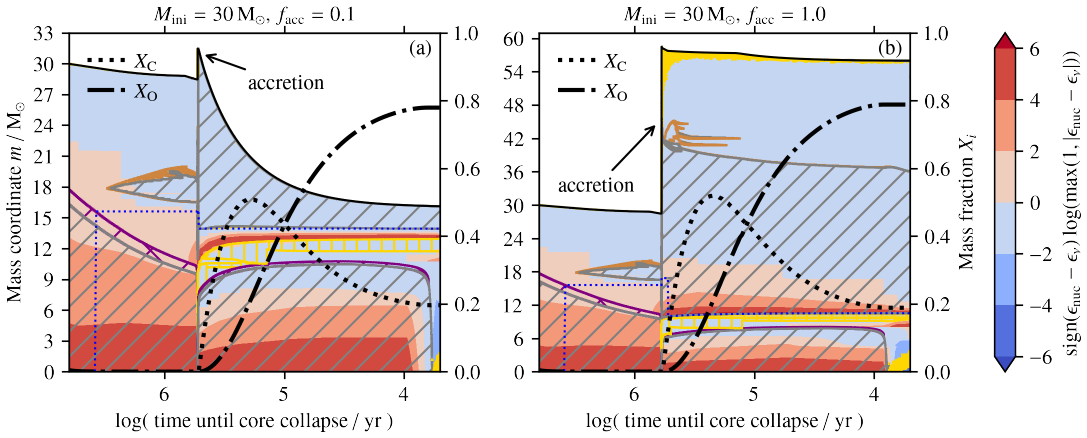}
    \caption{Same as Fig.~\ref{fig:kipp-comparison-m13} but for initially $30\,\msun$ stars accreting (a) 10\% and (b) 100\% of their initial mass as early \caseb systems.}
    \label{fig:kipp-comparison-m30}
\end{figure*}

\begin{figure*}
    \centering
    \includegraphics{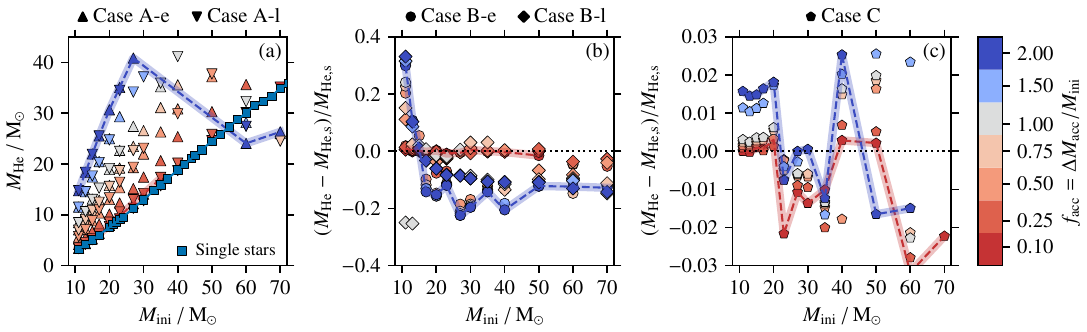}
    \caption{Same as Fig.~\ref{fig:MCO-vs-Mini} but for the helium core masses $M_\mathrm{He}$.}
    \label{fig:MHe-vs-Mini}
\end{figure*}

\begin{figure*}
    \centering
    \includegraphics{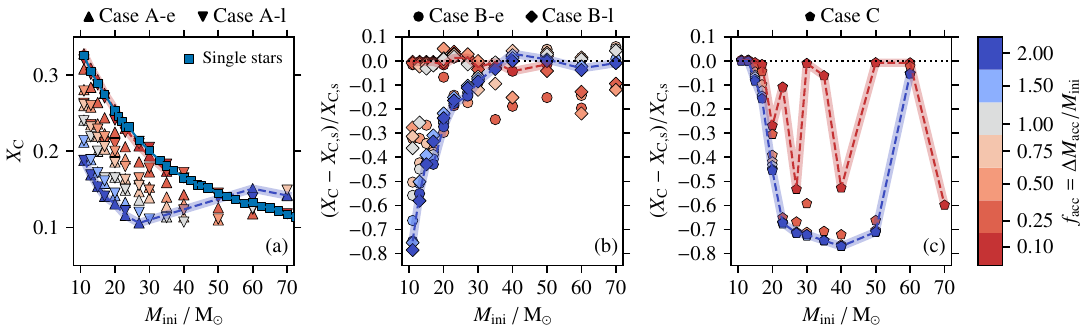}
    \caption{Same as Fig.~\ref{fig:XC-vs-MCO} but as a function of the initial mass $M_\mathrm{ini}$.}
    \label{fig:XC-vs-Mini}
\end{figure*}

Complementary HR diagrams to Fig.~\ref{fig:hrd-case-be-m13} of accreting stars with an initial mass of $13\,\msun$ for early and late \casea, and late \caseb and \casec are shown in Figs.~\ref{fig:hrd-case-ae-al-m13} and~\ref{fig:hrd-case-bl-c-m13}, respectively. An overview of the final fates of the models in our grid is provided in Fig.~\ref{fig:mini-facc-overview}. An additional Kippenhahn diagram for initially $30\,\msun$ accretors can be found in Fig.~\ref{fig:kipp-comparison-m30}. The helium core masses and central carbon mass fractions after core-helium burning are shown in Figs.~\ref{fig:MHe-vs-Mini} and~\ref{fig:XC-vs-Mini} as a function of initial mass. A table with key properties of all our accretor models is provided in Tab.~\ref{tab:models}.

\begin{figure*}
    \centering
    \includegraphics[width=\linewidth]{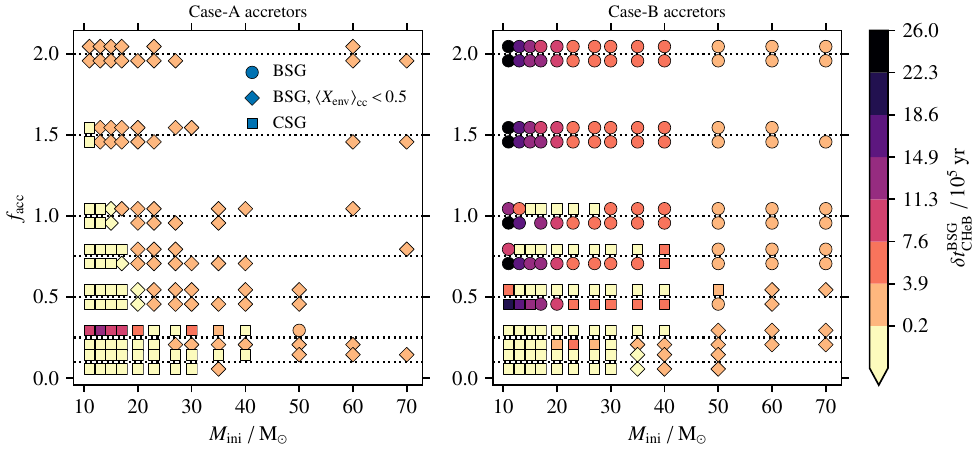}
    \caption{Similar to Fig.~\ref{fig:mini-facc-overview}, but showing the time $\delta t_\mathrm{CHeB}^\mathrm{BSG}$ spend as BSGs at $\log T_\mathrm{eff}\,/\,\mathrm{K}>3.9$ during core helium burning of our \casea and~B accretors in the $M_\mathrm{ini}$--$f_\mathrm{acc}$ plane (\cf Table~\ref{tab:models}). By definition, \casec accretors have finished core helium burning before the accretion event and are thus not shown here. The core-helium-burning BSG lifetimes $\delta t_\mathrm{CHeB}^\mathrm{BSG}$ naturally scale with the total mass of stars (lower mass stars having longer lifetimes) and \casebe accretors are more easily converted into long-lived BSGs as described in the main text (see Sect.~\ref{sec:bsg-structures}).}
    \label{fig:mini-facc-bsg-lifetime}
\end{figure*}

\onecolumn
\begin{landscape}
\tabcolsep=0.10cm
\begin{ThreePartTable}
    \renewcommand{\LTcapwidth}{\linewidth}
\begin{longtable}{ccccccccccccccccccccccc}
        \caption{\label{tab:models}Properties of the single and accretor models. The data cover the initial mass $M_\mathrm{ini}$, the model case (single or accretor; in the case of the latter, indication of evolutionary state of the accretor), the fraction $f_\mathrm{acc}$ of accreted mass in terms of $M_\mathrm{ini}$, the lifetime $t_\mathrm{cc}$ until core collapse, the final mass $M_\mathrm{final}$ of the models, the He, CO and iron core masses $M_\mathrm{He}$, $M_\mathrm{CO}$ and $M_\mathrm{Fe}$, respectively, at core collapse, the compactness $\xi_\mathrm{2.5}$, the radial mass derivative $\mu_4$ and mass $M_4$ at a specific entropy of $s=4$, the central entropy $s_\mathrm{c}$ at core collapse, the SN type (BH indicates direct collapse to BH), the gravitational compact-object remnant mass $M_\mathrm{rm}$, the SN ejecta mass $M_\mathrm{ej}$, a flag indicating the likely occurrence of SN fallback, the luminosity $\log\,L_\mathrm{cc}$, effective temperature $T_\mathrm{eff,cc}$ and radius $R_\mathrm{*}$ at core collapse, average hydrogen mass fraction $\left<X_\mathrm{env}\right>_\mathrm{cc}$ in the hydrogen-rich envelope at core collapse, time $\delta t_\mathrm{CHeB}^\mathrm{BSG}$ spent as a BSG at $\log T_\mathrm{eff}/\mathrm{K}>3.9$ during core-helium burning, and the time $\delta t_\mathrm{LBV}$ spent in the LBV-regime in the HR diagram.}\\
        \toprule
        $M_\mathrm{ini}$ & Case & $f_\mathrm{acc}$ & $t_\mathrm{cc}$ & $M_\mathrm{final}$ & $M_\mathrm{He}$ & $M_\mathrm{CO}$ & $M_\mathrm{Fe}$ & $\xi_\mathrm{2.5}$ & $\mu_\mathrm{4}$ & $M_\mathrm{4}$ & $s_\mathrm{c}$ & SN & $M_\mathrm{rm}$ & $M_\mathrm{ej}$ & Fall- & $\log\,L_\mathrm{cc}$ & $\log\,T_\mathrm{eff,cc}$ & $R_*$ & $\left<X_\mathrm{env}\right>_\mathrm{cc}$ & $\delta t_\mathrm{CHeB}^\mathrm{BSG}$ & $\delta t_\mathrm{LBV}$ \\
        ($\msun$) &  &  & (Myr) & ($\msun$) & ($\msun$) & ($\msun$) & ($\msun$) &  &  &  & ($N_\mathrm{A} k_\mathrm{B}$) & Type & ($\msun$) & ($\msun$) & back? & ($\log\,\lsun$) & ($\log\,\mathrm{K}$) & ($\rsun$) &  & $10^5\,\mathrm{yr}$ & $10^4\,\mathrm{yr}$ \\
        \midrule
        \endfirsthead
        \caption{continued.}\\
        \toprule
        $M_\mathrm{ini}$ & Case & $f_\mathrm{acc}$ & $t_\mathrm{cc}$ & $M_\mathrm{final}$ & $M_\mathrm{He}$ & $M_\mathrm{CO}$ & $M_\mathrm{Fe}$ & $\xi_\mathrm{2.5}$ & $\mu_\mathrm{4}$ & $M_\mathrm{4}$ & $s_\mathrm{c}$ & SN & $M_\mathrm{rm}$ & $M_\mathrm{ej}$ & Fall- & $\log\,L_\mathrm{cc}$ & $\log\,T_\mathrm{eff,cc}$ & $R_*$ & $\left<X_\mathrm{env}\right>_\mathrm{cc}$ & $\delta t_\mathrm{CHeB}^\mathrm{BSG}$ & $\delta t_\mathrm{LBV}$ \\
        ($\msun$) &  &  & (Myr) & ($\msun$) & ($\msun$) & ($\msun$) & ($\msun$) &  &  &  & ($N_\mathrm{A} k_\mathrm{B}$) & Type & ($\msun$) & ($\msun$) & back? & ($\log\,\lsun$) & ($\log\,\mathrm{K}$) & ($\rsun$) &  & $10^5\,\mathrm{yr}$ & $10^4\,\mathrm{yr}$ \\
        \midrule
        \endhead
        \bottomrule
        \endfoot
         11.0 & Single & -- & 22.4 & 10.0 & 3.2 & 2.0 & 1.51 & 0.01 & 0.02 & 1.53 & 0.78 & II & 1.30 & 8.6 & 0 & 4.76 & 3.52 & 718 & 0.68& 0.0 & 0.0 \\
\midrule
         11.0 & \caseae & 0.10 & 21.6 & 11.0 & 3.4 & 2.1 & 1.50 & 0.02 & 0.04 & 1.55 & 0.81 & II & 1.33 & 9.6 & 0 & 4.80 & 3.52 & 757 & 0.69& 0.0 & 0.0 \\
         11.0 & \caseae & 0.25 & 21.6 & 12.0 & 4.2 & 2.8 & 1.57 & 0.10 & 0.05 & 1.65 & 0.86 & II & 1.42 & 10.4 & 0 & 4.95 & 3.52 & 912 & 0.69& 0.0 & 0.0 \\
         11.0 & \caseae & 0.50 & 21.5 & 13.0 & 5.6 & 3.9 & 1.59 & 0.14 & 0.05 & 1.71 & 0.90 & II & 1.45 & 11.3 & 0 & 5.12 & 3.53 & 1059 & 0.68& 0.0 & 0.0 \\
         11.0 & \caseae & 0.75 & 21.0 & 13.3 & 7.1 & 5.3 & 1.67 & 0.25 & 0.12 & 1.66 & 0.91 & II & 1.46 & 11.6 & 0 & 5.28 & 3.57 & 1076 & 0.66& 0.1 & 0.0 \\
         11.0 & \caseae & 1.00 & 20.6 & 13.1 & 8.5 & 6.6 & 2.09 & 0.65 & 0.22 & 2.46 & 1.14 & BH & 13.1 & 0.0 & 0 & 5.40 & 3.61 & 1013 & 0.64& 0.1 & 0.0 \\
         11.0 & \caseae & 1.50 & 20.1 & 14.1 & 11.5 & 9.3 & 1.87 & 0.22 & 0.09 & 1.88 & 0.95 & II & 1.59 & 12.3 & 0 & 5.62 & 3.73 & 758 & 0.56& 0.1 & 1.4 \\
         11.0 & \caseae & 2.00 & 19.6 & 16.0 & 14.8 & 12.4 & 1.96 & 0.37 & 0.11 & 2.04 & 1.05 & II & 1.73 & 14.0 & 0 & 5.76 & 4.12 & 148 & 0.32& 0.5 & 48.7 \\
\end{longtable}
    \end{longtable}
\end{ThreePartTable}
\end{landscape}
\twocolumn

\end{appendix}

\end{document}